\documentclass[10pt,leqno]{amsart}

\usepackage{amsmath,amssymb,amsthm}
\usepackage{mathtools}
\usepackage{bm}

\usepackage{graphicx}
\usepackage[subrefformat=parens,labelformat=parens]{subfig}
\usepackage{booktabs}
\usepackage{multirow}
\usepackage{tabularx}
\usepackage{array}
\usepackage{dcolumn}

\usepackage{algorithm}
\usepackage{algpseudocode}

\usepackage{xcolor}
\usepackage{siunitx}
\usepackage{enumitem}
\usepackage{nicefrac}
\usepackage{comment}
\usepackage{csquotes}
\usepackage{url}

\usepackage[numbers]{natbib}

\usepackage[mathlines]{lineno}

\usepackage[
    colorlinks=true,
    linkcolor=black,
    citecolor=black,
    urlcolor=blue
]{hyperref}

\newcommand{\bb}[1]{\mathbf{#1}}
\newcommand{\bs}[1]{\boldsymbol{#1}}

\newcommand{\Comsol}{COMSOL Multiphysics\textsuperscript{\textregistered}}

\begin{document}

\title{Analytical Prediction of Buckling Limits in Isotropic Pentamode
Lattices} %%%%%%%%%%%%
% \author{Marco Verbicaro}
% \address{Department of Mechanical Engineering, Politecnico di Milano,
% Via La Masa 1, 20156 Milano, Italy}

% \author{Francesco Braghin}
% \address{Department of Mechanical Engineering, Politecnico di Milano,
% Via La Masa 1, 20156 Milano, Italy}

% \author{Richard V. Craster}
% \address{Departments of Mathematics and Mechanical Engineering,
% Imperial College London, South Kensington, London SW7 2AZ, UK}
\author{Marco Verbicaro$^{*}$}
\author{Francesco Braghin$^{*}$}
\author{Richard V. Craster$^{\dagger}$}

\address{$^{*}$Department of Mechanical Engineering,
Politecnico di Milano, Italy}
\address{$^{\dagger}$Departments of Mathematics and Mechanical Engineering,
Imperial College London, UK}

\email{marco.verbicaro@polimi.it}

\date{\today}

\begin{abstract}
Pentamode lattices are architected materials characterized by an elastic response dominated by a single deformation mode. 
The low stiffness of the other deformation modes, however, may result in limited stability under compressive prestress, thereby limiting their mechanical performance. 
In this work, we develop a general analytical framework to determine the buckling limits of prestressed 2D and 3D isotropic pentamode lattices with generic link geometries.
Two distinct instability mechanisms are considered: a global instability associated with the loss of stability of the effective medium, and a local instability associated with finite-wavelength modes at the scale of the lattice cell. 
Closed-form expressions for the corresponding critical pressures are derived, and the buckling limit is obtained as the lower of the two.
The analytical predictions are validated through numerical simulations of different PM lattices, showing good agreement over the configurations considered.
The results reveal a trade-off between approaching the pentamode limit and maintaining sufficient buckling resistance, 
but also show that appropriate link designs can substantially increase the buckling limit while retaining the desired pentamode behavior. 
\end{abstract} %%%%%%%%%

\maketitle

\bigskip

\section{Introduction}
\label{sec:intro}

In recent decades, architected materials have emerged as a promising class of materials whose effective mechanical properties can be tailored through the design of their underlying microstructure. 
Pentamode (PM) lattices, first introduced by Milton \textit{et~al.} \cite{Milton1995WhichRealizable}, are a particular type of architected materials characterized by five (\textit{penta}) out of six easy modes of deformation, leading to their designation as "meta-fluids". This feature has made them attractive for advanced acoustic applications including wave guiding \cite{Chen2022BroadbandMetamaterials, Guo2017TopologicalStructures}, cloaking \cite{Quadrelli2021ExperimentalAcoustics, Buckmann2014AnMetamaterials, Norris2015AcousticCloaking}, and lenses \cite{Su2017BroadbandLens}, and has subsequently motivated their use in vibration and noise reduction \cite{Fraternali2018InnovativeReview, Zou2023BandgapSide} and lightweight structural applications \cite{Ocak2026AApplications} among others.

Investigations into their effective structural behavior have been carried out using different homogenization procedures. 
Gibson and Ashby \cite{Gibson1982TheMaterials} reported an early result in this regard for a two-dimensional (2D) hexagonal honeycomb, which can be considered the simplest 2D PM,
\footnote{Strictly speaking, the term \textit{pentamode} is not appropriate in 2D. Indeed, there are two easy modes of deformation out of three possible modes; hence \textit{bimode} is more appropriate \cite{Milton1995WhichRealizable}. However, both 2D bimodes and 3D pentamodes are characterized by a single non-zero eigenvalue. Because of this analogy, 2D bimodes are commonly referred to as 2D pentamodes in the literature. In this work, we adopt this convention and refer to bimodes as 2D pentamodes.}
while later works studied different three-dimensional (3D) elastic networks, including PMs \cite{Norris2014MechanicsNetworks, Gad2021AConstraints, Hedayati2024Closed-formMetamaterials}. In other studies, the homogenization process was conducted within the framework of wave propagation \cite{Gonella2008HomogenizationLattices, Phani2006WaveLattices}.

% Extensive studies have investigated the influence of lattice geometry, defining the effect of the design of the links characterizing their structure, showing that different bulk-to-shear ratios can be obtained. 
Extensive studies have investigated how lattice geometry, particularly the design of the links, influences the effective mechanical properties, showing that different bulk-to-shear ratios can be achieved.
Kadic \textit{et~al.} \cite{Kadic2012OnMetamaterials} investigated the classic diamond unit cell with bi-conical links: 
in this architecture the links meet at their smaller diameter, thereby improving the bulk-to-shear ratio. However, this also dramatically reduces the bulk modulus, which remains far below that of liquids. 
Remarkably, the proposed geometry was successfully fabricated and experimentally tested. 
Other lattice geometries have been proposed to overcome some limitations of PMs, including limited decoupling between bulk modulus and density, restricted flexibility in the tiling, and limited control over anisotropy.   
Both alternative topologies, such as hexagonal\cite{Li2019Three-dimensionalCells} or truncated octahedron\cite{Mohammadi2024Pentamodes:Properties, Wei2023AMetamaterial} architectures, and modified link designs \cite{Zou2023BandgapSide, Kadic2014PentamodeDensity, Cai2022DesignWaves} have been proposed, although their stiffness remains low compared with that of liquids. A solution to this problem was proposed by
Sigmund \cite{Sigmund2000AComposites}, who designed a 2D PM using topology optimization that attains Hashin-Shtrikman bounds\cite{hashin1963variational}. These bounds define the mathematical limits of the effective elastic moduli for composite materials. The proposed geometry exploited appropriately arranged laminated structures, yielding a material with the lowest possible shear modulus while maintaining the highest bulk modulus achievable with that material-to-void ratio. This concept has been extended to 3D geometries by Milton \textit{et~al.} \cite{Milton2017OnMaterials}, although only theoretically, proving that in 3D the Hashin-Shtrikman bound can be reached using an array of beams, i.e., a sheaf of thin fibers. Brambilla \textit{et~al.} \cite{Brambilla2025HighWater} went one step further, proving that the classical bi-conical link shape cannot attain high bulk moduli; hence, they designed and numerically tested a geometry with links composed of a sheaf of thin fibers, showing that the density and bulk modulus of water can be achieved.

Currently, the stability of PM structures is an active topic of investigation. Indeed, because PM lattices rely on thin connections to maximize the bulk-to-shear modulus ratio, they are intrinsically susceptible to buckling. 
% Buckling of lattice structures has already been investigated, although mainly for two dimensional lattices. 
% Guo \textit{et~al.}\cite{Guo2024AnalyticalModel} used a micropolar elasticity model of different unit cell topologies, showing that the constitutive relationships are largely affected by the preload. \todo[inline]{not sure about citing this one. Is not strictly related to PM}
% Strictly speaking of PMs, 
Adhikari \cite{Adhikari2021TheLattices} developed an analytical procedure to evaluate the equivalent elastic moduli and Poisson's ratios of externally loaded 2D honeycomb lattices, demonstrating a softening effect under compression.
As shown through simplified analytical models by Karli{\v{c}}i{\'{c}} \textit{et~al.} \cite{Karlicic2021WaveLattices} and Pal \textit{et~al.} \cite{Pal2016EffectLattices}, and numerically by Zhang \textit{et~al.} \cite{Zhang2018EffectLattices}, this softening has not only structural consequences but also affects wave propagation.
Progress has also been made towards identifying the critical buckling conditions. Adhikari \cite{Adhikari2021TheSolutions} demonstrated that for the homogenized material, a necessary and sufficient condition for buckling is the presence of at least a zero eigenvalue of the elasticity matrix.
However, Bordiga \textit{et~al.} \cite{Bordiga2021DynamicsLocalization} showed, through numerical and experimental investigations of a 2D grillage, that the dynamics of the equivalent homogenized material alone are insufficient to predict buckling collapse, as local instabilities may also develop within individual unit cells. This highlights the need to distinguish between instabilities associated with the macroscopic, effective response and those arising from the local geometry of the lattice.
For 3D lattices, we highlight the theoretical study of Sinha \textit{et~al.} \cite{Sinha2025EffectiveProgramming}, who investigated the effective elastic properties of prestressed non-PM 3D lattices. In parallel,
numerical and experimental studies have addressed the collapse of PM lattices subjected to compressive loads, notably by Lymperopoulos \textit{et~al.} \cite{Lymperopoulos2026BucklingCharacteristics, Lymperopoulos2024AdditiveMetamaterials} and Amendola \textit{et~al.} \cite{Amendola2016ExperimentalPlates, Amendola2017Non-linearPlates}.

Despite this extensive body of work on the mechanical response and stability of lattice materials, a general framework for PM structures with different link geometries is still lacking, particularly in three dimensions. Moreover, existing studies generally do not account simultaneously for the two distinct instability mechanisms highlighted by Bordiga \textit{et~al.} \cite{Bordiga2021DynamicsLocalization}: macroscopic instability of the homogenized material and local instability occurring at the unit-cell level.
In this work, we take a first step towards a general framework by considering isotropic PM structures composed of equal links with generic geometries and subjected to isotropic prestress.
The formulation is developed for both 2D and 3D lattices, thereby providing a description of their mechanical stability. 
In Section~\ref{sec:LatticeDefinition}, we first introduce representative geometries, which are subsequently evaluated numerically, and derive the generic models for both the individual links and the resulting lattices. 
Section~\ref{sec:Instability} then establishes the stability limits associated with the two buckling mechanisms. 
Finally, Section~\ref{sec:4_NumericalValidation} compares the analytical predictions with numerical results and discusses their implications, while Section~\ref{sec:conclusions} concludes the study with the main findings.

\section{Modeling of the Lattice}
\label{sec:LatticeDefinition}

In this section, we develop the mechanical models used to describe the dynamics and stability of the lattices considered in this work. We first formulate a general model for the individual lattice links and derive their mass and stiffness matrices in terms of a set of parameters that depend on the link geometry and material properties. These matrices are then used to formulate the equations of motion of the periodic lattice and derive its dynamic response. The resulting formulation provides the basis for the analytical determination of the buckling limits presented in Section~\ref{sec:Instability}.

\begin{figure}
    \centering
    \subfloat[]{
        \includegraphics[
            width=0.18\textwidth, 
            trim=2cm 0.15cm 3cm 0.25cm, 
            clip
        ]{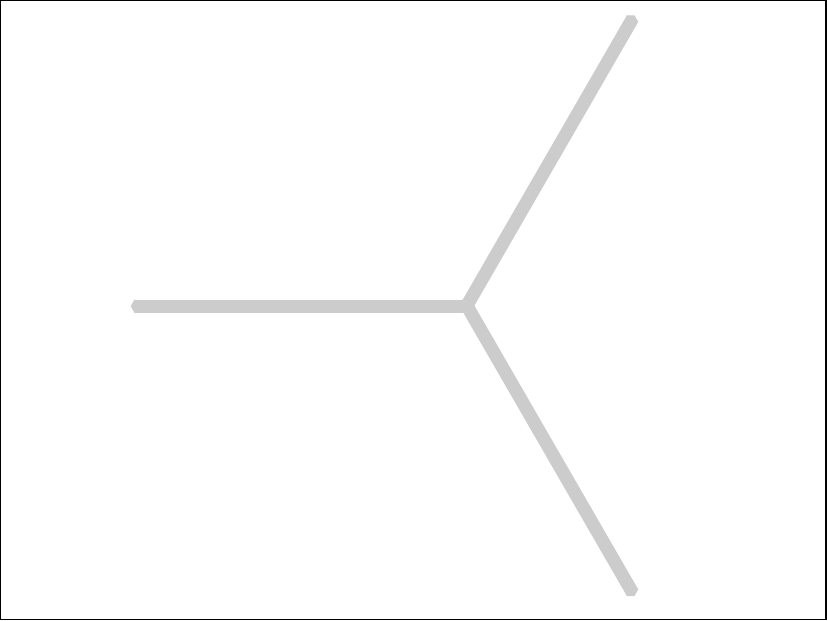}
        \label{fig:2DUnitCell}
    } 
    \subfloat[]{
        \raisebox{0cm}{\includegraphics[
            width=0.21\textwidth, 
            trim=0cm 0.15cm 1.5cm 0cm, 
            clip
        ]{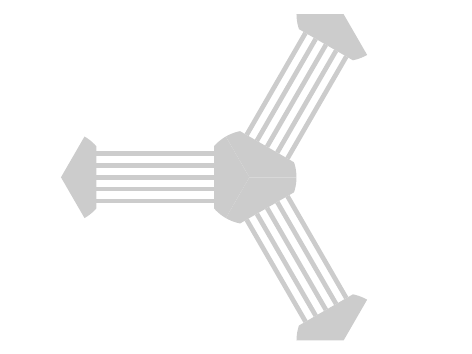}
        }
        \label{fig:2DUnitCellFasci}
    }
    \subfloat[]{
        \raisebox{0cm}{
            \includegraphics[
                width=0.24\textwidth, 
                trim= 0cm 0cm -1cm 0cm, 
                clip
            ]{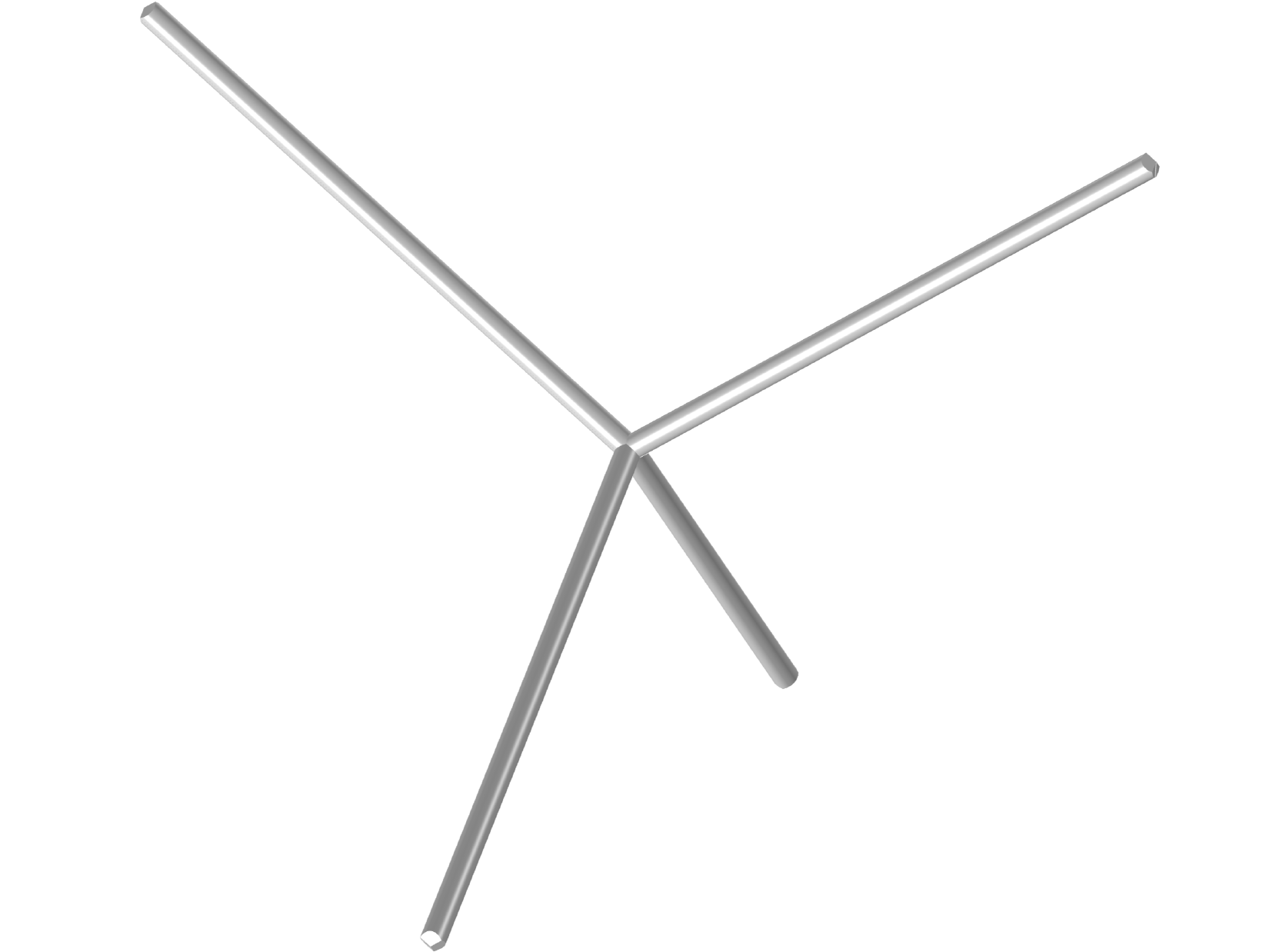}
            }
        \label{fig:3DUnitCell}
    }
    \subfloat[]{
        \includegraphics[
            width=0.24\textwidth
        ]{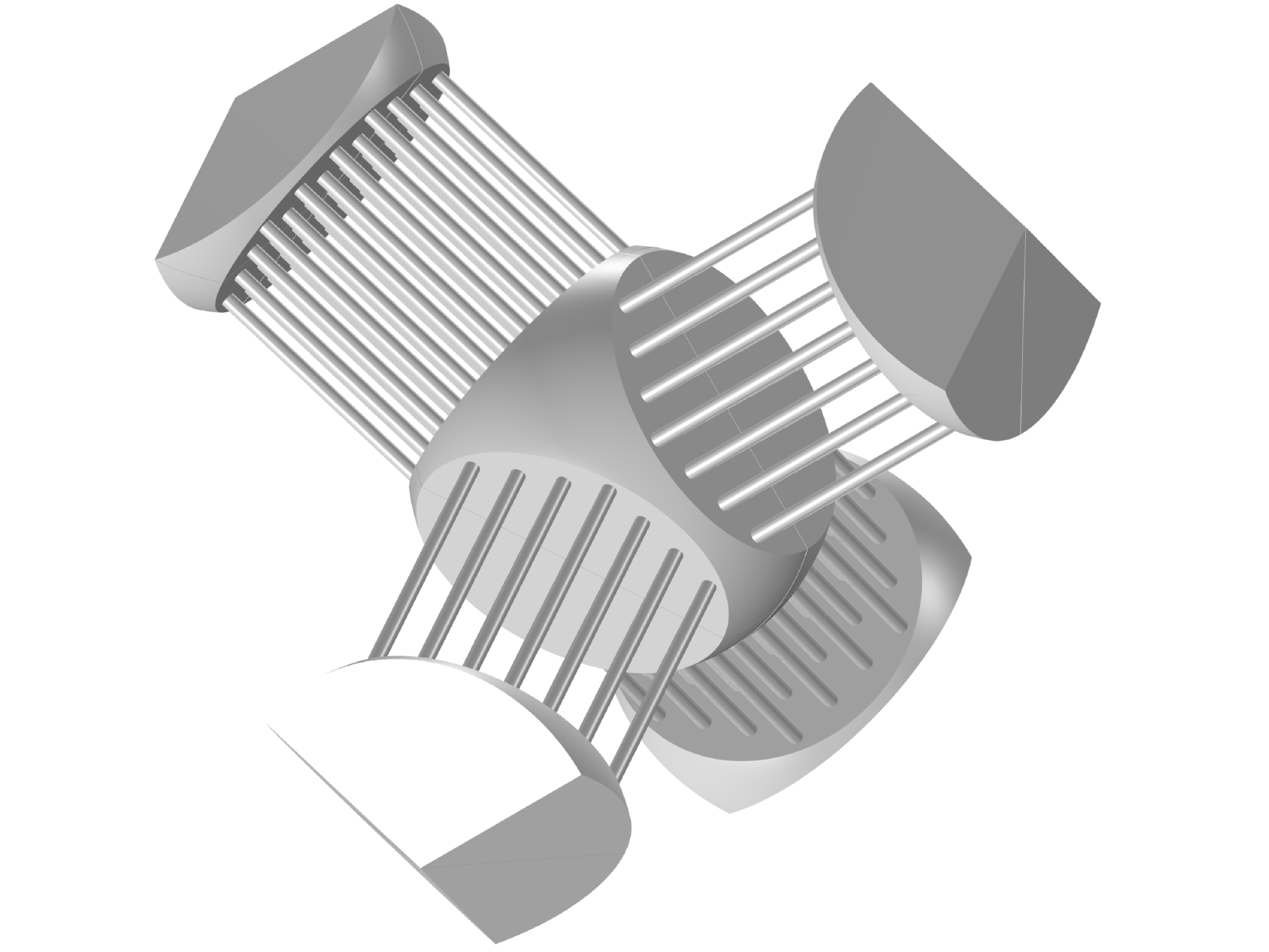}
        \label{fig:3DUnitCellFasci}
    }
    \\
    \subfloat[]{
        \includegraphics[
            width=0.38\textwidth,
             trim=1cm 1cm 1cm 1.2cm, 
            clip
        ]{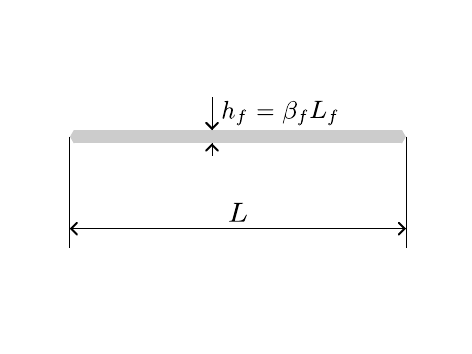}
        \label{fig:2DLink}
    }
    \hfill
    \subfloat[]{
        \includegraphics[
            width=0.46\textwidth, 
            trim=0cm 1cm 1cm 1.8cm, 
            clip
        ]{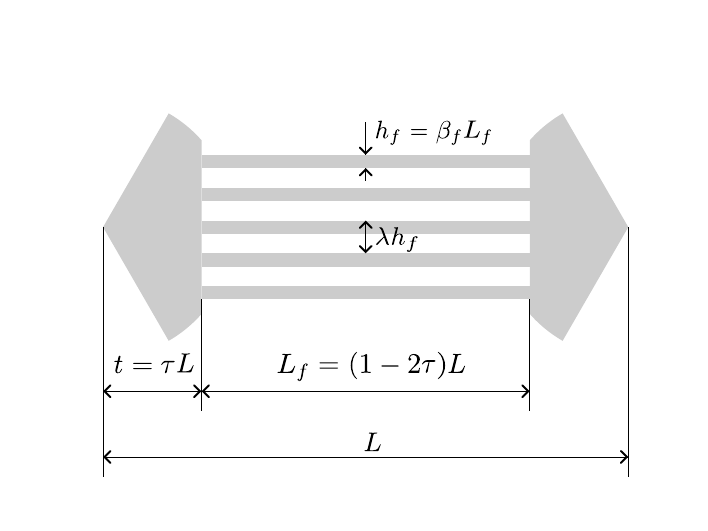}
        \label{fig:2DLInkFasci}
    }
    \\
    \subfloat[]{
    \raisebox{0.7cm}{
        \includegraphics[
            width=0.4\textwidth, 
            trim=2cm 1cm 2.5cm 1cm, 
            clip
        ]{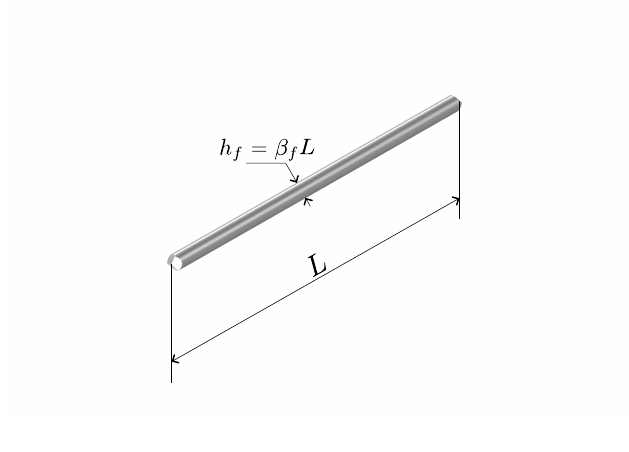}}
        \label{fig:3DLink}
    }
    \hfill
    \subfloat[]{
        \includegraphics[
            width=0.45\textwidth, 
            trim=1cm 0cm 2cm 0.1cm, 
            clip
        ]{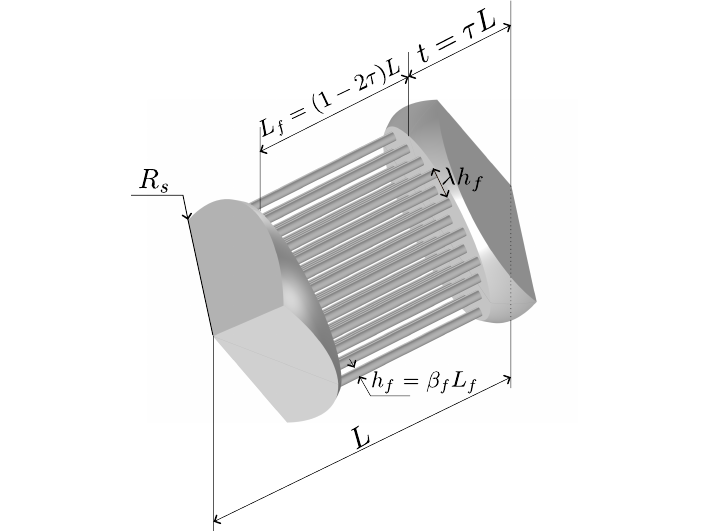}
        \label{fig:3DLinkFasci}
    }
    \caption{Representative examples of the unit-cell models studied in this work and their corresponding links. The 2D geometries are shown in \protect\subref{fig:2DUnitCell} and \protect\subref{fig:2DUnitCellFasci}, while the 3D geometries are shown in \protect\subref{fig:3DUnitCell} and \protect\subref{fig:3DUnitCellFasci}. The link models and the definitions of the variables governing their geometry are presented in \protect\subref{fig:2DLink}--\protect\subref{fig:3DLinkFasci}, respectively. }
    \label{fig:Link_and_cell_Models}
\end{figure}

We consider 2D and 3D isotropic pentamode lattices. As representative examples, Figure~\ref{fig:Link_and_cell_Models}\subref{fig:2DUnitCell}-\subref{fig:2DUnitCellFasci} shows hexagonal honeycomb geometries in 2D, while Figure~\ref{fig:Link_and_cell_Models}\subref{fig:3DUnitCell}-\subref{fig:3DUnitCellFasci} illustrates the corresponding 3D architecture. Related 2D and 3D lattice geometries have been studied in \cite{Gonella2008HomogenizationLattices, Phani2006WaveLattices, Sigmund2000AComposites} and \cite{Kadic2012OnMetamaterials, Brambilla2025HighWater}, respectively.
The 3D lattices considered here possess cubic symmetry and are therefore characterized by three independent elastic constants, whereas an isotropic solid is described by two. We nevertheless refer to these lattices as isotropic because, in the ideal pentamode limit, the shear stiffnesses vanish. Therefore, in this limit, the cubic constitutive relation reduces to the isotropic form, with the bulk modulus remaining as the only nonzero independent elastic parameter.

The following discussion treats the lattice links as black boxes connecting their two end nodes (see Figure~\ref{fig:BeamConventions}). No specific assumptions are made about their shape beyond the symmetry requirements discussed below. Consequently, the resulting formulation can also be applied to other link geometries with the same connectivity. We further identical links of length $L$, arranged with equal angular spacing at each node, and an externally applied isotropic pressure.
Finally, we consider an infinite lattice. This approximation is less restrictive than it may initially appear: pentamode lattices are typically designed to manipulate waves in the low-frequency regime, where the wavelength is much larger than the unit-cell size; therefore, a single wavelength encompasses many unit cells, making the infinite-lattice approximation appropriate.

\subsection{Modeling of the Links}
To construct the mass and stiffness matrices required to analyze the lattice dynamics and identify its buckling limits, we first consider its individual links. We derive a generic formulation for the link matrices by starting from their most general form and progressively reducing the number of independent terms through symmetry conditions and force and moment balance. The symmetry conditions are introduced to simplify the formulation and reflect the geometrical characteristics of the links considered in this work. 
We first consider the more general 3D case, while the 2D formulation is subsequently obtained as a particular case.

\begin{figure}
    \centering
    \includegraphics[width=0.8\linewidth]{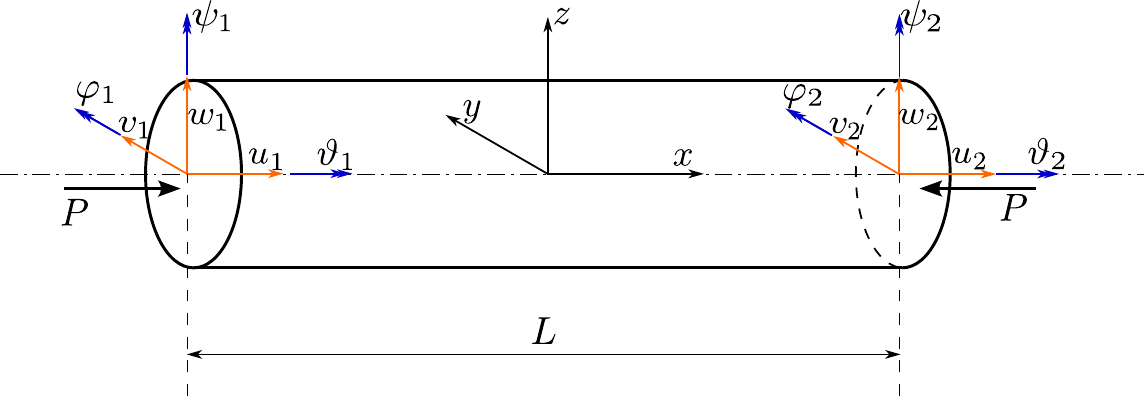}
    \caption{Definition of the used degree of freedom for a 3-dimensional beam. $P$ represents an external load acting on the beam which may lead to buckling.}
    \label{fig:BeamConventions}
\end{figure}

We consider each link as a two-node structural element in equilibrium under an externally applied axial load $P$, positive in compression. This load is transmitted between adjacent links through the nodes as a result of the applied external pressure. The link is characterized by the degrees of freedom
\begin{equation}
    \bb u = \{u_1,v_1, w_1, \vartheta_1,\varphi_1, \psi_1, u_2,v_2, w_2, \vartheta_2,\varphi_2, \psi_2\}^\top,
\end{equation}
which represent the translations and rotations of both nodes, as shown in Figure~\ref{fig:BeamConventions}. These degrees of freedom are defined as perturbations about the equilibrium configuration.
Since the beam geometry is not specified, its stiffness and mass matrices, $\mathsf K$ and $\mathsf M$ respectively, are, in principle, full symmetric $12\times12$ matrices, characterized by $78$ independent parameters. We seek to simplify these matrices by setting some of these coefficients to zero. To this end, we impose geometric constraints that are naturally satisfied by a broad class of beam geometries. Specifically, we consider the beam to feature symmetries about the $xy$, $xz$, and $yz$ planes of a properly chosen coordinate system, as shown in Figure~\ref{fig:BeamConventions}. Additionally, for the 3D case, we assume the existence of at least one rotation $\alpha\in(0,\pi)$ about the $x$ axis that leaves the beam unchanged; this condition is satisfied, for instance, by beams with a regular-polygon cross section.

For each of these symmetries, we can define a change of coordinates $\bb u = \mathsf S_i \bb u'$, where $\mathsf S_i$ are formulated as
\begin{equation}
\begin{aligned}
    \mathsf S_{xy} &= 
    \begin{bmatrix}
        \mathsf S_{p,xy}&[0]\\
        [0] & \mathsf S_{p,xy}
    \end{bmatrix}, &
    \mathsf S_{xz} &= 
    \begin{bmatrix}
        \mathsf S_{p,xz}&[0]\\
        [0] & \mathsf S_{p,xz}
    \end{bmatrix}, &
    \mathsf S_{yz} &= 
    \begin{bmatrix}
        [0] & \mathsf S_{p,yz}\\
        \mathsf S_{p,yz}&[0]
    \end{bmatrix},  &
    \mathsf S_{\alpha} &= 
    \begin{bmatrix}
        \mathsf S_{p,\alpha}&[0]\\
        [0] & \mathsf S_{p,\alpha}
    \end{bmatrix}, 
\end{aligned}
\end{equation}
with
\begin{equation}
    \begin{aligned}
        \mathsf S_{p,xy} &= 
    \text{diag}(1, 1, -1, -1, -1, 1); &
        \mathsf S_{p,xz} &= 
    \text{diag}(1, -1, 1, -1, 1, -1); \\
        \mathsf S_{p,yz} &= 
    \text{diag}(-1, 1, 1, 1, -1, -1); &
        \mathsf S_{p,\alpha} &= 
    \begin{bmatrix}
        1 & 0 & 0 & 0 & 0 & 0 \\
        0 & \cos\alpha & \sin\alpha & 0 & 0 & 0 \\
        0 & -\sin\alpha & \cos\alpha & 0 & 0 & 0 \\
        0 & 0 & 0 & 1 & 0 & 0 \\
        0 & 0 & 0 & 0 & \cos\alpha & \sin\alpha \\
        0 & 0 & 0 & 0 & -\sin\alpha & \cos\alpha
    \end{bmatrix}.
\end{aligned}
\end{equation}
A symmetry requires the mass and stiffness matrices to remain invariant under the corresponding coordinate transformation. Therefore,
\begin{equation}
    \begin{aligned}
        \mathsf X_{\mathrm{loc}}-\mathsf S_{xy}^\top \mathsf X_{\mathrm{loc}} \mathsf S_{xy}  &= [0], & 
        \mathsf X_{\mathrm{loc}}-\mathsf S_{xz}^\top \mathsf X_{\mathrm{loc}} \mathsf S_{xz}  &= [0], & 
        \mathsf X_{\mathrm{loc}}-\mathsf S_{yz}^\top \mathsf X_{\mathrm{loc}} \mathsf S_{yz}  &= [0], & 
        \mathsf X_{\mathrm{loc}}-\mathsf S_{\alpha}^\top \mathsf X_{\mathrm{loc}} \mathsf S_{\alpha}  &= [0],
    \end{aligned}
    \label{eq:symmetryConditions}
\end{equation}
where $\mathsf X_{\mathrm{loc}}$ can be either the mass or the stiffness matrix expressed in this local coordinate system.
The matrix equations in Eq.~\eqref{eq:symmetryConditions} are expanded componentwise, producing a set of scalar equations for the individual matrix entries. Many of these equations are redundant, and only the independent constraints are retained. Enforcing these constraints determines the following sparse form of the full $12\times12$ matrix:
\begin{equation}
    \label{eq:M_or_K_beam_3D}
    \mathsf X_{\mathrm{loc}} = \begin{bmatrix}
        a_X & 0   & 0   & 0   & 0   & 0   & e_X & 0   & 0   & 0   & 0   & 0   \\
            & b_X & 0   & 0   & 0   & f_X & 0   & g_X & 0   & 0   & 0   & j_X \\
            &     & b_X & 0   &-f_X & 0   & 0   & 0   & g_X & 0   &-j_X & 0   \\
            &     &     & c_X & 0   & 0   & 0   & 0   & 0   & h_X & 0   & 0   \\
            &     &     &     & d_X & 0   & 0   & 0   & j_X & 0   & i_X & 0   \\
            &     &     &     &     & d_X & 0   &-j_X & 0   & 0   & 0   & i_X \\
            &     &     &     &     &     & a_X & 0   & 0   & 0   & 0   & 0   \\
            &     &     &     &     &     &     & b_X & 0   & 0   & 0   &-f_X \\
            &     &     & sym &     &     &     &     & b_X & 0   & f_X & 0   \\
            &     &     &     &     &     &     &     &     & c_X & 0   & 0   \\
            &     &     &     &     &     &     &     &     &     & d_X & 0   \\
            &     &     &     &     &     &     &     &     &     &     & d_X \\
    \end{bmatrix}
\end{equation}
We can further simplify the matrices by considering the force and moment balances. To this end, we define the generalized external forces condensed at the nodes of the link as 
\begin{equation}
\label{eq:ForcesDefinition}
    \bb F = \{F_1, F_2 \dots, F_{12}\}^\top = \mathsf M_{\mathrm{loc}} \ddot{\bb u} + \mathsf K_{\mathrm{loc}} \bb u.
\end{equation} 
The vector $\bb F$ includes both forces and moments, as in classical beam theories such as Timoshenko or Euler-Bernoulli. It represents the generalized forces associated with the perturbation about the equilibrium established by the external load $P$. The term $\mathsf M_{\mathrm{loc}} \ddot{\bb u}$ represents the condensed generalized inertia forces, while $\mathsf K_{\mathrm{loc}} \bb u$ is the condensed generalized elastic forces. We then use the force and moment balance conditions to identify the relations among the remaining independent coefficients of the stiffness and mass matrices. 

To isolate the effect of stiffness, we consider a static deformation of the link superimposed on the equilibrium configuration established by $P$. Taking into account this external load, the static equilibrium of forces and moments can be expressed as
\begin{equation}
    \left\{
    \begin{aligned}
        F_1+ F_7  &= 0 \\
        F_2+ F_8  &= 0 \\
        F_3+ F_9  &= 0 \\
        F_4+ F_{10} &= 0 \\
        F_5+ F_{11} + F_{3}L + P(w_1-w_2)  &= 0 \\
        F_6+ F_{12} + F_{8}L + P(v_2-v_1) &= 0,
    \end{aligned}
    \right.
\end{equation}
These conditions must hold for any admissible $\bb u$. 
Thus, we consider displacement vectors in which only one component of $\bb u$ is nonzero at a time, with all other components set to zero. This allows us to identify the independent relations among the matrix coefficients.
% Thus, we activate a single component of $\bb u$  at a time, setting the others to zero. 
Some of the resulting equations are redundant, leaving:
\begin{equation}
% \label{eq:simpl_K}
    \left\{
    \begin{aligned}
        e_K &= -a_K\\
        g_K & = -b_K\\
        f_K & = j_K = \frac 1 2 \left(b_KL+P\right)\\
        h_K & = -c_K\\
        i_K &= -d_K +\frac 1 2 (b_KL+P)L.
    \end{aligned}
    \right.
\end{equation}
Using these relations, $\mathsf K$ is simplified as follows:
\begin{align}
\label{eq:K_beam_3D}
    \mathsf K_{\mathrm{loc}} = \begin{bmatrix}
        a_K & 0   & 0   & 0   & 0   & 0   &-a_K & 0   & 0   & 0   & 0   & 0   \\
            & b_K & 0   & 0   & 0   & f_K & 0   &-b_K & 0   & 0   & 0   & f_K \\
            &     & b_K & 0   &-f_K & 0   & 0   &     &-b_K & 0   &-f_K & 0   \\
            &     &     & c_K & 0   & 0   & 0   & 0   & 0   &-c_K & 0   & 0   \\
            &     &     &     & d_K & 0   & 0   & 0   & f_K & 0   & i_K & 0   \\
            &     &     &     &     & d_K & 0   &-f_K & 0   & 0   & 0   & i_K \\
            &     &     &     &     &     & a_K & 0   & 0   & 0   & 0   & 0   \\
            &     &     &     &     &     &     & b_K & 0   & 0   & 0   &-f_K \\
            &     &     & sym &     &     &     &     & b_K & 0   & f_K & 0   \\
            &     &     &     &     &     &     &     &     & c_K & 0   & 0   \\
            &     &     &     &     &     &     &     &     &     & d_K & 0   \\
            &     &     &     &     &     &     &     &     &     &     & d_K \\
    \end{bmatrix},
\end{align}
with $f_K =\frac 1 2 \left(b_KL+P\right)$, $i_K = -d_K+f_KL$, leaving four independent coefficients, $a_K$, $b_K$, $c_K$, and $d_K$. Although their dependence on the applied load $P$ is not made explicit at this stage, these coefficients may in general depend on $P$. The applied load therefore affects the tangent stiffness both explicitly through the equilibrium conditions and implicitly through these coefficients. This load-dependent change in stiffness is what ultimately leads to buckling, as discussed later in Section~\ref{sec:Instability}.

Similarly, the mass matrix $\mathsf M_{\mathrm{loc}}$ can be further simplified by imposing rigid-body motions. We first consider a rigid translation in the $x$ direction, such that $\bb u = \{u1 = u, 0, \dots, u_2 = u, 0, \dots \}^\top$, with both nodes undergoing the same acceleration $\ddot u \hat{\bs{\imath}}$. The corresponding total inertia force of the link is
\begin{equation}
\bb f_{\mathrm{inertia}} = -M_{\mathrm{link}}\ddot u\hat{\bs{\imath}},
\end{equation}
where $M_{\mathrm{link}}$ is the total mass of the link. The external forces must balance the inertial force. From Eq.~\eqref{eq:ForcesDefinition}, we therefore obtain
\begin{equation}
    \bb f_{\mathrm{inertia}}
    = -(F_1+F_7)\hat{\bs{\imath}}
    = -2(a_M+e_M)\ddot u\hat{\bs{\imath}},
\end{equation}
which gives
\begin{equation}
    a_M+e_M=\frac{M_{\mathrm{link}}}{2}.
\end{equation}
The same reasoning can be applied to rigid-body translations in the $y$ and $z$ directions, yielding
\begin{equation}
    b_M+g_M=\frac{M_{\mathrm{link}}}{2}.
\end{equation}
These relations ensure that the condensed mass matrix reproduces the total translational inertia of the beam. They will be used later to simplify the results of the homogenization procedure.

As anticipated, the 2D case can be readily derived as a special case of the 3D analysis, yielding: 
\begin{align}
    \label{eq:2D_dof}
    \bb u &= \{u_1,v_1, \psi_1, u_2,v_2, \psi_2\}^\top,\\
    \label{eq:K_beam_2D}
    \mathsf K_{\mathrm{loc}} &= \begin{bmatrix}
        a_K &0      & 0     &-a_K &   0      & 0      \\
                &b_K & f_K & 0       &-b_K  & f_K\\
                &       & d_K & 0        &- f_K& -d_K+ f_KL  \\
                &       &         & a_K   & 0         & 0     \\
                & sym&         &           &  b_K   &- f_K\\
                &       &         &           &            & d_K \\
    \end{bmatrix}\\
    \label{eq:M_beam_2D}
    \mathsf M_{\mathrm{loc}} &= \begin{bmatrix}
        a_M &0      & 0     & e_M &   0      & 0      \\
                &b_M & f_M & 0       &g_M  &  j_M \\
                &       & d_M & 0        &-j_M  & i_M  \\
                &       &         & a_M   & 0         & 0     \\
                & sym&         &           &  b_M   &-f_M \\
                &       &         &           &            & d_M \\
    \end{bmatrix}.
\end{align}
% This completes the derivation of the generic local mass and stiffness matrices for both the 3D and 2D link models.
\subsection{Assembly of the Lattice Matrices}
With the link model established, the matrices of a representative unit cell can be derived and subsequently used to formulate the dynamics of the entire lattice. To this end, we first define the nodal coordinates and connectivity of the unit cell. We then introduce a local reference system for each link and the corresponding rotation matrices to transform the link matrices into the global coordinate system. Finally, connectivity matrices are used to assemble the contributions of the individual links into the mass and stiffness matrices of the unit cell. These matrices are then used to formulate the dynamics of the periodic lattice and derive its dispersion relation in terms of the coefficients of the generic link model.

Although the procedure is conceptually identical for both the 2D and 3D configurations, the derivation is presented separately, starting with the 2D lattice and then extending the formulation to the 3D case. This order introduces the fundamental concepts in the simpler 2D setting before addressing the additional geometrical complexity of the 3D topology.

\begin{figure}
\centering
    \subfloat[]{
    \raisebox{5 mm}{
        \includegraphics[width=0.3\textwidth]{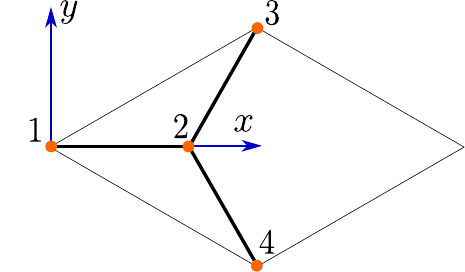}
        \label{fig:2DUnitCellSchematic}}
    }
    % \hfill
    \subfloat[]{
        % \raisebox{15 mm}{
            \includegraphics[width=0.4\linewidth]{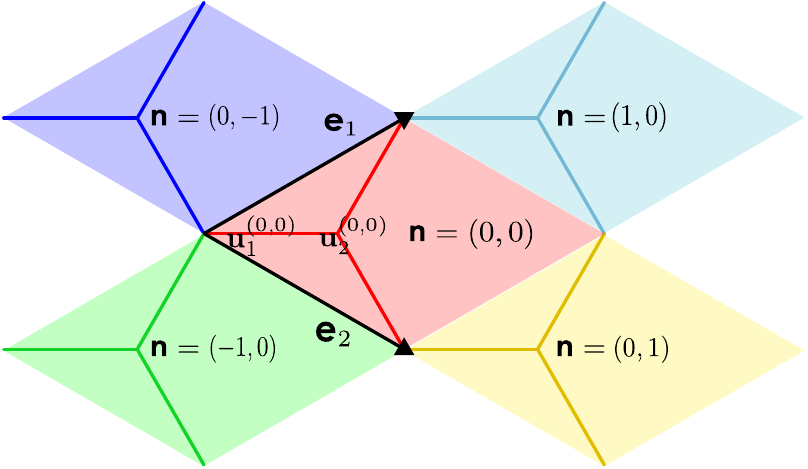}
            \label{fig:2D}}
        % }
    \subfloat[]{
        \raisebox{3 mm}{
            \includegraphics[width=0.28\linewidth]{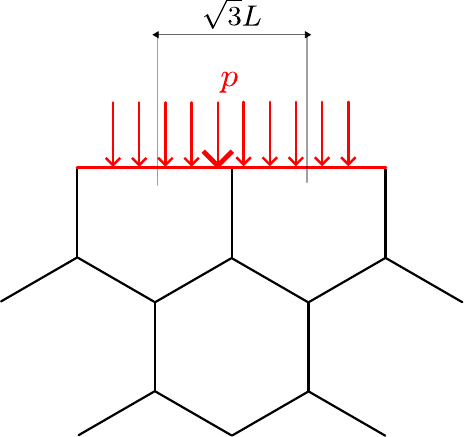}
            \label{fig:Load2D}}
    }
    \caption{Schematic of the generic 2D unit cells (a), its arrangement in the lattice topology (b), and the interaction between the external pressure and the lattice (c).}
    \label{fig:UnitCell_2D_Topology}
\end{figure}

Let us consider a 2D pentamode lattice. Figure~\ref{fig:UnitCell_2D_Topology}\subref{fig:2DUnitCellSchematic} shows the unit cell of a pentamode lattice composed of generic beams connected at the nodes, highlighted in orange. This geometry corresponds to a regular hexagonal honeycomb with identical links equally spaced by \SI{120}{\degree}, which guarantees isotropy in the plane \cite{Gonella2008HomogenizationLattices}. 
We define the distance between two connected nodes as $L$. The nodal coordinates are
\begin{equation}
    \left\{
    \begin{aligned}
        \bb c_1 & = \left(0,0\right),\\
        \bb c_2 & = \left(L,0\right),\\
        \bb c_3 & = \left(\frac{3}{2}L,\frac{\sqrt3}{2}L\right),\\
        \bb c_4 & = \left(\frac{3}{2}L,-\frac{\sqrt3}{2}L\right)
    \end{aligned}
    \right.
\end{equation}

The mass and stiffness matrices derived for each link are expressed in their respective local coordinate systems. To assemble the contributions of the different links in a common global reference system, these matrices are first transformed using rotation matrices that relate each local coordinate system to the global reference system. The rotation matrices are determined by the orientation of the corresponding link with respect to the global reference system. 
We define a local coordinate system for the link connecting nodes $i$ and $j$ by aligning its $x$-axis with the direction of the link. In the global reference system, this direction is given by
\begin{equation}
    \bb v_{ij,x}=
    \frac{\bb c_j-\bb c_i}
    {\|\bb c_j-\bb c_i\|},
\end{equation}
where $\|\cdot\|$ denotes the Euclidean norm. The vector $\bb v_{ij,y}$ is then defined to complete an orthonormal reference frame. Due to the imposed symmetries, its choice does not affect the resulting formulation.
The corresponding rotation matrices are given by:
\begin{equation}
    \begin{aligned}
        \mathsf R_{ij,2D} &=
        \begin{bmatrix}
            \mathsf R^*_{ij,2D} & [0]\\
            [0] & \mathsf R^*_{ij,2D}
        \end{bmatrix}, 
        & \textrm{with }
        \mathsf R^*_{ij,2D} &=
        \begin{bmatrix}
            \bb v_{ij,x}^{\top} & 0\\
            \bb v_{ij,y}^{\top} & 0\\
            \begin{matrix}
            0 & 0
            \end{matrix} & 1
        \end{bmatrix}.
    \end{aligned}
\end{equation}

To assemble the global matrices, we introduce the connectivity matrices $\mathsf L_{ij}$, Boolean matrices that identify the degrees of freedom associated with each link within the global displacement vector:
\begin{equation}
    \bb u_{ij}=\mathsf L_{ij}\bb u_{\mathrm{glob}},
\end{equation}
where $\bb u_{ij}$ collects the generalized displacements of nodes $i$ and $j$, while $\bb u_{\mathrm{glob}}$ contains the generalized displacements of all the nodes in the unit cell.
The contribution of each link to the global matrices is obtained as
\begin{equation}
    \mathsf{ \tilde X}_{ij} = 
    \mathsf L_{ij}^{\top}
    \mathsf R_{ij}^{\top}
    \mathsf X_{\mathrm{loc}}
    \mathsf R_{ij}
    \mathsf L_{ij}.
\end{equation}
The matrices of the unit cell are then obtained by summing the contributions of all the beams,
\begin{equation}
\hat{\mathsf X}=\sum_{ij}\mathsf{\tilde  X}_{ij}.
\end{equation}
The resulting stiffness and mass matrices feature the block structure
\begin{align}
\label{eq:UnitCellMatrix}
    \hat{\mathsf X}_{2D}=
    \begin{bmatrix}
        \hat{\mathsf X}_{11} & \hat{\mathsf X}_{12} & [0] & [0]\\
        \hat{\mathsf X}_{12}^{\top} & \hat{\mathsf X}_{22} & \hat{\mathsf X}_{23} & \hat{\mathsf X}_{24}\\
        [0] & \hat{\mathsf X}_{23}^{\top} & \hat{\mathsf X}_{33} & [0]\\
        [0] & \hat{\mathsf X}_{24}^{\top} & [0] & \hat{\mathsf X}_{44}
    \end{bmatrix},
\end{align}
where each block $ \hat{\mathsf X}_{ij}\in\mathbb R^{3\times3}$. Equation~\eqref{eq:UnitCellMatrix} shows that all nodes are connected only to node $\#2$, the other coupling terms being  equal to $0$. This block structure will be considered in the following, where the connections between different cells have to be taken into account to study the lattice topology.

\begin{figure}
\centering
    \subfloat[]{
    \raisebox{7 mm}{
        \includegraphics[width=0.25\textwidth]{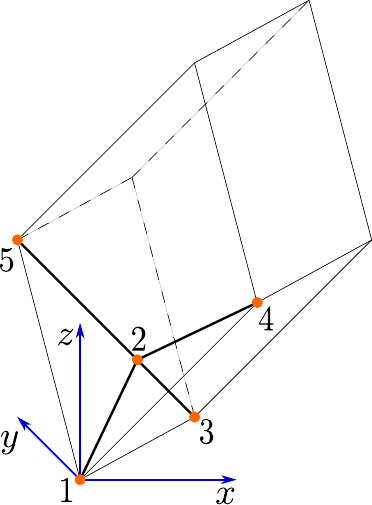}
        \label{fig:3DUnitCellSchematic}}
     }
    % \hfill
    \subfloat[]{
        \includegraphics[width=0.33\linewidth]{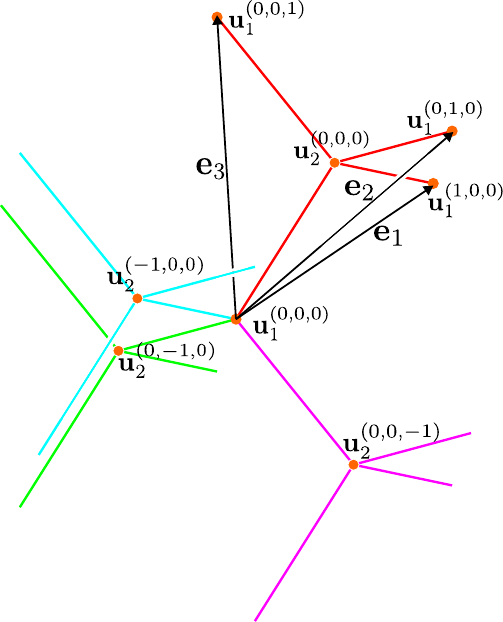}
        \label{fig:3D}
    }
    \\
    \subfloat[]{
        \includegraphics[width=0.3\linewidth]{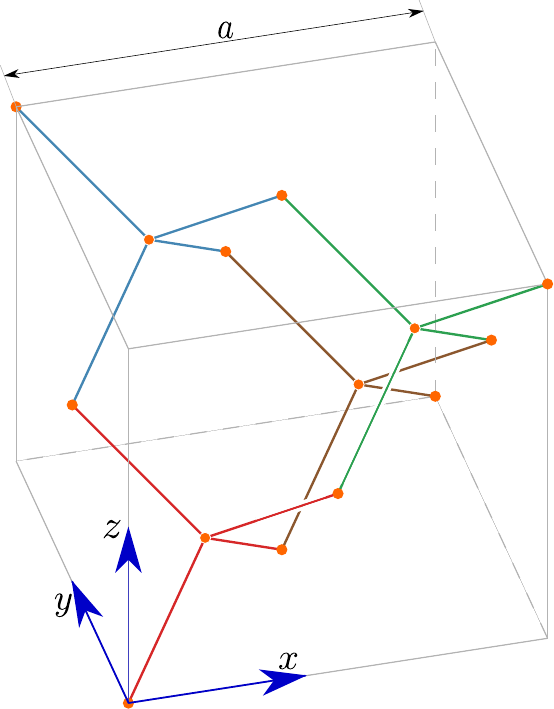}
        \label{fig:cubic_symmetry}
    }
    \subfloat[]{
        \includegraphics[width=0.55\linewidth]{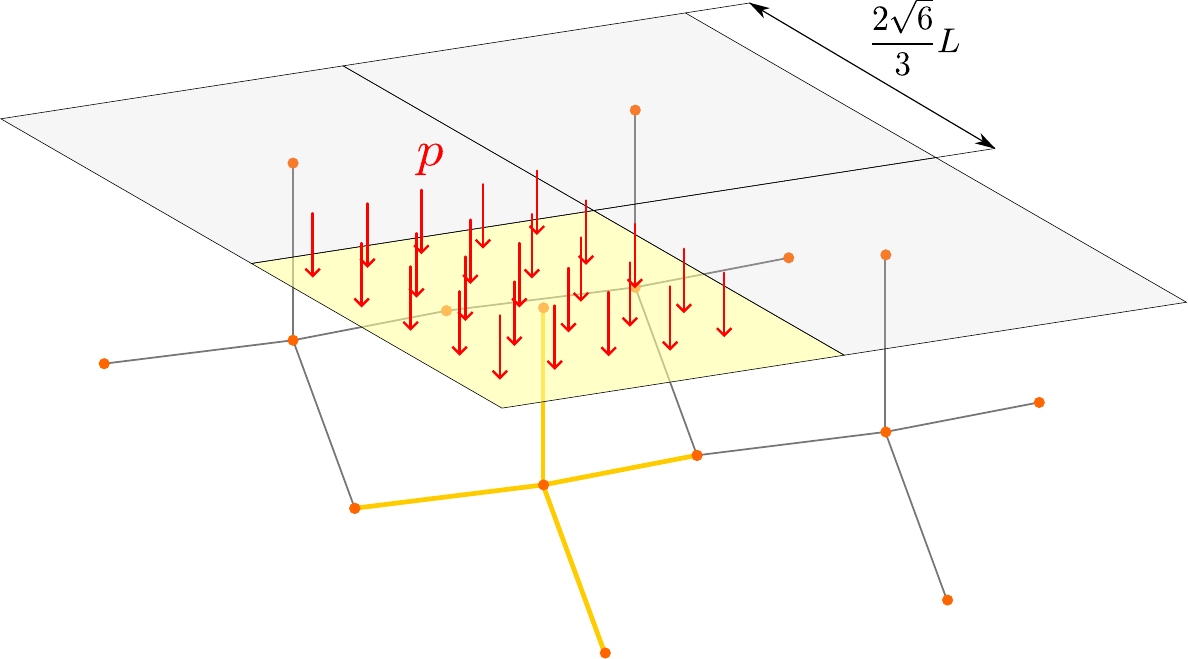}
        \label{fig:Load3D}
    }
    \caption{Schematic of the generic 3D pentamode unit cell (a), its arrangement in the lattice topology (b),  its cubic symmetry (c), and a schematic of the interaction of the lattice with the external pressure (d)}
    \label{fig:UnitCell_3D_Topology}
\end{figure}

Having defined the stiffness and mass matrices for a single unit cell, we can now extend the formulation to the entire periodic lattice, following \cite{Colquitt2015HighLattices}.
The simplified geometry of the two-dimensional lattice is shown in Figure~\ref{fig:UnitCell_2D_Topology}\subref{fig:2D}. Each unit cell is identified by the index $\bb n\in\mathbb Z^2$ and contains two independent nodes located at positions $\bb c_i^{\bb n}$ and characterized by the degrees of freedom $\bb u_i^{\bb n}$, with $i=1,2$. These correspond to nodes $\#1$ and $\#2$ of Figure~\ref{fig:UnitCell_2D_Topology}\subref{fig:2DUnitCellSchematic}; the remaining nodes are equivalent to node $\#1$ in the neighboring unit cells and therefore need not be introduced as independent degrees of freedom.
We define the basis vectors of the lattice $(\bb e_1,\bb e_2)$, with $\bb e_1=\bb c_3$ and $\bb e_2=\bb c_4$. The position of a generic node belonging to cell $\bb n$ can therefore be written as
\begin{equation}
\bb c_i^{\bb n}=\bb c_i+\mathsf E\bb n,
\end{equation}
where $\mathsf E=[\bb e_1,\bb e_2]$ is the lattice transformation matrix. We also define the displacement vector of cell $\bb n$ as
\begin{align}
\bb u(\bb n)&=
\begin{bmatrix}
\bb u_1^{\bb n}\\
\bb u_2^{\bb n}
\end{bmatrix},
\end{align}
where $\bb u_i^{\bb n}$ contains the degrees of freedom of the $i$-th node of the $\bb n$-th cell.
We now consider the connections between a representative unit cell $\bb n$ and the surrounding cells. Owing to periodicity, every cell has the same connectivity and therefore obeys the same equation, up to a translation. It is thus sufficient to formulate the equation for one representative cell, while retaining its interactions with cells at relative positions $\bb p$. We represent the degrees of freedom of such a cell by $\bb u(\bb n+\bb p)$ and define the set of relevant relative positions as
\begin{align}
\mathcal P&=\{-1,0,1\}^2.
\end{align}
We denote by $\mathsf X(\bb p)$ the contribution of the cell $\bb n+\bb p$ to the equation governing cell $\bb n$, where $\mathsf X$ denotes either $\mathsf K$ or $\mathsf M$. Since the interactions are limited to the cells in $\mathcal P$, $\mathsf X(\bb p)=0$ for $\bb p\in\mathbb Z^2\setminus\mathcal P$.
With this formulation and assuming time-harmonic motion, the equation that governs the dynamics of the unit cell $\bb n$ is
\begin{align}
    \sum_{\bb p\in\mathcal P}
    \left[
    \left(
    \mathsf K(\bb p)-\omega^2\mathsf M(\bb p)
    \right)
    \bb u(\bb n+\bb p)
    \right]
    =0.
    \label{eq:EoM}
\end{align}
By assembling the matrices, the corresponding coupling matrices are obtained as follows:
\begin{align}
\begin{aligned}
        \mathsf X([-1,0])& =
    \begin{bmatrix}
    [0] & \hat{\mathsf X}_{23}^{\top}\\
    [0] & [0]
    \end{bmatrix},
    &
    \mathsf X([0,-1])& =
    \begin{bmatrix}
    [0] & \hat{\mathsf X}_{24}^{\top}\\
    [0] & [0]
    \end{bmatrix},
    \\
    \mathsf X([0,0])& =
    \begin{bmatrix}
    \hat{\mathsf X}_{11}+\hat{\mathsf X}_{33}+\hat{\mathsf X}_{44} & \hat{\mathsf X}_{12}\\
    \hat{\mathsf X}_{12}^{\top} & \hat{\mathsf X}_{22}
    \end{bmatrix},
    &
    \mathsf X([0,1])& =
    \begin{bmatrix}
    [0] & [0]\\
    \hat{\mathsf X}_{24} & [0]
    \end{bmatrix},
    \\
    \mathsf X([1,0])& =
    \begin{bmatrix}
    [0] & [0]\\
    \hat{\mathsf X}_{23} & [0]
    \end{bmatrix}
    &
    \mathsf X([\pm 1,\pm 1])& =
    \begin{bmatrix}
    [0] & [0]\\
    [0] & [0]
    \end{bmatrix}.
\end{aligned}
\end{align}
We introduce the Fourier transform in space
\begin{align}
    \bb u^F(\bs\kappa)
    =
    \sum_{\bb n\in\mathbb Z^2}
    \bb u(\bb n)
    e^{-i\bs\kappa\cdot\mathsf E\bb n},
\end{align}
where $\bs\kappa$ is the wavevector associated with the lattice wave.
Applying this transform to Eq.~\eqref{eq:EoM}, the dispersion relation of the perfect infinite lattice is obtained:
\begin{align}
    \sum_{\bb p\in\mathcal P}
    \left[
    \left(
    \mathsf K(\bb p)
    % e^{i\bs\kappa\cdot\mathsf E\bb p}
    -
    \omega^2
    \mathsf M(\bb p)
    \right)
    e^{i\bs\kappa\cdot\mathsf E\bb p}
    \bb u^F(\bs\kappa)
    \right]
    =0,
    \label{eq:EoM_kappa}
\end{align}
or, in a more compact form, 
\begin{align}
    \left(
    \sigma(\bs \kappa)
    -
    \omega^2
    \mu(\bs\kappa)
    \right)
    \bb u^F(\bs\kappa)
    &=0,
    \label{eq:EoM_kappa_compact}
\end{align}
with 
\begin{align}
    \sigma(\bs \kappa)&=  \sum_{\bb p\in\mathcal P} \mathsf K(\bb p)e^{i\bs\kappa\cdot\mathsf E\bb p}
    &
    \mu(\bs\kappa)&= \sum_{\bb p\in\mathcal P}
    \mathsf M(\bb p)e^{i\bs\kappa\cdot\mathsf E\bb p},
\end{align}
The matrices $\mathsf K(\bb p)$, and therefore $\sigma(\boldsymbol \kappa)$, depend on the axial load $P$ through the stiffness coefficients in Eq.~\eqref{eq:K_beam_2D}, which are in turn determined by the external pressure $p$. We therefore establish the relation between $P$ and $p$, which will be used below to express the lattice response and identify the buckling condition in terms of the applied pressure. Referring to the loading configuration shown in Figure~\ref{fig:UnitCell_2D_Topology}\subref{fig:Load2D},
% \begin{figure}
%     \centering
%     \subfloat[]{%
%         \includegraphics[
%             width=0.3\linewidth,
%             % trim=4.5cm 0.15cm 5cm 0.15cm,
%             % clip
%         ]{Figures/Modeling/2DLoadDefinition}%
%         \label{fig:LoadDefinition2D}
%     }
%     \hfill
%     \subfloat[]{%
%         \includegraphics[
%             width=0.5\linewidth,
%             % trim=5.4cm 0cm 5cm 0.15cm,
%             % clip
%         ]{Figures/Modeling/3DLoadDefinition}%
%         \label{fig:LoadDefinition3D}
%     }
%     \caption{\cbox{Caption}}
%     \label{fig:LoadDefinition}
% \end{figure}
for the two-dimensional lattice
\begin{align}
    P=p\sqrt{3}L.
    \label{eq:relation_p_P_2D}
\end{align}
This relation can be independently verified by computing the effective stress generated in the lattice when the load $P$ is applied to each link. Using the formulation derived in \cite{Norris2014MechanicsNetworks},
\begin{equation}
    \sigma_{\mathrm{eff}} = -\frac {LP} {V_{\mathrm{cell}}}\sum_{i=1,i\neq 2}^4 \bb v_{2i,x}\otimes\bb v_{2i,x} = -p I, 
    \label{eq:sigmaNorris}
\end{equation}
where $V_{\mathrm{cell}} = 3\sqrt{3}/2L^2$ is the unit cell volume, $\bb v_{2i,x}$ the unit vector directed along the links from the central node $\#2$ to node $\# i$, and $I$ the $2\times2$ identity matrix. The result is the hydrostatic stress $-pI$. Therefore, these axial forces in the links reproduce the externally applied pressure correctly at the macroscopic level.
% The framework established in this section will be employed in the following section to analyze the stability of the lattice.
\\
\\
% \subsubsection{3D Lattice}
We now consider the 3D pentamode lattices schematized in Figure~\ref{fig:UnitCell_3D_Topology}\subref{fig:3DUnitCellSchematic}. 
Following \cite{Milton1995WhichRealizable, Kadic2012OnMetamaterials, Brambilla2025HighWater}, the unit cell possesses diamond symmetry.
Assuming that all node-to-node distances are equal to $L$, four unit cells can be inscribed in a cube of side length
\begin{align*}
    a=\frac{4L}{\sqrt3},
\end{align*}
as shown in Figure~\ref{fig:UnitCell_3D_Topology}\subref{fig:cubic_symmetry}.
The nodal coordinates are therefore
\begin{align}
    \left\{
        \begin{aligned}
            \bb c_1 &= (0,0,0),\\
            \bb c_2 &= \left(\frac a4,\frac a4,\frac a4\right),\\
            \bb c_3 &= \left(\frac a2,\frac a2,0\right),\\
            \bb c_4 &= \left(\frac a2,0,\frac a2\right),\\
            \bb c_5 &= \left(0,\frac a2,\frac a2\right).
        \end{aligned}
    \right.
\end{align}
The beam direction vector is still defined as
\begin{align}
    \bb v_{ij,x} =
    \frac{\bb c_j-\bb c_i}{\|\bb c_j-\bb c_i\|}.
\end{align}
The vectors $\bb v_{ij,y}$ and $\bb v_{ij,z}$ are chosen to complete an orthonormal basis. Due to the symmetries imposed to the link, any orthonormal choice of $\bb v_{ij,y}$ and $\bb v_{ij,z}$ yields the same final result.
The corresponding rotation matrices become
\begin{align}
\mathsf R_{ij,3D}
&=
\begin{bmatrix}
\mathsf R^*_{ij,3D} &[0] &[0] &[0]\\
[0] & \mathsf R^*_{ij,3D} &[0] &[0]\\
[0] &[0] & \mathsf R^*_{ij,3D} &[0]\\
[0] &[0] &[0] & \mathsf R^*_{ij,3D}
\end{bmatrix}, \textrm{ with}
&
\mathsf R^*_{ij,3D}
&=
\begin{bmatrix}
    \bb v_{ij,x}^{\top}\\
    \bb v_{ij,y}^{\top}\\
    \bb v_{ij,z}^{\top}
\end{bmatrix}.
\end{align}
Following the same assembly procedure adopted for the 2D case, the unit-cell matrices takes the form
\begin{align}
\label{eq:3DUnitCellMatrix}
\hat{\mathsf X}_{3D}=
\begin{bmatrix}
\hat{\mathsf X}_{11} & \hat{\mathsf X}_{12} & [0] & [0] & [0]\\
\hat{\mathsf X}_{12}^{\top} & \hat{\mathsf X}_{22} & \hat{\mathsf X}_{23} & \hat{\mathsf X}_{24} & \hat{\mathsf X}_{25}\\
[0] & \hat{\mathsf X}_{23}^{\top} & \hat{\mathsf X}_{33} & [0] & [0]\\
[0] & \hat{\mathsf X}_{24}^{\top} & [0] & \hat{\mathsf X}_{44} & [0]\\
[0] & \hat{\mathsf X}_{25}^{\top} & [0] & [0] & \hat{\mathsf X}_{55}
\end{bmatrix},
\end{align}
where $\hat{\mathsf X}_{ij}\in\mathbb R^{6\times6}$.
For the 3D lattice, the basis vectors are $(\bb e_1,\bb e_2,\bb e_3)$, where $\bb e_i = \bb c_{i+2}$,  and each unit cell, identified by the index $\bb n\in\mathbb Z^3$, contains two independent nodes as shown in Figure~\ref{fig:UnitCell_3D_Topology}\subref{fig:3D}. Their positions are given by
\begin{align}
    \bb c_i^{\bb n}=\bb c_i+\mathsf E\bb n,
\end{align}
with $\mathsf E=[\bb e_1,\bb e_2,\bb e_3]$.

The governing equations retain the form of Eq.~\eqref{eq:EoM}, with $\mathcal P=\{-1,0,1\}^3$. The non-vanishing coupling matrices are
\begin{align}
\begin{aligned}
        \mathsf X([0,0,-1])& =
    \begin{bmatrix}
    [0] & \hat{\mathsf X}_{25}^{\top}\\
    [0] & [0]
    \end{bmatrix},
    &
    \mathsf X([0,-1,0])& =
    \begin{bmatrix}
    [0] & \hat{\mathsf X}_{24}^{\top}\\
    [0] & [0]
    \end{bmatrix},
    \\
    \mathsf X([-1,0,0])& =
    \begin{bmatrix}
    [0] & \hat{\mathsf X}_{23}^{\top}\\
    [0] & [0]
    \end{bmatrix},
    &
    \mathsf X([0,0,0])& =
    \begin{bmatrix}
    \hat{\mathsf X}_{11}+\hat{\mathsf X}_{33}+\hat{\mathsf X}_{44}+\hat{\mathsf X}_{55}
    &
    \hat{\mathsf X}_{12}\\
    \hat{\mathsf X}_{12}^{\top}
    &
    \hat{\mathsf X}_{22}
    \end{bmatrix},
    \\
    \mathsf X([1,0,0])& =
    \begin{bmatrix}
    [0] & [0]\\
    \hat{\mathsf X}_{23} & [0]
    \end{bmatrix},
    &
    \mathsf X([0,1,0])& =
    \begin{bmatrix}
    [0] & [0]\\
    \hat{\mathsf X}_{24} & [0]
    \end{bmatrix},
    \\
    \mathsf X([0,0,1])& =
    \begin{bmatrix}
    [0] & [0]\\
    \hat{\mathsf X}_{25} & [0]
    \end{bmatrix}.
\end{aligned}
\end{align}
The Fourier-transformed problem follows directly from the two-dimensional case, again leading to Eq.~\eqref{eq:EoM_kappa}. 

Finally, considering Figure~\ref{fig:UnitCell_3D_Topology}\subref{fig:Load3D} the relation between the axial load and the externally applied pressure is
\begin{align}
    P=p\frac{4}{\sqrt{3}}L^2.
    \label{eq:relation_p_P_3D}
\end{align}
As in the two-dimensional case, Eq.~\eqref{eq:sigmaNorris} with $V_{\mathrm {cell}} = a^3/4 = 16L^3/(3\sqrt 3)$ and extended up to node~$\#5$ can be used to verify this relation. Substituting the corresponding link directions and the relation between $P$ and $p$ into Eq.~\eqref{eq:sigmaNorris} yields the hydrostatic stress $-pI$, where $I$ denotes the $3\times3$ identity matrix.

\section{Buckling Instability of the lattice}
\label{sec:Instability}

% In this section, we determine the limit pressure for the lattices considered previously. Two possible collapse mechanisms can occur at different scales. The lattice structure can collapse collectively due to a global instability of the effective medium, or it can experience a local instability of the representative unit cell. The following two subsections are devoted, respectively, to the study of these two mechanisms, providing a closed-form expression for the collapse pressure in each case. The critical limit pressure is then taken as the minimum of the two values. 
In this section, we determine the buckling limit pressure for the lattices considered previously. Two instability mechanisms can occur at different scales. The lattice can lose stability through a global instability of the effective medium, or through a local instability associated with wavelengths comparable to the unit-cell size. The following two subsections are devoted to these two mechanisms, respectively, and provide a closed-form expression for the corresponding buckling pressure. The critical limit pressure is then given by the minimum of the two values.

To identify the onset of instability, we consider the dynamic response of the lattice as the applied pressure increases. In particular, the instability threshold is associated with the appearance of an additional zero-frequency mode, beyond those corresponding to rigid-body motions. The latter are always present at $\boldsymbol\kappa=\boldsymbol 0$ and do not indicate an instability. At the instability threshold, the frequency of an additional mode reaches zero; beyond this point, its squared frequency becomes negative, corresponding to an imaginary eigenfrequency and an unstable mode \cite{Bokaian1988NaturalLoads}.
Since the lattices considered here are infinite periodic structures, their dynamics are characterized by infinitely many modes. Rather than analyzing the full lattice directly, we use the dispersion relation derived in the previous section, which provides the natural frequencies as functions of the wavevector and the applied load.

As in the previous sections, we first consider the simpler two-dimensional structures and then extend the analysis to the more complex three-dimensional case.

\subsection{Global (Homogenized) Instability}
 
The global instability is associated with the homogenized response of the lattice, obtained by considering the dispersion relation in the long-wavelength limit, i.e. $\boldsymbol\kappa\rightarrow\boldsymbol 0$. In this limit, the acoustic branches are linear, with slopes determined by the corresponding wave propagation velocities. Therefore, beside the rigid-body modes already present at $\boldsymbol\kappa=\boldsymbol 0$, an additional zero-frequency mode can arise when the propagation velocity vanishes for a given direction. In the homogenized description, this condition corresponds to the vanishing of at least one eigenvalue of the effective elasticity matrix. As shown in \cite{Adhikari2021TheSolutions}, this is a necessary and sufficient condition for buckling.
We therefore first derive the effective elasticity matrix from the long-wavelength limit of the dispersion relation, following the homogenization framework of \cite{Colquitt2015HighLattices}. The global buckling limit is then determined by identifying the applied pressure at which one of the eigenvalues of the effective elasticity matrix becomes zero.

Physically, homogenization describes the response of the lattice when observed at a length scale much larger than the size of its unit cell. The lattice exhibits two distinct spatial scales: a short scale associated with the periodic unit-cell structure, and a long scale associated with the variation of the displacement over many unit cells. The short-scale behavior is described by the discrete cell index $\bb n$, while the slow variation over the long scale is described by introducing the variable
\begin{equation}
\bs\eta=\varepsilon\mathsf E\bb n,
\end{equation}
where $\mathsf E$ maps the cell coordinates onto the Cartesian coordinate system
and $0<\varepsilon\ll1$ characterizes the separation between the two scales. Indeed, while moving from cell $\bb n$ to cell $\bb n+\bb p$ corresponds to a finite change $\bb p$ in the short-scale variable, the corresponding change in the long-scale variable $\bs\eta$ is equal to $\varepsilon\mathsf E\bb p$, i.e., of order $ \varepsilon$. The small parameter $\varepsilon$ therefore quantifies the variation of the long-scale variable over a finite number of unit cells.
For example, if $N\gg1$ denotes the number of unit cells contained within one wavelength, we can take $\varepsilon=1/N$. In this case, the two scales describe variations at the level of a single unit cell and at the level of the wavelength, respectively.

To account for these two distinct types of variation, we treat the displacement as depending on both the discrete cell index and the long-scale variable, namely, $\bb u (\bb n ) = \bb u (\bb n, \bs \eta)$. Equation~\eqref{eq:EoM} then becomes
\begin{align}
    \sum_{\bb p \in \mathcal P}\left[\left(\mathsf K(\bb p) -\omega^2 \mathsf M(\bb p)\right) \bb u(\bb n +\bb p, \bs \eta+\varepsilon\mathsf E \bb p) \right]=0.
    \label{eq:EoM_twoVar}
\end{align}
Following \cite{Colquitt2015HighLattices}, we expand $\bb u$ in a Taylor series with respect to the long-scale variable $\bs \eta$, obtaining an asymptotic expansion in $\varepsilon$. We distinguish the different orders $\varepsilon^a$ in the expansion, with higher order terms providing progressively smaller contributions. The expansion yields: 
\begin{align}
    \left[ 
        \left(
            \sum_{a=0}^\infty\varepsilon^a\sigma_{(a)}
        \right) 
        -\omega^2
        \left(
            \sum_{a= 0}^\infty \varepsilon^a\mu_{(a)} 
        \right)
    \right]
    \bb u\left(\bb n, \bs \eta\right)
\end{align}
where 
\begin{align}
\label{eq:sigma_order_a}
    \sigma_{(a)} &= \sum_{\bb p \in \mathcal P}\frac{1}{a!}\mathsf K(\bb p )\left((\mathsf E \bb p)\cdot\nabla\right)^a, \\
\label{eq:mu_order_a}
    \mu_{(a)} &= \sum_{\bb p \in \mathcal P}\frac{1}{a!}\mathsf M(\bb p )\left((\mathsf E \bb p)\cdot\nabla\right)^a,
\end{align}
and $\nabla =\nabla_{\bs\eta}$ denotes the gradient with respect to the long-scale variable $\bs\eta$.
At the homogenized scale, the variations within individual unit cells are not resolved, and only the slow variation over many unit cells is retained. Accordingly, at leading order, the displacement is taken to depend only on the long-scale variable, so that $\bb u(\bb n, \bs \eta) = \bb u (\bs \eta)$, with the dependence on the short-scale variable $\bb n$ removed.

For $\bb u (\bs \eta)$ and $\omega^2$ we adopt the following ansatz: 
\begin{align}
    \omega^ 2 & = \sum_{a = 0}^\infty\varepsilon^a\omega_{(a)}^2, \\
    \bb u(\bs \eta ) & = \sum_{a = 0}^\infty\varepsilon^a\bb u_{(a)}(\bs \eta ).
\end{align}
This corresponds to representing their solutions in terms of leading-order contributions, $\omega_{(0)}^2$ and $\bb u_{(0)}$, which describe the response at the homogenized scale, together with higher-order contributions, $\omega_{(a)}^2$ and $\bb u_{(a)}(\bs\eta)$, which provide corrections at order $\varepsilon^a$.

We can solve the problem sequentially, collecting terms at each order in $\varepsilon$, starting from the leading-order terms and proceeding to higher-order corrections.  Since we are interested in the long-wavelength, linear dispersion, the frequency scales linearly with the wavenumber, $\omega\sim\|\bs\kappa\|$, and therefore $\omega^2$ is quadratic in $\|\bs\kappa\|$. It is therefore sufficient to retain terms up to order $\varepsilon^2$. Collecting the terms at each order gives
\begin{align}
    \left\{
    \begin{aligned}
        \mathcal O(\varepsilon^0): && [\sigma_{(0)} -\omega_{(0)}^2\mu_{(0)}]\bb u_{(0)}(\bs \eta) &= 0\\
        \mathcal O(\varepsilon^1): && [\sigma_{(0)} -\omega_{(0)}^2\mu_{(0)}]\bb u_{(1)}(\bs \eta) &= -[\sigma_{(1)} -\omega_{(1)}^2\mu_{(0)} -\omega_{(0)}^2\mu_{(1)}]\bb u_{(0)}(\bs \eta)\\
        \mathcal O(\varepsilon^2): && [\sigma_{(0)} -\omega_{(0)}^2\mu_{(0)}]\bb u_{(2)}(\bs \eta)
        &= \begin{aligned}[t]
            &-[\sigma_{(1)} -\omega_{(1)}^2\mu_{(0)} -\omega_{(0)}^2\mu_{(1)}]\bb u_{(1)}(\bs \eta)\\
            &\qquad -[\sigma_{(2)} -\omega_{(2)}^2\mu_{(0)} -\omega_{(1)}^2\mu_{(1)} -\omega_{(0)}^2\mu_{(2)}]\bb u_{(0)}(\bs \eta)
        \end{aligned}
    \end{aligned}
    \right.
\end{align}

At order $\mathcal O(\varepsilon^0)$, we obtain a classical eigenvalue problem. The matrices $\sigma_{(0)}$ and $\mu_{(0)}$ are, respectively, $\sum_{\bb p \in \mathcal P}\mathsf K(\bb p)$ and  $\sum_{\bb p \in \mathcal P}\mathsf M(\bb p)$. These are the same matrices that would be obtained by imposing rigid connections between the boundary nodes ($\#1$, $\#3$, and $\#4$) of the single unit cell shown in Figure~\ref{fig:UnitCell_2D_Topology}\subref{fig:2DUnitCellSchematic}, with the unit-cell matrices given in Eq.~\eqref{eq:UnitCellMatrix}. The $\mathcal O(\varepsilon^0)$ problem therefore admits two zero eigenvalues, $\omega_{(0),i}^2$, corresponding to the two allowed rigid-body motions. The associated eigenvectors are 
\begin{align}
    \bb v_i& = [\delta_{1i}, \delta_{2i}, 0,\delta_{1i},\delta_{2i}, 0]^\top, &i \in \{1,2\}
\end{align}
where $\delta_{ij}$ is the Kronecker delta function. 
We do not consider the remaining eigenvalues, as we are interested in the low-frequency behavior. The leading-order displacement can therefore be written as
\begin{align}
        \label{eq:u_order0}
        \bb u_{(0)} &= \mathsf V\boldsymbol \varphi(\bs \eta)
\end{align}
where      
\begin{align}
    \mathsf V &= [\bb v_1, \bb v_2] \\
    \bs \varphi(\bs \eta) &= [\varphi_1(\bs \eta), \varphi_2(\bs \eta)]^\top.
\end{align}
This solution describes the leading-order motion: the vectors $\bb v_i$ represent rigid motions at the short scale, while the functions $\varphi_i(\bs\eta)$ describe their variation over the long scale. Thus, at leading order, the unit cell moves rigidly, while its macroscopic displacement varies over the long scale.

We next consider the problem at order $\mathcal O (\varepsilon^1)$. Since $\omega_{(0)}^2=0$ for the acoustic modes of interest, the governing equation reduces to
\begin{align}
    \sigma_{(0)}\bb u_{(1)}(\bs \eta) = -[\sigma_{(1)}-\omega_{(1)}^2\mu_{(0)}]\bb u_{(0)}.
\end{align}
We impose solvability using the Fredholm alternative (see e.g. \cite{benisrael1974generalized}). Because $\sigma_{(0)}$ is singular, the Fredholm alternative requires the right-hand side to be orthogonal to the null space of $\sigma_{(0)}$. In the present case, this null space is spanned by the columns of $\mathsf V$, and therefore
\begin{align}
    \mathsf V^\top\left[\sigma_{(1)} -\omega_{(1)}^2\mu_{(0)}\right]\mathsf V\bs \varphi(\bs \eta) = 0.
\end{align}
Since in our case $\mathsf V^\top \sigma_{(1)}\mathsf V = 0$ and $\mathsf V^\top \mu_{(0)}\mathsf V $ is non-singular, this condition yields $\omega_{(1)}^2 = 0$. 
This result is consistent with the linear acoustic dispersion in the long-wavelength limit, for which $\omega\propto\|\bs\kappa\|$ and therefore $\omega^2$ starts at order $\|\bs\kappa\|^2$, corresponding here to order $\varepsilon^2$.
The solution of the first-order problem can then be written as
\begin{align}
    \bb u_{(1)} = -\sigma_{(0)}^+\sigma_{(1)}\mathsf V\bs \varphi(\bs\eta) + \mathsf V\bs \psi(\bs\eta),
\end{align}
where $(\cdot)^+$ denotes the Moore-Penrose pseudoinverse. The term $\sigma_{(1)}$ is a first-order differential operator with respect to the long-scale variable $\bs\eta$. Thus, the first term represents the correction of order $\varepsilon$ to the rigid-cell motion induced by the spatial variation of the leading-order macroscopic displacement field $\bs\varphi(\bs\eta)$. The second term belongs to the null space of $\sigma_{(0)}$ and hence has the same rigid-motion structure as the leading-order solution.

Order $\mathcal O (\varepsilon^2)$ is reduced to
\begin{align}
    \sigma_{(0)}\bb u_{(2)} = -\sigma_1\bb u_{(1)}-\left[\sigma_{(2)} -\omega_{(2)}^2\mu_{(0)} \right]\bb u_{(0)}.
\end{align}
Applying the Fredholm solvability condition yields the following second-order differential equation for the macroscopic displacement field $\bs\varphi(\bs \eta)$:
\begin{align}
\label{eq:homogenizedEoM}
    \left[\mathsf V^\top \sigma_{(1)}\sigma_{(0)}^+\sigma_{(1)}\mathsf V -\mathsf V^\top \sigma_{(2)}\mathsf V  \right]\bs\varphi(\bs \eta)+\omega_{(2)}^2\left[\mathsf V^\top \mu_{(0)}\mathsf V \right]\bs \varphi(\bs \eta) = 0.
\end{align}
Since $\omega_{(0)}^2=\omega_{(1)}^2=0$, the leading nonzero contribution to the frequency is $\omega^2=\varepsilon^2\omega_{(2)}^2$. We therefore refer to $\omega_{(2)}^2$ simply as $\omega^2$. Eq.~\eqref{eq:homogenizedEoM} thus provides the governing equation for the wave propagation in the homogenized lattice.
% The solution of Eq.~\eqref{eq:homogenizedEoM} describes the motion of the homogenized lattice under wave propagation. Having $\mu_{(0)}$ expressed in Eq.~\eqref{eq:mu_order_a}, we can compute the term $\mathsf V ^\top\mu_{(0)}\mathsf V$.
Since $\mu_{(0)}$ is given by Eq.~\eqref{eq:mu_order_a}, the effective mass matrix can be obtained directly as $\mathsf V ^\top\mu_{(0)}\mathsf V$.
For the 2D lattices considered here, this term yields the $2\times2$ matrix
\begin{align}
    \mathsf V ^\top\mu_{(0)}\mathsf V =    3(a_M+e_M+b_M+g_M)      
         \begin{bmatrix}
             1 & 0\\
             0 & 1
         \end{bmatrix} = M_{\mathrm{tot}}[\delta_{ij}]
\end{align}
where we have used $(a_M+e_M)=(b_M+g_M)=M_{\mathrm{link}}/2$ and denote the total mass of a single unit cell by $M_{\mathrm{tot}}= 3M_{\mathrm{link}}$.

Using  the unit cell volume $V_{\mathrm{cell}} = 3\sqrt3/2L^2$,
we define the lattice density as $\rho_{\mathrm{cell}} = M_{\mathrm{tot}}/V_{\mathrm{cell}}$. Dividing Eq.~\eqref{eq:homogenizedEoM} by $V_{\mathrm{cell}}$, then gives
\begin{align}
\label{eq:homogenizedEoM_normalized}
   \frac{1}{V_{\mathrm{cell}}} \left[\mathsf V^\top \sigma_{(1)}\sigma_{(0)}^+\sigma_{(1)}\mathsf V -\mathsf V^\top \sigma_{(2)}\mathsf V  \right]\bs\varphi(\bs \eta)+\omega^2\rho_{\mathrm{cell}}\bs \varphi(\bs \eta)=0.
\end{align}
We can then compare this equation with the general elastodynamic (Navier--Cauchy) equation $\nabla\cdot(\mathbb C\vcentcolon\nabla\bb u) +\omega^2\rho\bb u  = 0$ \cite{Graff1991WaveSolids}, and identify the corresponding effective elasticity tensor $\mathbb C$ for the lattice under consideration. 

% The Navier-Cauchy equation can be expanded as follows
% \begin{align}
%     \sum_{i,j = 1}^3 \mathsf A_{ij}\frac{\partial^2}{\partial\eta_{i}\partial\eta_j} \bb u +\omega^2\rho \bb u=0
% \end{align}
% where $\mathsf A_{ij}$ are six different $3\times 3$ matrices depending on the coefficients $\mathbb C_{ijkl}$ of the constitutive matrix. 
In 2D, the Navier--Cauchy equation can be simplified by considering plane strain and neglecting variations along the third dimension, such that $\partial\bb u/\partial\eta_3=\bb 0$. The equation therefore involves only the two spatial variables $\eta_1$ and $\eta_2$, and takes the same form as the homogenized equation derived above.
Expressing $\mathbb C$ in Voigt notation, the 2D constitutive matrix under these assumptions is given by
\begin{align}
    \mathbb C_{Voigt} = 
        \begin{bmatrix}
             C_{1111} &  C_{1122} & 0 \\
             C_{1122 } &  C_{2222} & 0 \\
            0 & 0 & 2 C_{1212}
        \end{bmatrix} 
\end{align}
where
\begin{align}
    \begin{aligned}
        C_{1111} = C_{2222} & = \frac{a_K }{2\sqrt 3}\frac{a_K+3b_K}{a_K+b_K},\\
        C_{1122}  & = \frac{a_K}{2\sqrt 3}\frac{a_K-b_K}{a_K+b_K}+\frac{P}{\sqrt 3L}, \\
        2C_{1212} & = \frac{ 2a_K b_K}{\sqrt 3(a_K+b_K)}- \frac{P}{\sqrt 3L}.
    \end{aligned}
\end{align}
These expressions show that $\mathbb C$ depends solely on the load acting on each beam $P$, and on the stiffness coefficient $a_K,b_K$ of the link stiffness matrix expressed in Eq.~\eqref{eq:K_beam_2D}. The stiffness coefficients themselves depend on the applied load through the contribution of the geometric stiffness.

The eigenvalues of $\mathbb C$ are
\begin{align}
    \label{eq:lambda1_2D}
    \sqrt 3 \lambda_1 &= a_K+\frac{P}{L}; \\
    \label{eq:lambda23_2D}
    \sqrt 3 \lambda_2 = \sqrt 3 \lambda_3 &= \frac{2a_Kb_K}{a_K+b_K}-\frac PL  .
\end{align}
These expressions are consistent with the isotropic effective response of the honeycomb, as only two independent eigenvalues are present. 
% For the compressive loading considered here, as is commonly the case in such structures, $a_K>b_K$, and the global buckling limit is associated with the vanishing of $\lambda_2=\lambda_3$.

For linear buckling analysis, we assume that the stiffness matrix depends linearly on the external load, such that $\mathsf K = \mathsf K_{e}-P\mathsf K_{g}$, where $\mathsf K_e $ is the elastic stiffness matrix and $\mathsf K_g$ is the geometric stiffness matrix, which depends only on the geometry of the element. This assumption implies $a_K = a_{K,e}-Pa_{K,g}$ and $b_K = b_{K,e}-Pb_{K,g}$. In most cases, including those shown in Figure~\ref{fig:Link_and_cell_Models}, $a_{K,g}= 0$. % and  $a_K>b_K$

The global buckling limit is reached when one of the eigenvalues of the effective elasticity matrix vanishes. Considering  $a_K>b_K$, the relevant condition is $\lambda_2=\lambda_3=0$. Substituting the load-dependent expression for $b_K$ into Eq.~\eqref{eq:lambda23_2D} yields an equation for the critical load $P$. Using the relation between $P$ and the external pressure $p$, the corresponding buckling pressure can be obtained.

For the lattices considered here, we typically have $a_K\gg b_K$. Under this assumption, Eq.\eqref{eq:lambda23_2D} simplifies to
\begin{align}
    \sqrt 3 \lambda_2 = \sqrt 3 \lambda_3 &\approx 2b_K-\frac PL  
\end{align}
The corresponding global buckling pressure is therefore
\begin{align}
    \label{eq:p_lim_hom_2D_generic_simplified}
    \frac{p_{lim,hom}\sqrt{3}}{a_K} = \frac{2b_{K,e}/a_K}{2b_{K,g}L+1}
\end{align}

If the approximation $a_K\gg b_K$ is not adopted, the same procedure can be applied directly to Eq.~\eqref{eq:lambda23_2D}. Solving the resulting nonlinear equation for the critical load gives the closed-form expression
\begin{align}
\label{eq:p_lim_hom_2D_generic}
    \frac{p_{lim, hom}\sqrt{3}}{a_{K}} &= H\left(1-\sqrt{1-\frac{\frac{b_{K,e}}{a_{K}}\frac{2}{b_{K,g}L}}{H^2}}\right) &\qquad
    H & = \left(1+\frac{1}{2b_{K,g}L}\left(1+\frac{b_{K,e}}{a_{K}}\right)\right)
\end{align}
Equations~\eqref{eq:p_lim_hom_2D_generic_simplified} and~\eqref{eq:p_lim_hom_2D_generic} provide closed-form expressions for the global buckling pressure of the 2D lattices considered here, respectively under the approximation $a_K\gg b_K$ and without this approximation. These expressions relate the global buckling pressure directly to the mechanical properties of the individual links and hence provide the desired analytical description of the global instability limit for the 2D lattices.

As an example, this result can be applied to the reference two-dimensional structures considered in this work.
Let us consider the more generic structure in Figure~\ref{fig:Link_and_cell_Models}\subref{fig:2DUnitCellFasci}. The links between nodes consist of  two circles, cut at a distance $t = \tau L$, with $\tau \in (0,1/2)$, from the center of each node and connected to each other through a sheaf of $N_f$ fibers of length $L_f = L(1-2\tau)$ and height $h_f = \beta L_f$. With this convention, the classical honeycomb in Figure~\ref{fig:Link_and_cell_Models}\subref{fig:2DUnitCell} is recovered by setting $\tau = 0, N_f = 1$. Here, we consider the circular regions to be rigid. This assumption is reasonable, since their stiffness is much larger than that of the fiber bundles, particularly for transverse deformations. The resulting approximation is not expected to significantly affect the final result, as verified in Section \ref{sec:4_NumericalValidation}. However, if a more detailed model is required, the contribution of the circular regions can instead be included, for example by modeling them as beams with an effective length or by computing the stiffness coefficients of the links using a finite-element model.

We consider slender fibers ($\beta\ll 1$) and model them as Euler--Bernoulli beams. This assumption is consistent with the limited shear-to-axial stiffness ratio required for the links in these lattices to operate in the desired PM regime.
In Appendix~\ref{sec:AppendixSheafOfBeams} we derive $a_K = N_f E_{ps} \beta$, $b_{K,e} = N_fE_{ps}\beta^3$ and $b_{K,g}=6/(5L_f)$. 
Thus, the global buckling pressure for these honeycomb structures is
\begin{equation}
    \begin{aligned}
    \label{eq:p_lim_exact_2D}
        \frac{p_{lim, hom}\sqrt3}{N_fE_{ps}\beta } &= H\left(1 -\sqrt{1-\frac{\frac53(1-2\tau)\beta^2}{H^2}}\right)&\qquad
        H &= \left(1+\frac{5}{12}(1-2\tau)(1+\beta^2 )\right).
    \end{aligned}
\end{equation}
This closed-form expression provides the global buckling pressure directly in terms of the geometrical and material parameters of the links.
We can better understand the influence of the parameters by expanding the square root in a Taylor series about $\beta = 0$, or equivalently by substituting the expressions for the stiffness coefficients into Eq.\eqref{eq:p_lim_hom_2D_generic_simplified}, which gives
\begin{align}
    \frac{p_{lim, hom}\sqrt3}{N_fE_{ps}\beta } = \frac{10(1-2\tau)}{17-10\tau}\beta^2+o(\beta^3).
\end{align}
This result shows that the global buckling pressure depends on:
\begin{itemize}
    \item the axial stiffness of the links $a_K = N_f E_{ps}\beta$ linearly;
    \item the square of the beam aspect ratio $\beta ^2$, when the axial stiffness is kept fixed;
    \item the parameter $\tau$, with increasing values having a detrimental effect on stability. Here, $0<\tau<1/2$.
\end{itemize}
A small aspect ratio $\beta$ is a key parameter for achieving the desired PM behavior, but reducing it also has a detrimental effect on the global buckling pressure. This effect can, however, be mitigated by increasing the stiffness of the links, for example by increasing the number of fibers $N_f$. Thus, the analytical result provides a direct means of balancing the requirements for PM behavior and structural stability.\\
\\
The same procedure is applied to the 3D pentamode structures. 
The only difference at the leading order is the number of rigid-body modes associated with the null eigenvalues of the $\mathcal O(\varepsilon^0)$ problem. In this case, there are three independent rigid-body modes, associated with eigenvalues $\omega_{(0),i }^2 = 0$ and with corresponding eigenvectors
\begin{align}
    \bb v_i &= [\delta_{i1},\delta_{i2},\delta_{i3},0,0,0,\delta_{i1},\delta_{i2},\delta_{i3},0,0,0]^\top & i&\in\{1,2,3\}
\end{align}
The full Navier-Cauchy equation is considered, hence obtaining the constitutive matrix in Voigt notation: 
\begin{align}
    \mathbb C_{Voigt} = 
        \begin{bmatrix}
            C_{1111} & C_{1122} & C_{1133} & 0& 0 & 0 \\
            C_{1122 } & C_{2222} & C_{2233} & 0& 0 & 0 \\
            C_{1133 } & C_{2233} & C_{3333} & 0& 0 & 0 \\
            0 & 0 & 0 & 2 C_{2323} & 0 & 0\\
            0 & 0 & 0 & 0 & 2 C_{1313} & 0\\
            0 & 0 & 0 & 0 & 0 & 2 C_{1212}
        \end{bmatrix} 
\end{align}
where 
\begin{align}
    C_{1111} = C_{2222} = C_{3333}  & =\frac 4 9 \frac{(a_K+2b_K)L^2}{V_{\mathrm{cell}}},\\
    C_{1122} =C_{1133} = C_{2233} & = \frac 4 9 \frac{\left((a_K-b_K)L^2+3PL\right)}{V_{\mathrm{cell}}}, \\
    2 C_{2323} = 2 C_{1313} = 2 C_{1212} & =\frac{4}{V_{\mathrm{cell}}}\left(\frac{ a_Kb_KL^2}{a_K+2b_K} -\frac{ PL}{3}\right).
\end{align}
and $V_{\mathrm{cell}} = 16L^3/(3\sqrt 3)$.

When needed, the elasticity tensor can be reformulated as the fourth-order tensor $C_{ijkl}$, which is the appropriate representation for constructing the Christoffel tensor and determining the wave propagation velocities along a given direction~\cite{Graff1991WaveSolids}. 
Specifically, for a unit propagation direction $\mathbf{n}$, the Christoffel tensor is defined as
$
\mathsf \Gamma_{ik} = C_{ijkl} n_j n_l,
$
and the wave propagation velocities are obtained by solving the Christoffel eigenvalue problem
$
\left( \mathsf \Gamma - c ^2 \rho \mathsf I \right) \bb u = 0,
\label{eq:Christoffel}
$
where $c$ is the wave velocity, $\rho$ is the material density, $\mathsf I$ the identity matrix, and $\mathbf{u}$ is the corresponding polarization vector. These velocities are then used for the numerical validation presented in Section~\ref{sec:4_NumericalValidation}.

The geometries considered have a cubic symmetry, which
% is anisotropic and 
yields a stiffness matrix with three distinct eigenvalues: 
\begin{align}
\label{eq:lambda1_3D}
    \lambda_1 &=  \frac{\sqrt{3}}{4}\left(\frac{a_K}{L}+2\frac{P}{L^2}\right) \\
    \label{eq:lambda23_3D}
    \lambda_2  = \lambda_3&=  \frac{\sqrt{3}}{4}\left(\frac{b_K}{L}-\frac{P}{L^2}\right) \\
    \label{eq:lambda456_3D}
    \lambda_4  = \lambda_5=\lambda_6&=  \frac{\sqrt{3}}{4}\left(\frac3L\frac{a_Kb_K}{a_K+2b_k}-\frac{P}{L^2}\right) 
\end{align}
Both $\lambda_{2,3}$ and $\lambda_{4,5,6}$ can become negative with the application of an external pressure. However, it is readily verified that $\lambda_{4,5,6}>\lambda_{2,3}$ for $a_K>b_K$, as is also the case for practical 3D lattices. Therefore, the global buckling pressure is obtained by setting $\lambda_{2,3}(P)=0$. Using $b_K = b_{K,e}-Pb_{K,g} $, we obtain
\begin{align}
\label{eq:p_lim_hom_3D_generic}
    \frac{p_{lim, hom}}{a_K/L} = \frac{\sqrt{3}}{4}\frac{b_{K,e}/a_K}{Lb_{K,g}+1}.
\end{align}
Thus, also in the 3D case, the global buckling pressure is directly related to the mechanical properties of the individual links.
% \todo[inline]{concluding remark as in 2D}

As an example, we again consider the 3D lattices shown in Figure~\ref{fig:Link_and_cell_Models}. Each link is composed of two  spheres, properly cut at a distance $t = \tau L$, which are modeled as rigid bodies, and connected to each other by $N_f$ cylindrical fibers of length $L_f = (1-2\tau)L$ and diameter $d_f = \beta L_f$. Using the expressions derived in Appendix~\ref{sec:AppendixSheafOfBeams}, $a_K = \pi E \beta ^2L_fN_f/4$, $b_{K,e} =  3\pi E \beta ^4L_fN_f/16$, and $b_{K,g} = 6/(5L_f)$ and recalling Eq.~\eqref{eq:relation_p_P_3D}, we obtain
\begin{align}
    \frac{p_{lim, hom}}{a_K/L} = \frac{15\sqrt{3}}{16}\frac{(1-2\tau)}{6+5(1-2\tau)}\beta^2.
\end{align} 
% Similar conclusions to those drawn for the 2D case can therefore be obtained for the 3D lattices.
Overall, the homogenization analysis provides a direct analytical relation between the global buckling pressure and the mechanical properties of the individual links. In particular, it highlights the trade-off between the low transverse stiffness required for PM behavior and the resulting susceptibility to global instability, providing a useful guideline for the design of stable PM lattices.

% The analysis ultimately provides the critical buckling load as a function of the homogenized elastic response of pentamode lattices. As this load is approached, the underlying stiffness matrix leads to a progressive reduction of selected eigenvalues, which vanish at the onset of instability. This result not only yields the limit pressure in closed-form, but also suggests that controlled preload can be used as a tuning parameter. By varying the applied compressive load below the instability threshold, one can, in principle, modulate the effective mechanical response of the structure, adjusting how closely it approaches an ideal pentamode behavior. This points toward the possibility of achieving reversible and adaptive control of the metamaterial properties through pre-stress.

\subsection{Local (Unit-Cell) Instability}

If the homogenization procedure shown in the previous subsection focuses on the behavior of the lattice about the origin of the dispersion diagram, i.e., in the long-wavelength limit ($|\bs\kappa|\ll1$), the local instability analysis concerns shorter wavelengths, with characteristic lengths comparable to the size of the unit cell. 
The governing instability mechanism is therefore determined by comparing the limit pressures associated with the global and local instability conditions

A single link or an isolated unit cell with fixed boundary conditions cannot be considered for this purpose, as the additional constraints required to isolate these systems modify their stiffness compared with that of the corresponding cells in the infinite lattice. We therefore perform the local instability analysis starting from Eq.~\eqref{eq:EoM_kappa}, which accounts for the interaction between neighboring unit cells. To determine the local limit, we set in Eq.~\eqref{eq:EoM_kappa} the most critical wavevector $\bs \kappa^\star$. 
The matrices $\mathsf K(\bb p)$ are then expressed explicitly as functions of the applied load $P$. The resulting generalized eigenvalue problem is solved to determine the critical load $P_{lim}$ and, consequently, the corresponding limit pressure $p_{lim}$.

% In 2D, we consider the critical wavevector as $\bs \kappa^\star = \{{2 \pi/(3 L), 0}\}$. This corresponds to the longest wave considered on the boundary of the irreducible Brillouin zone (IBZ). At this wavevector, the stiffness matrix can be written as
We focus on finite wavevectors at the boundary of the irreducible Brillouin zone (IBZ), where the wavelengths are comparable to the unit-cell size. Among these wavevectors, the longest wavelength represents a natural candidate for the critical condition, in analogy with classical buckling problems in which the lowest buckling mode is associated with the longest admissible wavelength. Accordingly, for the considered 2D lattices, we consider the wavevector $\bs\kappa^\star = {{2 \pi/(3 L), 0}}$. The dispersion relations reported in Appendix~\ref{sec:AppendixDispersions} confirm that this wavevector indeed corresponds to the first local instability as the applied pressure is increased. At this wavevector, the stiffness matrix can be written as
\begin{align}
    \sigma(\bs \kappa^\star) &= \sum_{\bb p \in \mathcal P}\mathsf K(\bb p)e^{i \bs \kappa^\star  \mathsf E \bb p} = 
    \begin{bmatrix}
        [\sigma_{11}(\bs \kappa^\star)] &[\sigma_{12}(\bs \kappa^\star)]\\
        [\sigma_{12}(\bs \kappa^\star)]^\top &[\sigma_{22}(\bs \kappa^\star)]
    \end{bmatrix}
\end{align}
where 
\begin{align}
    \begin{aligned}
       [ \sigma_{11}(\bs \kappa^\star)] &= [\sigma_{22}(\bs \kappa^\star)]  = 
       \begin{bmatrix}
           \frac{3}{2}(a_K+b_K) & 0 & 0 \\
           0  & \frac{3}{2}(a_K+b_K) & 0 \\
          0 & 0 & d_K
       \end{bmatrix}
       \\
       [\sigma_{12}(\bs \kappa^\star)]  & = 
       \begin{bmatrix}
           \frac{1}{2}(-a_K+3b_K) & 0 & 0 \\
           0  & \frac{1}{2}(3a_K-b_K) & (b_KL+P) \\
          0 & -(b_KL+P) & d_K-\frac{1}{2}(b_KL+P)L
       \end{bmatrix}.
    \end{aligned}
\end{align}
The local instability condition is obtained by requiring $\det\left[\sigma(\bs\kappa^\star)\right]=0$, which depends on the load $P$ both directly and through $b_K = b_K(P)$ and $d_K = d_K(P)$. The determinant factorizes as
\begin{align}
    \begin{aligned}
        \det(\sigma(\bs \kappa^\star) )& =  2 a_K D_1 D_2D_3 =0\\
        D_1 &= (a_K+3b_K)\\
        D_2 &= 8b_Kd_K -(b_KL+P)( 2b_KL+P)\\
        D_3 &= (3a_K+b_K)\left(2d_K+\frac 12(b_KL+P)L\right)-(b_KL+P)^2.
    \end{aligned}
\end{align}
Assuming $ a_K>b_K >0$, $D_1$ is always positive and $D_3>D_2$. As discussed previously, each stiffness coefficient can be decomposed into elastic and geometric contributions, hence $b_K = b_{K,e}-Pb_{K,g}$ and $d_K = d_{K,e}-Pd_{K,g}$. At $P=0$, $D_3$ is always positive, whereas $D_2>0 \iff 4d_{K,e}>b_{K,e}L^2$. For a physically admissible link, this condition must be satisfied; otherwise, the stiffness matrix in Eq.~\eqref{eq:K_beam_2D} would not be positive semi-definite, indicating instability even in the absence of an applied load.
Consequently, the system is initially stable, with $D_2>0$ and $D_3>D_2$. As the pressure $p$ increases, the first local instability occurs when one of these factors vanishes. Since $D_3>D_2$, the relevant condition is $D_2=0$.
Therefore, the limit pressure can be obtained by solving $D_2=0$ for $P$. The resulting limit pressure is
\begin{align}
        \frac{p_{lim, loc}\sqrt{3}}{a_k} =  \frac{B}{2Aa_KL}\left(-1+\sqrt{1-\frac{4AC}{B^2}}\right)
        \label{eq:p_lim_cell-2D}
\end{align}
where 
\begin{equation}
    \begin{aligned}
        A  &  = 1-8b_{K,g}d_{K,g}-3b_{K,g}L+2b_{K,g}^2L^2,  &
        B  &  =  8b_{K,e}d_{K,g}+8b_{K,g}d_{K,e}+3b_{K,e}L-4b_{K,e}b_{K,g}L^2,\\
        C  &  = 2b_{K,e}(b_{K,e}L^2-4d_{K,e}).
    \end{aligned}
\end{equation}
The main difference between the local and homogenized instability limits is the explicit dependence of the local condition on the bending stiffness $d_K$. This parameter can therefore strongly affect which instability mechanism occurs first. In particular, a larger bending stiffness increases $D_2$ at $P=0$, thereby requiring a larger pressure for $D_2$ to vanish. In this case, the homogenized instability limit may be reached before the local one. This effect is particularly relevant for links composed of a sheaf of beams. As shown in \cite{Brambilla2025HighWater}, in the limiting case considered for such links, $d_{K,e}$ remains finite while $b_{K,e}$ tends to $0$. 

% We can substitute the values explicitly indicated in Appendix~\ref{sec:AppendixSheafOfBeams}. The limit obtained should then be compared to what is found in Eq.~\eqref{eq:p_lim_hom_2D_generic}; the lowest is the limit pressure for the 2D honeycomb. 
% \todo[inline]{maybe tell that is not done explicitly for the higher complexity}
% The expressions for the link coefficients derived in Appendix~\ref{sec:AppendixSheafOfBeams} can then be substituted into Eq.~\eqref{eq:p_lim_cell-2D}. Unlike the homogenized case, we do not report the fully expanded expression, as the resulting formula is considerably more cumbersome. The local limit can instead be evaluated directly from Eq.~\eqref{eq:p_lim_cell-2D} using the coefficients given in Appendix~\ref{sec:AppendixSheafOfBeams}, and compared with the homogenized limit in Eq.~\eqref{eq:p_lim_hom_2D_generic}. The smaller of the two determines the critical buckling pressure of the 2D lattice.
The expressions for the link coefficients derived in Appendix~\ref{sec:AppendixSheafOfBeams} can then be substituted into Eq.~\eqref{eq:p_lim_cell-2D}. Unlike the homogenized case, however, the resulting expression is considerably more cumbersome, making it more difficult to draw general conclusions on the effect of the individual link properties. Nevertheless, Eq.~\eqref{eq:p_lim_cell-2D} provides a closed-form expression for the local buckling pressure once the link coefficients are specified. The critical buckling pressure of the 2D lattice is then given by the lower of the local and homogenized limits.\\
\\
For the 3D lattices, the $\bs \kappa^\star_L = \pi/a\{1,1,1\}$ is initially considered because it corresponds to the longest wavelength among the finite wavevectors on the boundary of the IBZ, and therefore represents a natural candidate for the critical buckling mode. The dispersion relations reported in Appendix~\ref{sec:AppendixDispersions}, however, demonstrate that the mode at $\bs \kappa^\star_X = \{0,2\pi/a,0\}$ becomes unstable at a nearly identical load to that at $\bs \kappa^\star_L$. This can be qualitatively attributed to the fact that the critical condition depends not only on the wavelength but also on the orientation of the perturbation with respect to the lattice geometry. In particular, the $X$ mode propagates along a principal direction of the cubic lattice, which can favor local instability despite its shorter wavelength. Both critical wavevectors are therefore considered in the following analysis.

At the critical wavevectors, the stiffness matrix takes the form: 
\begin{align}
    \sigma(\bs \kappa^\star) &= \sum_{\bb p \in \mathcal P}\mathsf K(\bb p)e^{i \bs \kappa^\star  \mathsf E \bb p} = 
    \begin{bmatrix}
        [\sigma_{11}(\bs \kappa^\star)] &[0]&[\sigma_{13}(\bs \kappa^\star)] &[\sigma_{14}(\bs \kappa^\star)]\\
         &[\sigma_{22}(\bs \kappa^\star)] &[\sigma_{23}(\bs \kappa^\star)] &[\sigma_{24}(\bs \kappa^\star)] \\
        & &[\sigma_{33}(\bs \kappa^\star)] &[0]\\
        \text{sym.} &  &  &[\sigma_{44}(\bs \kappa^\star)]
    \end{bmatrix}
\end{align}

For $\bs \kappa^\star_L$ the nonzero blocks are
\begin{align}
        \begin{aligned}
         [ \sigma_{11}(\bs \kappa^\star_L)] &= [\sigma_{33}(\bs \kappa^\star_L)]  = 
       \frac{4}{3}(a_{K}+2b_{K})\begin{bmatrix}
           1 & 0 & 0 \\
           0  & 1 & 0 \\
          0 & 0 & 1
       \end{bmatrix}
       \\
       [ \sigma_{22}(\bs \kappa^\star_L)] &= [\sigma_{44}(\bs \kappa^\star_L)]  = 
       \frac{4}{3}(c_{K}+2d_{K})\begin{bmatrix}
           1 & 0 & 0 \\
           0  & 1 & 0 \\
          0 & 0 & 1
       \end{bmatrix}
       \\
       [\sigma_{13}(\bs \kappa^\star_L)] & =
         \frac23\begin{bmatrix}
           a_{K}+2b_{K} & b_{K}-a_{K} & b_{K}-a_{K} \\
            b_{K}-a_{K}  & a_{K}+2b_{K} &  b_{K}-a_{K} \\
           b_{K}-a_{K} &  b_{K}-a_{K} & a_{K}+2b_{K}
       \end{bmatrix}
       \\
       [\sigma_{24}(\bs \kappa^\star_L)] & =
         \frac23\begin{bmatrix}
           (c_{K}-2i_{K}) & -(c_{K}+i_{K}) & -(c_{K}+i_{K}) \\
            -(c_{K}+i_{K}) & (c_{K}-2i_{K}) &  -(c_{K}+i_{K}) \\
           -(c_{K}+i_{K}) &  -(c_{K}+i_{K}) & (c_{K}-2i_{K})
       \end{bmatrix}
       \\
       [\sigma_{14}(\bs \kappa^\star_L)] & = [\sigma_{23}(\bs \kappa^\star_L)] = 
         \frac{2}{\sqrt{3}}\begin{bmatrix}
           0 & f_{K} & -f_{K} \\
            -f_{K}  & 0 &  f_{K} \\
           f_{K} &  -f_{K} & 0
       \end{bmatrix},
    \end{aligned}
\end{align}
while for $\bs \kappa^\star_X$
\begin{align}
    \begin{aligned}
        [ \sigma_{11}(\bs \kappa^\star_X)] &= [\sigma_{33}(\bs \kappa^\star_X)]  =
        [ \sigma_{11}(\bs \kappa^\star_L)] %= [\sigma_{33}(\bs \kappa^\star_L)] 
       &
       [ \sigma_{22}(\bs \kappa^\star_X)] &= [\sigma_{44}(\bs \kappa^\star_X)]  = 
       [ \sigma_{22}(\bs \kappa^\star_L)] %= [\sigma_{44}(\bs \kappa^\star_L)] 
       \\
       [\sigma_{13}(\bs \kappa^\star_L)] & =
         -\frac43(a_K-b_K)
        \begin{bmatrix}
           0 & 0 & 1 \\
           0 & 0 & 0  \\
           1 & 0 & 0        
        \end{bmatrix}
       &
       [\sigma_{24}(\bs \kappa^\star_X)] & =
         -\frac43 (c_{K}+i_{K})\begin{bmatrix}
           0 & 0 & 1 \\
           0 & 0 & 0 \\
           1 & 0 & 0
       \end{bmatrix}
       \\
       [\sigma_{14}(\bs \kappa^\star_X)] & = [\sigma_{23}(\bs \kappa^\star_X)] = 
         \frac{4}{\sqrt{3}}f_{K}\begin{bmatrix}
           0 & 0 & -1 \\
           0 & 0 & 0\\
           1 & 0 & 0
       \end{bmatrix}
    \end{aligned}
\end{align}
where $f_K = (b_KL+P)/2$ and $i_K = -d_K+f_KL$.

We first consider $\bs \kappa^\star_L$. The determinant of $\sigma(\bs \kappa^\star_L)$ is 
\begin{align}
\begin{aligned}
        \det(\sigma(\bs \kappa^\star_L) )& =  2^3\left(\frac{2}{3}\right)^9 a_K 
        D_{L,1} 
        D_{L,2}
        D_{L,3} 
        {D_{L,4}}^2
        {D_{L,5}}^2
        =0\\
        D_{L,1} &=(a_K+8b_K)\\
        D_{L,2} &=(c_K+8d_K-2(b_KL+P)L)\\
        D_{L,3} &=(3c_K+2(b_KL+P)L)\\
        D_{L,4} &=b_K\left(4c_K+5d_K-\frac54(b_KL+P)L\right)-\frac 34(b_KL+P)P\\
        D_{L,5} &=(4a_K+5b_K)\left(3d_K + \frac 12 (b_KL+P)L\right)-\frac 94(b_KL+P)^2.
\end{aligned}
\end{align}
As in the 2D case, the stiffness coefficients are decomposed into elastic and geometric contributions, after which each factor can be solved for the corresponding critical pressure. In contrast to the 2D case, however, it is not straightforward to identify a single factor that is always the first to vanish. We therefore determine the candidate limit pressure associated with each factor and select the smallest positive value. The resulting limits are
\begin{align}
    \begin{aligned}
        \frac{p_{lim, loc, 1}}{a_K/L} &= \frac{\sqrt{3}}{4}\frac{1+8b_{K,e}/a_K}{8b_{K,g}L} \\
        \frac{p_{lim, loc, 2}}{a_K/L} &= \frac{\sqrt{3}}{4}\frac{c_{K,e}+2(4d_{K,e}-b_{K,e}L^2)}{(c_{K,g}+8d_{K,g}-2(b_{K,g}L-1)L)a_KL} \\
        \frac{p_{lim, loc, 3}}{a_K/L} &= \frac{\sqrt{3}}{4}\frac{3 c_{K,e}/(a_KL)+2 (b_{K,e}/a_K)L}{3 c_{K,g}+2(b_{K,g}L-1)L} \\
        \frac{p_{lim, loc, i}}{a_K/L} &= \frac{\sqrt{3}}{4}\frac{-B_i \pm\sqrt{B_i^2-4A_iC_i}}{4A_ia_K L}   & i&\in  \{4,5\}
    \end{aligned}
\end{align}
where
\begin{align}
    & \left\{
    \begin{aligned}
            A_4 & =  b_{K,g}(4c_{K,g}+5d_{K,g}+2L)-\frac 14(3+5b_{K,g}^2L^2)  \\
            B_4 & = -b_{K,e}(4c_{K,g}+5d_{K,g}+2L)-b_{K,g}\left(4c_{K,e}+5\left(d_{K,e}-\frac{1}{2}b_{K,e}L^2\right)\right) \\
            C_4 & =  b_{K,e}\left(4c_{K,e}+5\left( d_{K,e}-\frac 14 b_{K,e}L^2  \right)\right)
    \end{aligned}
    \right.\\
   & \left\{
    \begin{aligned}
            A_5 & =\frac{1}{4}(60b_{K,g}d_{K,g}+8b_{K,g}L+b_{K,g}^2L^2-9)\\
            B_5 & = -\left(2a_K(6d_{K,g}-L)+b_{K,e}(15d_{K,g}+2L)+\frac 12 b_{K,g}(30 d_{K,e}+(4a_K+b_{K,e})L^2)\right)\\
            C_5 &  =2a_K(6d_{K,e}+b_{K,e}L^2)+\frac 14 b_{K,e}(60d_{K,e}+b_{K,e}L^2) 
    \end{aligned}
    \right.
\end{align}
% We can avoid considering t
The solution $p_{lim, loc, 1}$ can be discarded since the global buckling limit found in Eq.~\eqref{eq:p_lim_hom_3D_generic} is always stricter. For the remaining conditions, it is not straightforward to identify a priori which one leads to the lowest buckling pressure, especially because it is complex to define a relation between the .

We proceed similarly for $X$. The determinant factorizes as
\begin{align}
    \begin{aligned}
        \det (\sigma(\bs \kappa^\star_X)) &= 2^{14}\left(\frac{2}{3}\right)^{10}
        {D_{X,1}}^2
        {D_{X,2}}^2
        {D_{X,3}}^2
        {D_{X,4}}^2 
        = 0 \\
        D_{X,1}&= a_K+2b_K \\
        D_{X,2}&= c_K+2d_K\\
        D_{X,3}&= 3b_K\left(d_K-\frac14b_KL^2\right)-\frac P4\left(P-4b_kL\right)\\
        D_{X,4}&= (2  a_K + b_K)  (2 c_K + d_K) + (b_K L + P)\left(a_KL-\frac 34 b_KL-\frac 14 P \right)
    \end{aligned}
\end{align}
yielding
\begin{align}
    \begin{aligned}
        \frac{p_{lim,loc,6}}{a_K/L}&=\frac{\sqrt{3}}{4} \frac{1+2b_{K,e}/a_K}{2b_gL} \\
        \frac{p_{lim,loc,7}}{a_K/L}&=\frac{\sqrt{3}}{4} \frac{c_{K,e}+2d_{K,e}}{a_KL(c_{K,g}+2d_{K,g})}\\
        \frac{p_{lim,loc,i}}{a_K/L}&= \frac{\sqrt{3}}{4}\frac{-B_i \pm\sqrt{B_i^2-4A_iC_i}}{4A_ia_K L} 
        & i&\in\{8,9\}
    \end{aligned}
\end{align}
where 
\begin{align}
    & \left\{
    \begin{aligned}
            A_8 & =  b_{K,g}(3d_{K,g}+L) -\frac 14(1+3b_{K,g}^2L^2) \\
            B_8 & =  -b_{K,e}(3d_{K,g}+L)-3b_{K,g}\left(d_{K,e}-\frac{1}{2}b_{K,e}L^2\right) \\
            C_8 & = 3 b_{K,e}\left( d_{K,e}-\frac 14 b_{K,e}L^2  \right)
    \end{aligned}
    \right.\\
   & \left\{
    \begin{aligned}
            A_9 & 
            =-\frac{3}{4}+b_{K,g}\left(\left(d_{K,g}-\frac{1}{4}b_{K,g}L^2\right)+2c_{K,g}+L\right)\\
            B_9 & 
            = -(2a_K + b_{K,e})(2c_{K,g} + d_{K,g}) + (a_K - b_{K,e})L - \frac12b_{K,g}(4c_{K,e} + 2d_{K,e} + (2a_K - b_{K,e})L^2)\\
            C_9 &
            =  (2a_K + b_{K,e})(2c_{K,e} + d_{K,e}) + \frac 14 b_{K,e}(4a_K - b_{K,e})L^2;\\
    \end{aligned}
    \right.
\end{align}
As in the analysis at $L$, the solution $p_{lim, loc,6}$, associated with $D_{X,1}$, can be discarded because it is always above the global buckling limit. The remaining candidate pressures are compared with those obtained at $L$ to determine the critical local buckling pressure.
The expressions for the link coefficients derived in Appendix~\ref{sec:AppendixSheafOfBeams} can then be substituted into the expressions above. As in the 2D case, the resulting expressions are considerably more cumbersome than those obtained for the global limit, and are further complicated in 3D by the additional dependence on the torsional stiffness coefficients. Nevertheless, the local buckling pressure is given in closed form once the link coefficients are specified.

\section{Numerical Validation}
\label{sec:4_NumericalValidation}

% To ease the analytical description of the geometries 
% % shown in Figure~\ref{fig:Link_and_cell_Models}\subref{fig:2DUnitCellFasci} and \subref{fig:3DUnitCellFasci} 
% featured by sheaf of beams
% we considered the nodal connections as lumped nodal masses. Particularly, we added the translational and rotational inertia terms $M_S$ and $I_S$ to nodes $\#1$ and $\#2$ of the lattice schematics in Figure~\ref{fig:UnitCell_2D_Topology}\subref{fig:2DUnitCellSchematic} and \ref{fig:UnitCell_3D_Topology}\subref{fig:3DUnitCellSchematic}. Nodes $\#3$ and $\#4$ were not considered because, owing to periodicity, they are equivalent to node $\#1$ of neighboring cells and therefore do not require additional lumped inertia contributions.
The analytical formulations developed in the previous sections are validated through numerical simulations of both 2D and 3D lattice structures. In each case, the links connecting adjacent nodes are modeled either as single beams or as sheaves of beams, as detailed in Appendix~\ref{sec:AppendixSheafOfBeams}.

The analytical predictions are compared with the corresponding numerical results to assess the accuracy and robustness of the proposed framework. All numerical analyses are conducted using \Comsol; the base material is assumed to be aluminum, with Young's modulus $E=\qty{70}{\giga\pascal}$, mass density $\rho=\qty{2700}{\kilogram\per\meter\cubed}$, and Poisson's ratio $\nu = 0.3$. The node-to-node length is set to $L = \qty{1}{\centi\meter}$.

The numerical validation is based on the dispersion relations of the considered structures, from which the homogenized wave velocities and the lowest eigenfrequency at the critical wavevector can be obtained. For each prescribed compressive load, a stationary analysis is first performed to determine the corresponding deformed equilibrium configuration. A prestressed eigenfrequency analysis is then performed by linearizing the equations about this configuration, and Bloch--Floquet periodic boundary conditions are imposed to compute the dispersion relation. The resulting dispersion diagrams are discussed in detail in Appendix~\ref{sec:AppendixDispersions}.

For the global limit, the wave velocities are obtained from the slopes of the acoustic branches in the vicinity of the origin of the Brillouin zone. For the local limit, the lowest eigenfrequency associated with the most critical wavevector $\bs\kappa^\star$ is evaluated for different load levels. In both cases, the resulting load-dependent quantities are used to identify the critical pressure, which is then compared with the corresponding analytical prediction.

Finally, to obtain a more accurate estimate of the numerical critical pressure without significantly increasing the computational cost, the squared wave velocity for the global limit and the squared eigenfrequency for the local limit are linearly fitted as functions of the applied load. The critical pressure is then identified as the intersection of the corresponding fitted curve with the zero axis.
\\
\\
For the 2D structures, plane strain conditions are assumed. Accordingly, the effective Young's modulus used in the analytical calculations is $E_{ps} = \qty{76.9}{\giga\pascal}$. For the sheaf-of-beams geometries, the numerical simulations are performed considering both rigid and compliant connections between the fibers and the bodies connecting them, in order to assess the effect of this modeling assumption.

\begin{figure}
    \centering
    \subfloat[]{%
        \raisebox{-4.5mm}{
            \includegraphics[
                width=0.37\linewidth,
                trim=4cm 0.15cm 5cm 0.15cm,
                clip
            ]{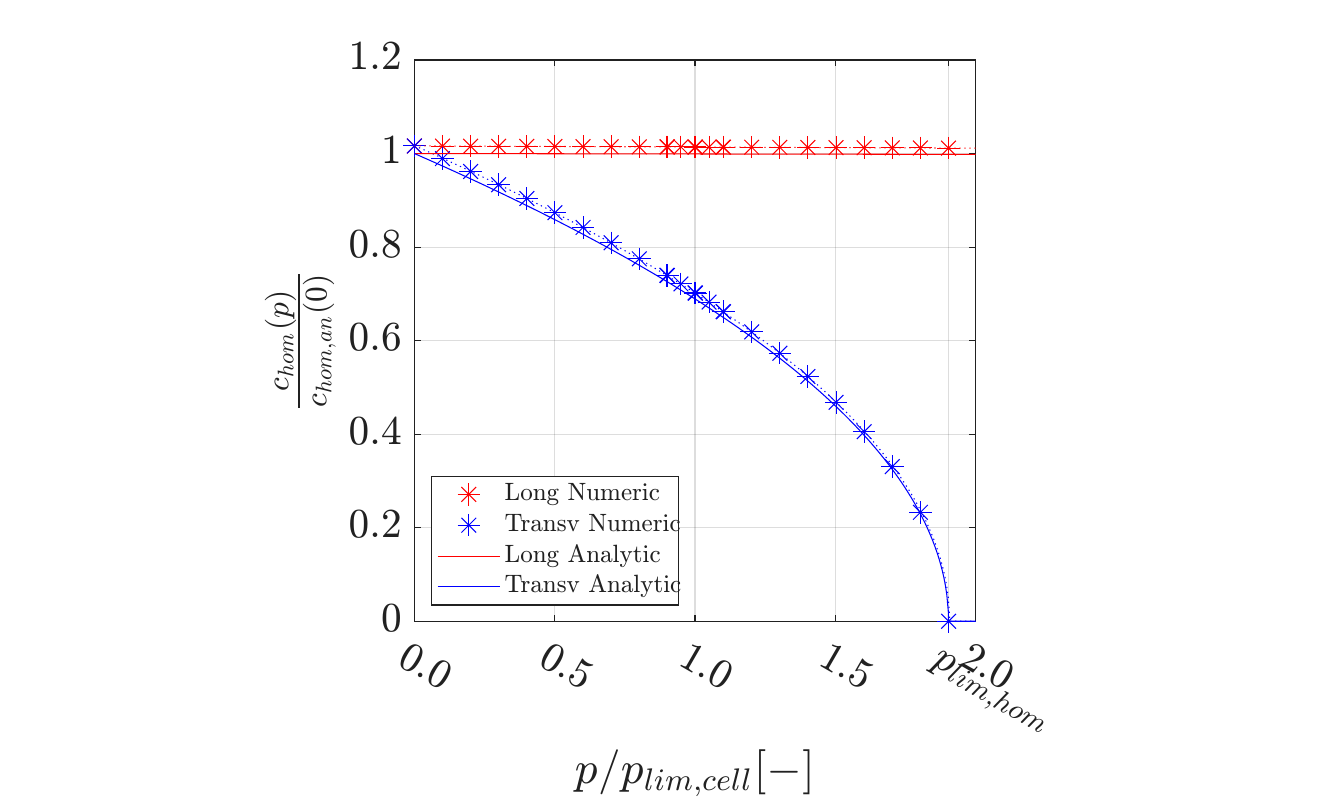}%
            \label{fig:2D_VelocitiesComparisonFibra}
        }
    }
    % \hfill
    \subfloat[]{%
        \includegraphics[
            width=0.30\linewidth,
            trim=5.4cm 0cm 5cm 0.15cm,
            clip
        ]{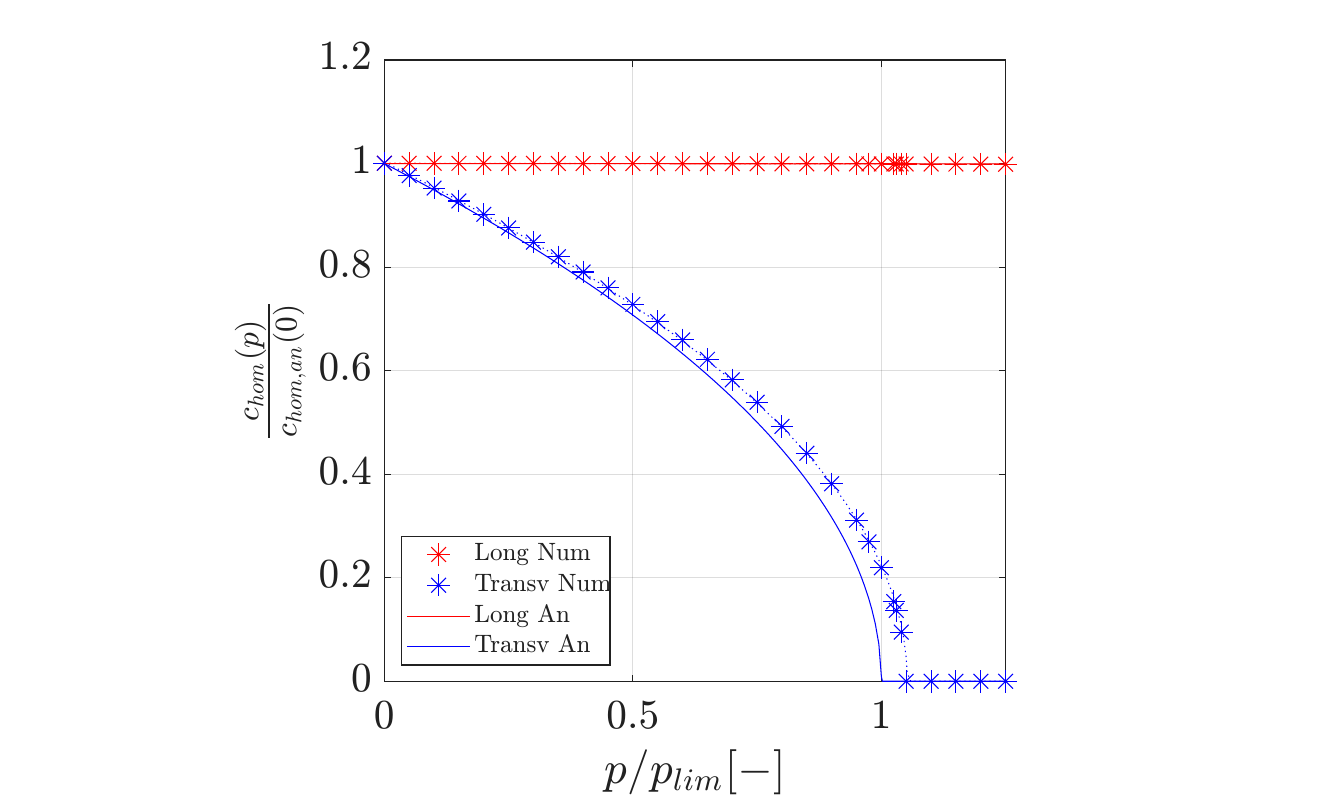}%
        \label{fig:2D_VelocitiesComparisonFascioRigid}
    }
    % \hfill
    \subfloat[]{%
        \includegraphics[
            width=0.30\linewidth,
            trim=5.4cm 0cm 5cm 0.15cm,
            clip
        ]{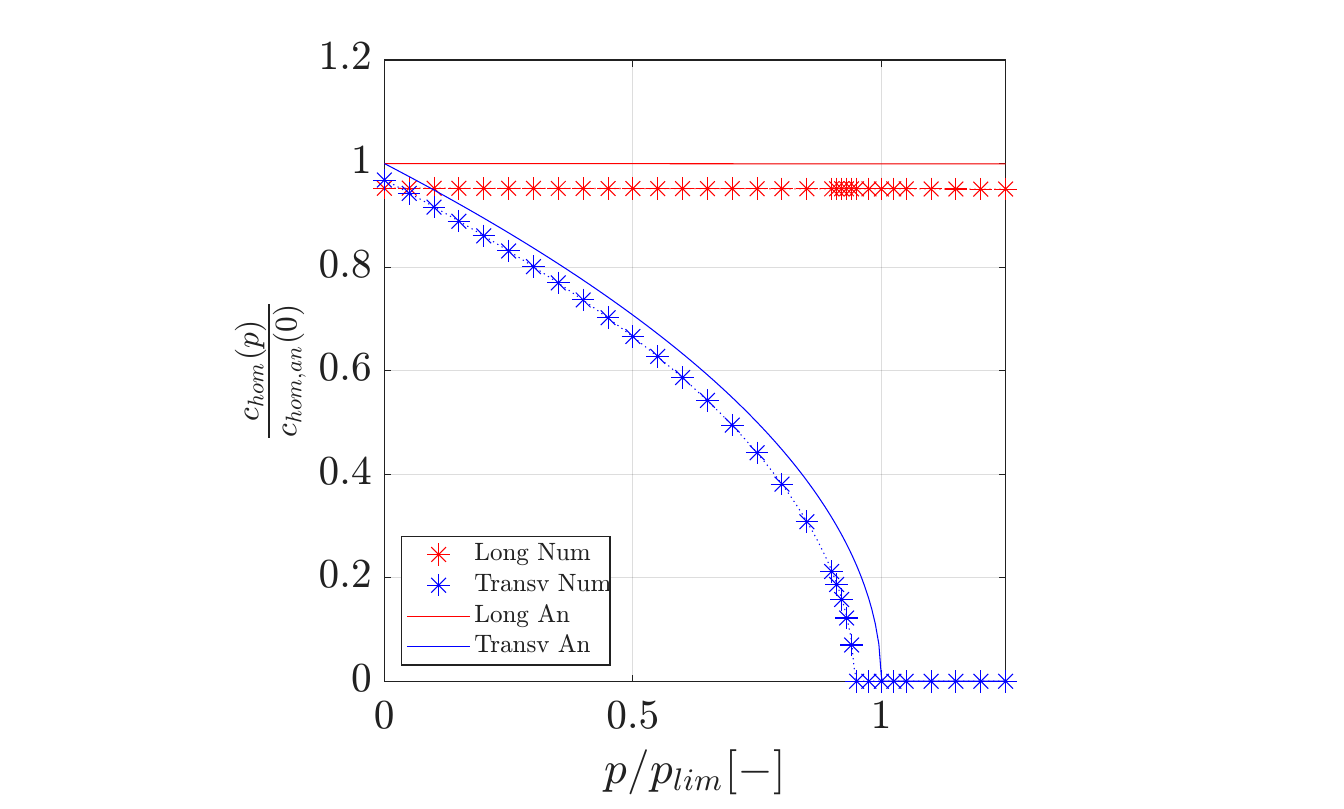}%
        \label{fig:2D_VelocitiesComparisonFascioCompl}
    }
    \caption{Longitudinal and transverse wave propagation velocities of the equivalent homogenized material as functions of the applied pressure for the 2D structures. In (a) results for the single-fiber geometry are shown, while (b) and (c) correspond to the sheaf-of-beams geometry with rigid and compliant connections, respectively. The results are normalized by the corresponding analytical velocities at zero applied pressure. The dotted lines represent the linear fits of the numerical results.}
    \label{fig:2D_VelocitiesComparison}
\end{figure}

The homogenized wave velocities obtained numerically are computed from finite differences of the dispersion relation in the vicinity of the origin, considering the $\kappa_x$ direction, and are compared with the corresponding analytical predictions. Since the considered geometry is transversely isotropic, the homogenized wave velocities are independent of the propagation direction \cite{Gonella2008HomogenizationLattices}. Figure~\ref{fig:2D_VelocitiesComparison}
shows the resulting longitudinal and transverse wave velocities as functions of the applied pressure. The results are normalized by the corresponding analytical values at zero applied pressure. The dotted lines represent the linear fits used to determine the numerical critical pressures. 

\begin{figure}
    \centering
    \subfloat[]{%
        \raisebox{2mm}{
            \includegraphics[
                width=0.355\linewidth,
                trim=4cm 0.15cm 5cm 0.15cm,
                clip
            ]{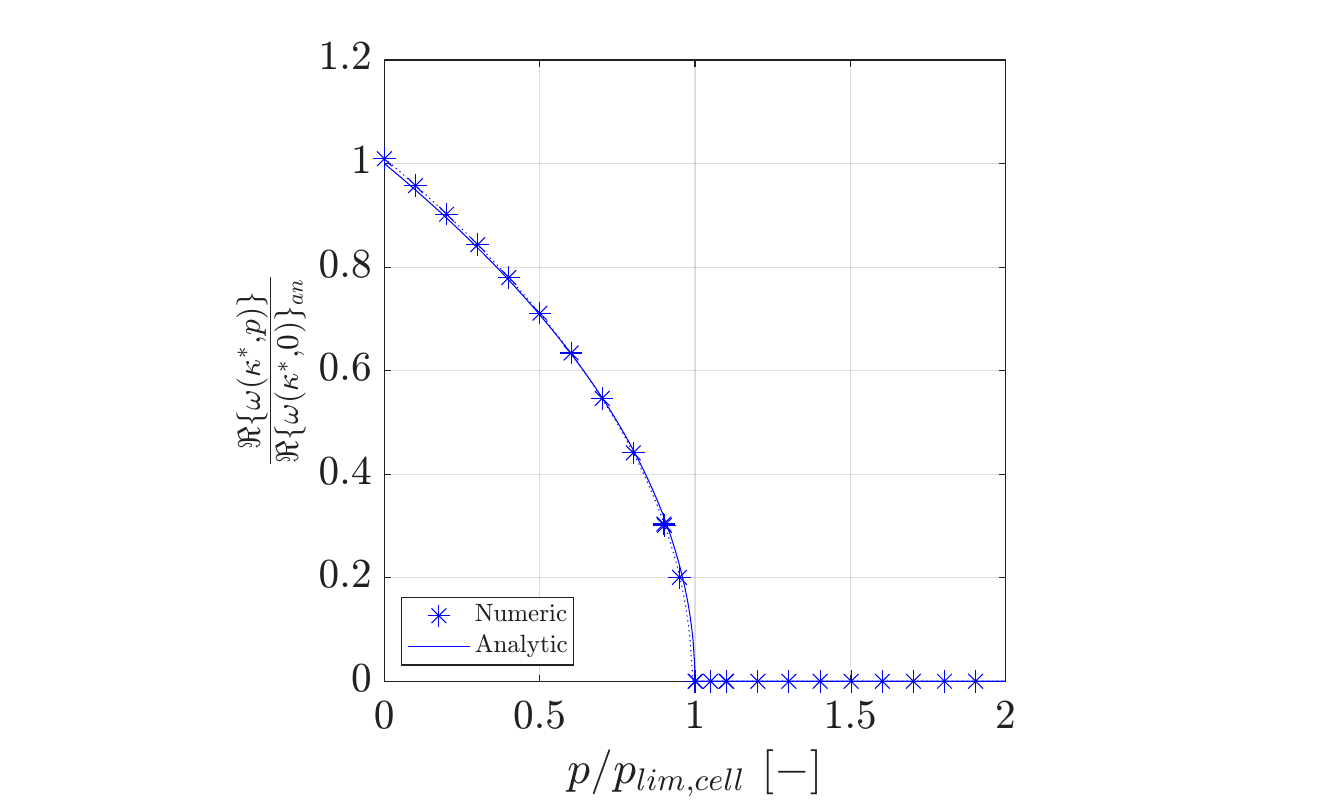}%
        }
    }
    \hfill
    \subfloat[]{%
        \includegraphics[
            width=0.31\linewidth,
            trim=5.4cm 0cm 5.5cm 0.15cm,
            clip
        ]{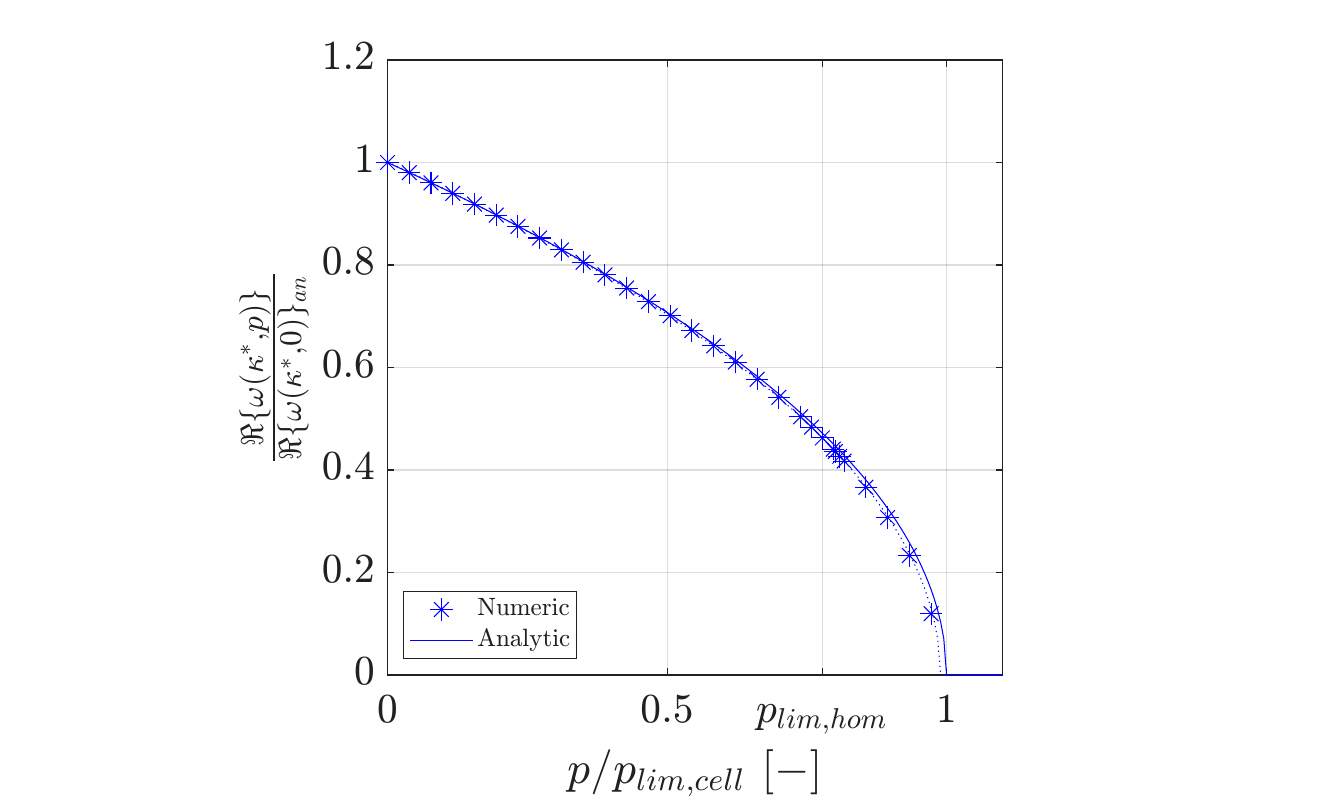}%
    }
    \hfill
    \subfloat[]{%
        \includegraphics[
            width=0.31\linewidth,
            trim=5.4cm 0cm 5.5cm 0.15cm,
            clip
        ]{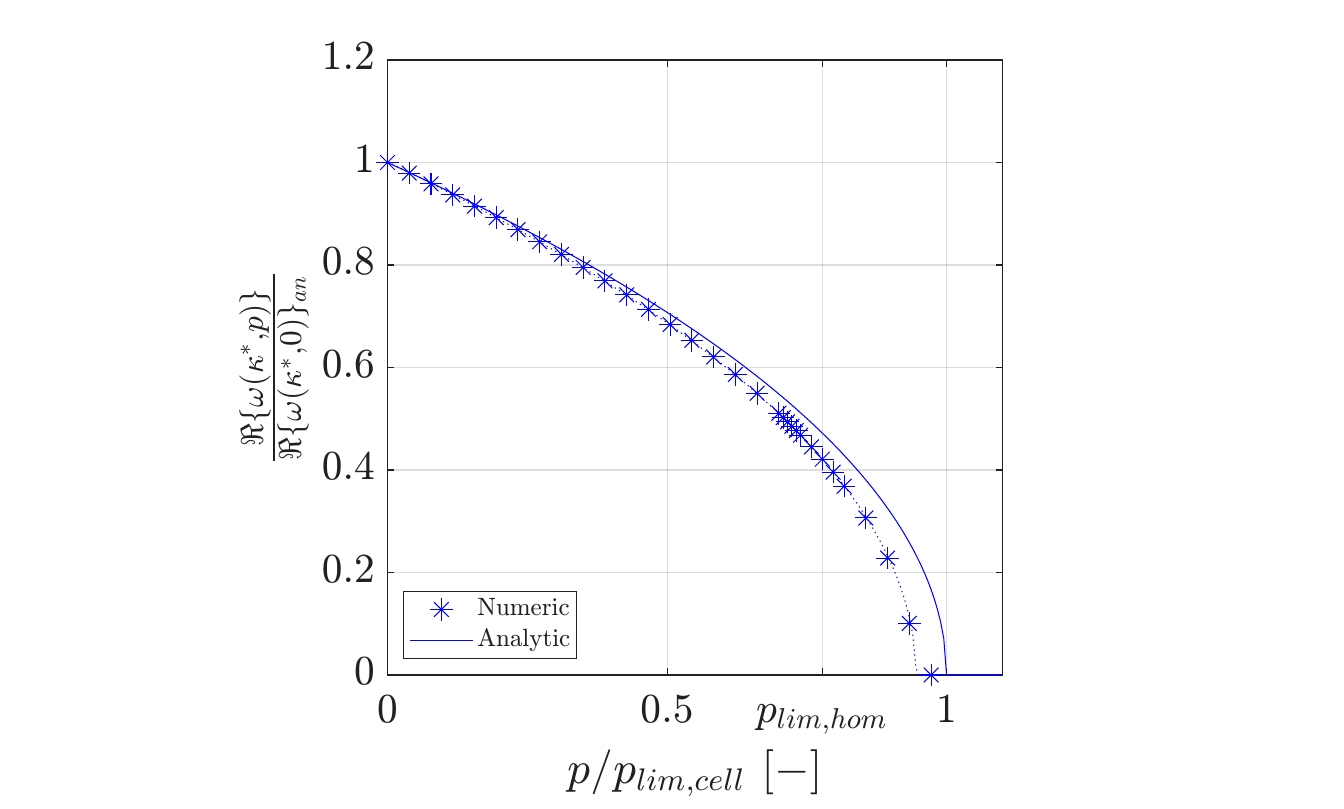}%
    }
    \caption{Lowest eigenfrequecy at the critical wavevector $\bs \kappa^\star$ as a function of the applied pressure for the 2D structures. In (a), results for the single-fiber geometry are shown, while (b) and (c) correspond to the sheaf-of-beams geometry with rigid and compliant connections, respectively. The results are normalized by the corresponding analytical eigenfrequency at zero applied pressure. The dotted lines represent the linear fits of the numerical results.}
    \label{fig:2D_OmegaComparison}
\end{figure}

Regarding the local instability limit, Figure~\ref{fig:2D_OmegaComparison}
shows the variation of the lowest eigenfrequency at the critical wavevector $\bs \kappa^\star$ as a function of the applied pressure. The analytical curves are computed using Eq.~\eqref{eq:EoM_kappa_compact}. The numerical eigenfrequencies are normalized by the corresponding analytical values at zero applied pressure, and the dotted lines represent the linear fits used to determine the numerical critical pressures.
% , where the definitions of the coefficients for the mass matrices are found in Appendix~\ref{sec:AppendixSheafOfBeams}.

Table~\ref{tab:2D_results}
\begin{table}[ht]
    \centering
    \caption{Geometric parameters, stiffness coefficients, and buckling pressures for the 2D single-fiber geometry (A) and the sheaf-of-beams geometry (B).}
    \label{tab:2D_results}
    \begin{tabular}{l||l||c|cc}
        \toprule
        \multicolumn{2}{c||}{Geometries} &  A& \multicolumn{2}{c}{ B}\\
        \midrule
        
        \multirow{5}{*}{
            \rotatebox[origin=c]{90}{
                \shortstack{Geometric\\coefficients}
            }
        }
        & $N_f$     & $1$ & \multicolumn{2}{c}{$5$}\\
        & $\beta$   & $0.05$& \multicolumn{2}{c}{$0.0286$}\\
        & $R_c/L$     & -- & \multicolumn{2}{c}{$0.2$}\\
        & $\tau$    & 0 & \multicolumn{2}{c}{$0.15$ }\\
        & $\lambda$ & -- & \multicolumn{2}{c}{ 3.06 }\\
        \midrule
        
        \multirow{5}{*}{%
            \rotatebox[origin=c]{90}{%
                \shortstack{Stiffness\\coefficients}%
            }%
        }
        & $a_K$ [\si{\mega\pascal}]& \num{3.85e3}& \multicolumn{2}{c}{\num{11.0e3}}\\
        & $b_{K,e}$ [\si{\mega\pascal}]& \num{9.62}& \multicolumn{2}{c}{\num{8.97}}\\
        & $b_{K,g}$ [\si{1 \per \meter}]& \num{1.20e2}& \multicolumn{2}{c}{\num{1.71e2}}\\
        & $d_{K,e}$ [\si{\newton}]& \num{0.321e3}& \multicolumn{2}{c}{\num{8.48e3}}\\
        & $d_{K,g}$ [\si{\meter}]  & \num{1.33e-3}& \multicolumn{2}{c}{\num{3.12e-3}}\\
        \midrule
        
         \multicolumn{2}{c||}{Link connection}& & Rigid & Compliant \\
        \midrule
        % \cmidrule(lr){2-5}
        \multirow{6}{*}{
            \rotatebox[origin=c]{90}{
                \shortstack{Pressure\\limits}
            }
        }
        & Global (analytic) [\si{\mega\pascal}]   & \num{3.26}& \multicolumn{2}{c}{\num{2.34}} \\
        & Global (numeric)  [\si{\mega\pascal}]   & \num{3.27}& 2.46 & 2.22 \\
        & Error              & $-0.24\%$&  $-4.8\%$ & $  +5.3\%$ \\
        \cmidrule(lr){2-5}
        & Local  (analytic)[\si{\mega\pascal}] & \num{1.72}& \multicolumn{2}{c}{\num{3.00}} \\
        & Local  (numeric) [\si{\mega\pascal}] & \num{1.70}& 2.98 & 2.85 \\
        & Error             & $+0.63\%$& $  +1.1\%$ & $+5.5\%$ \\
        
    \end{tabular}
\end{table}
summarizes the geometric parameters used to define the 2D lattices, the stiffness coefficients employed in the analytical calculations, and the resulting analytical and numerical buckling pressures.

The agreement between analytical and numerical predictions is excellent for the single-fiber geometry (A), with errors below $1\%$, while some discrepancies remain for the sheaf-of-beams geometry (B), although the errors remain below $5.5\%$ for both rigid and compliant connections. As expected, modeling the connections between the fibers as rigid bodies affects the overall stiffness of the structure and leads to lower homogenized wave velocities and eigenfrequencies.
This effect is particularly evident in the sheaf-of-beams geometry. For the case with rigid connections, the discrepancy in the homogenized wave velocity is negligible at zero applied pressure but increases noticeably with increasing pressure, as shown in Figure~\ref{fig:2D_VelocitiesComparison}~\subref{fig:2D_VelocitiesComparisonFascioRigid}. Since this behavior is not observed for the single-fiber geometry, the results suggest that the discrepancy mainly originates from the geometric contribution associated with the rigid connections rather than from the beam model itself.
The analytical predictions of the local buckling pressure show smaller errors than those observed for the homogenized wave velocities. This indicates that the modeling imperfections associated with the connections have a comparatively smaller influence on the predicted local instability limit.

\begin{figure}
    \centering
    \raisebox{-5mm}{
        \subfloat[]{%
            \includegraphics[
                width=0.37\linewidth,
                trim=4.5cm 0.15cm 5.3cm 0.15cm,
                clip
            ]{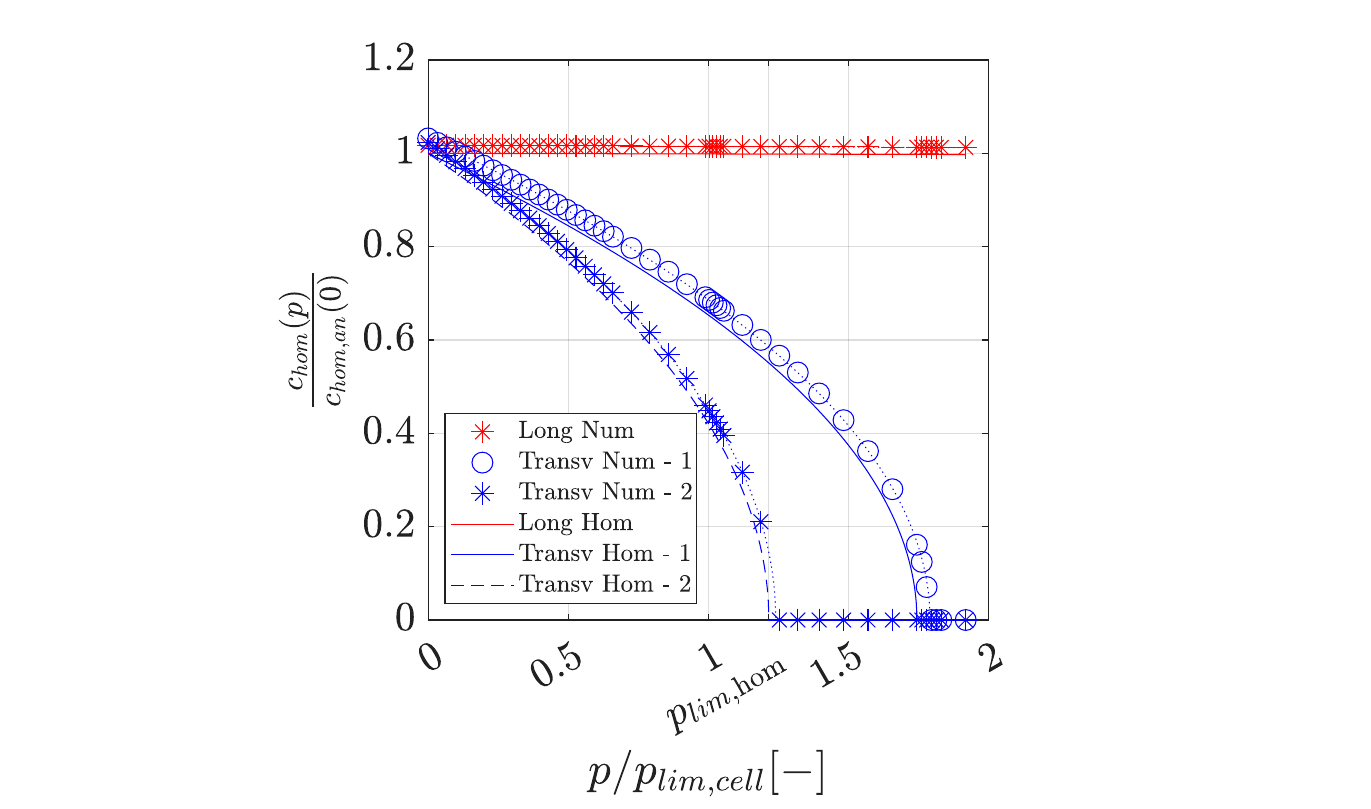}%
        }
    }
    % \hfill
    \subfloat[]{%
        \includegraphics[
            width=0.30\linewidth,
            trim=5.4cm 0cm 5.3cm 0.15cm,
            clip
        ]{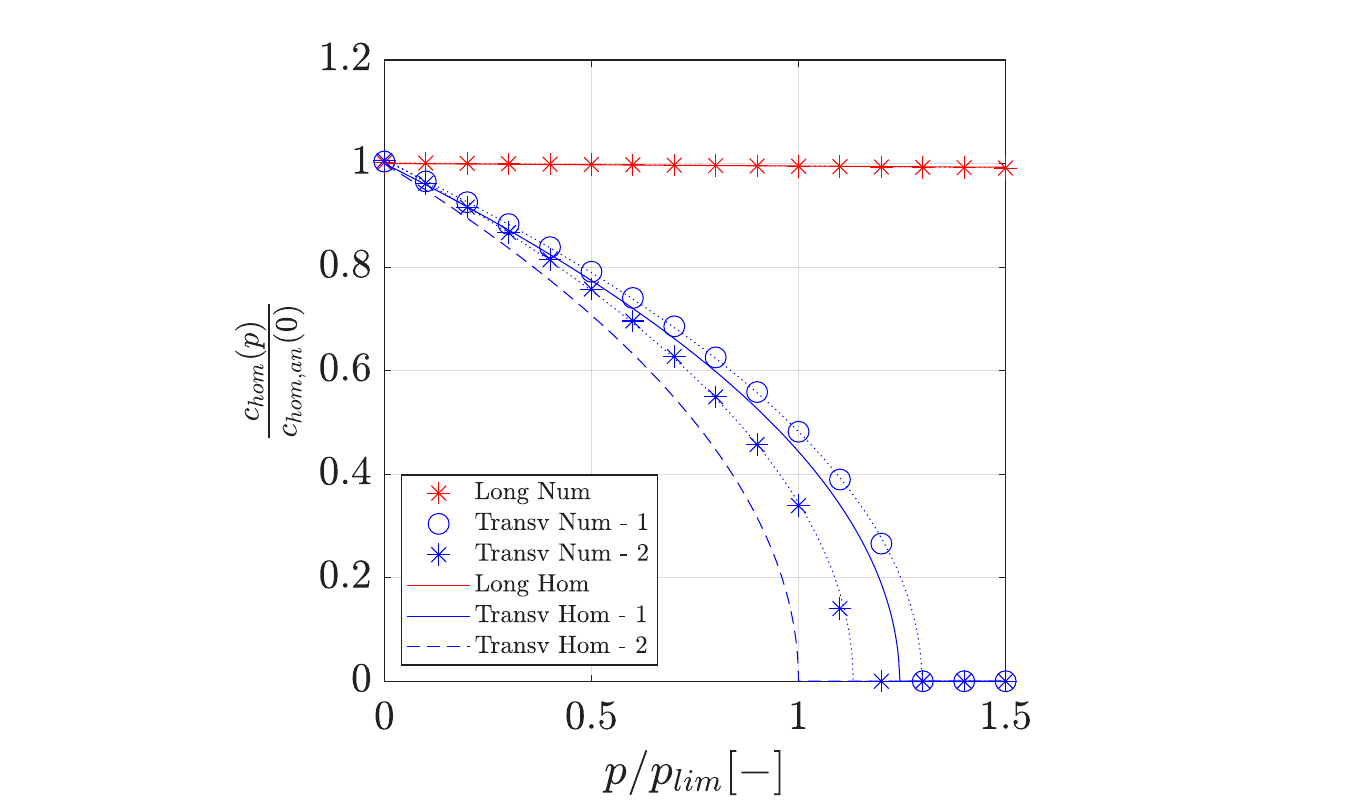}%
    }
    \hfill
    \subfloat[]{%
        \includegraphics[
            width=0.30\linewidth,
            trim=5.4cm 0cm 5.3cm 0.15cm,
            clip
        ]{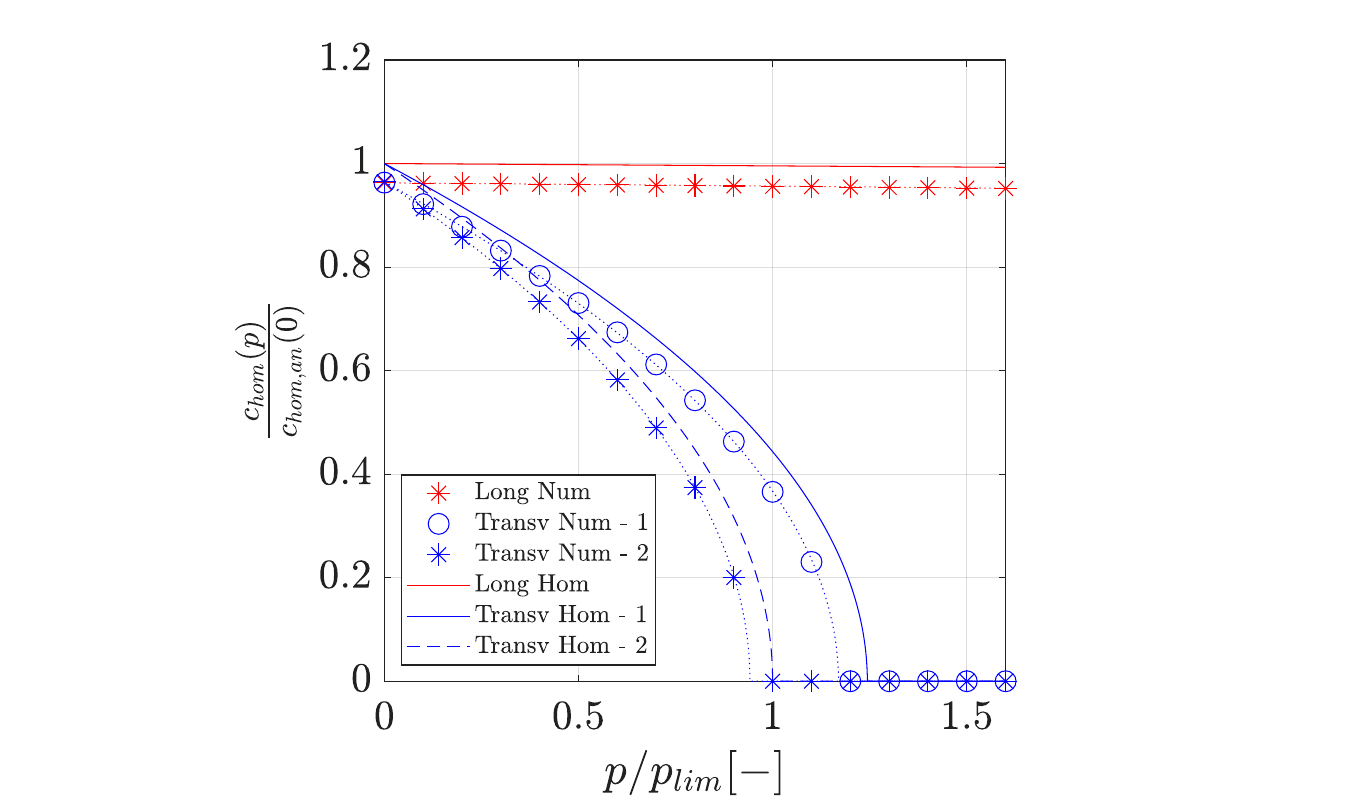}%
    }
    \caption{Longitudinal and transverse wave propagation velocities of the equivalent homogenized material as functions of the applied pressure for the 3D structures. In (a), the results for the single-fiber geometry are shown, while (b) and (c) correspond to thethe sheaf-of-beams geometry with rigid and compliant connections, respectively. The results are normalized by the corresponding analytical velocities at zero applied pressure. The dotted lines represent the linear fits of the numerical results. Due to the cubic symmetry of the lattices, the two transverse wave velocities depend on the propagation direction. The results shown here correspond to the $\Gamma-X$ direction, for which the most critical condition is found.}
    \label{fig:3D_VelocitiesComparison}
\end{figure}

\begin{figure}
    \centering
    \raisebox{-4mm}{
        \subfloat[]{%
            \includegraphics[
                width=0.355\linewidth,
                trim=4.5cm 0.15cm 5.6cm 0.15cm,
                clip
            ]{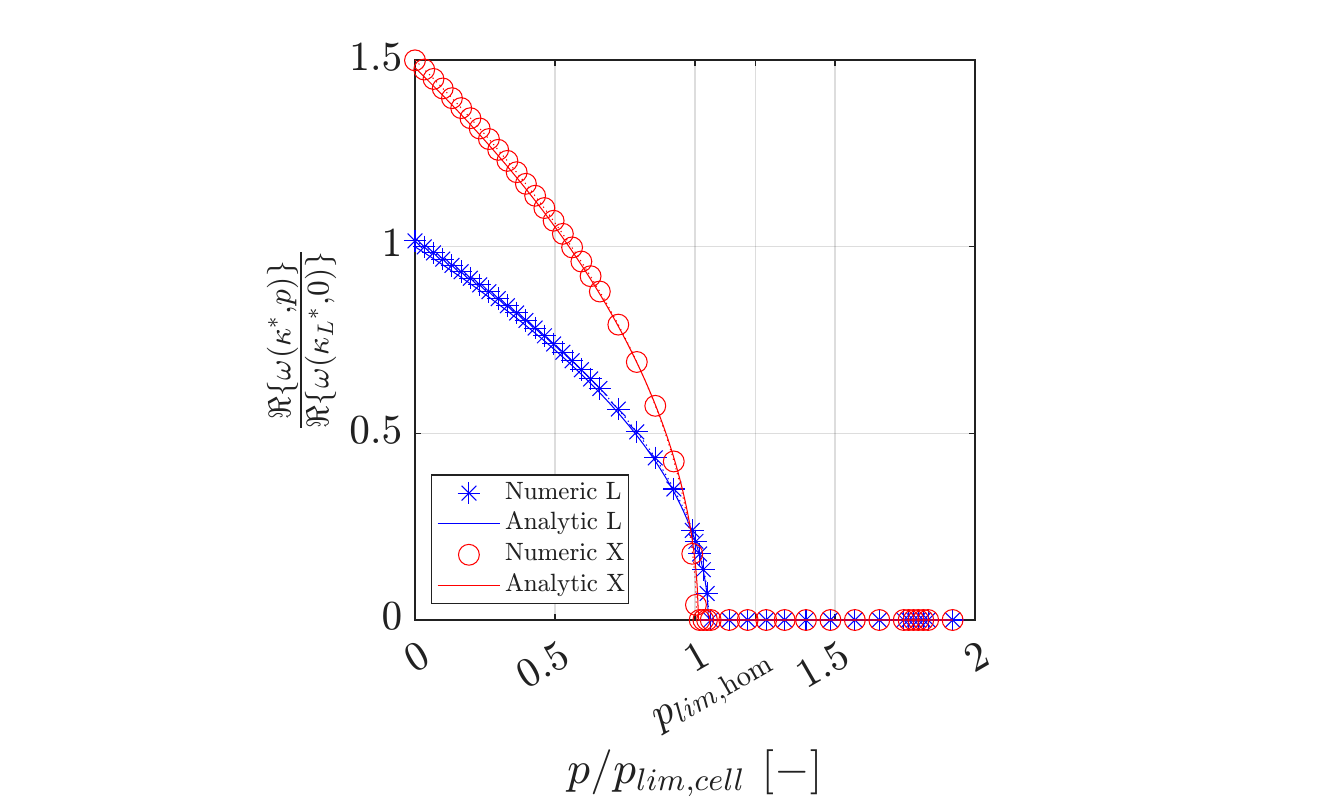}%
        }
    }
    % \hfill
    \subfloat[]{%
        \includegraphics[
            width=0.30\linewidth,
            trim=5.45cm 0cm 5.6cm 0.15cm,
            clip
        ]{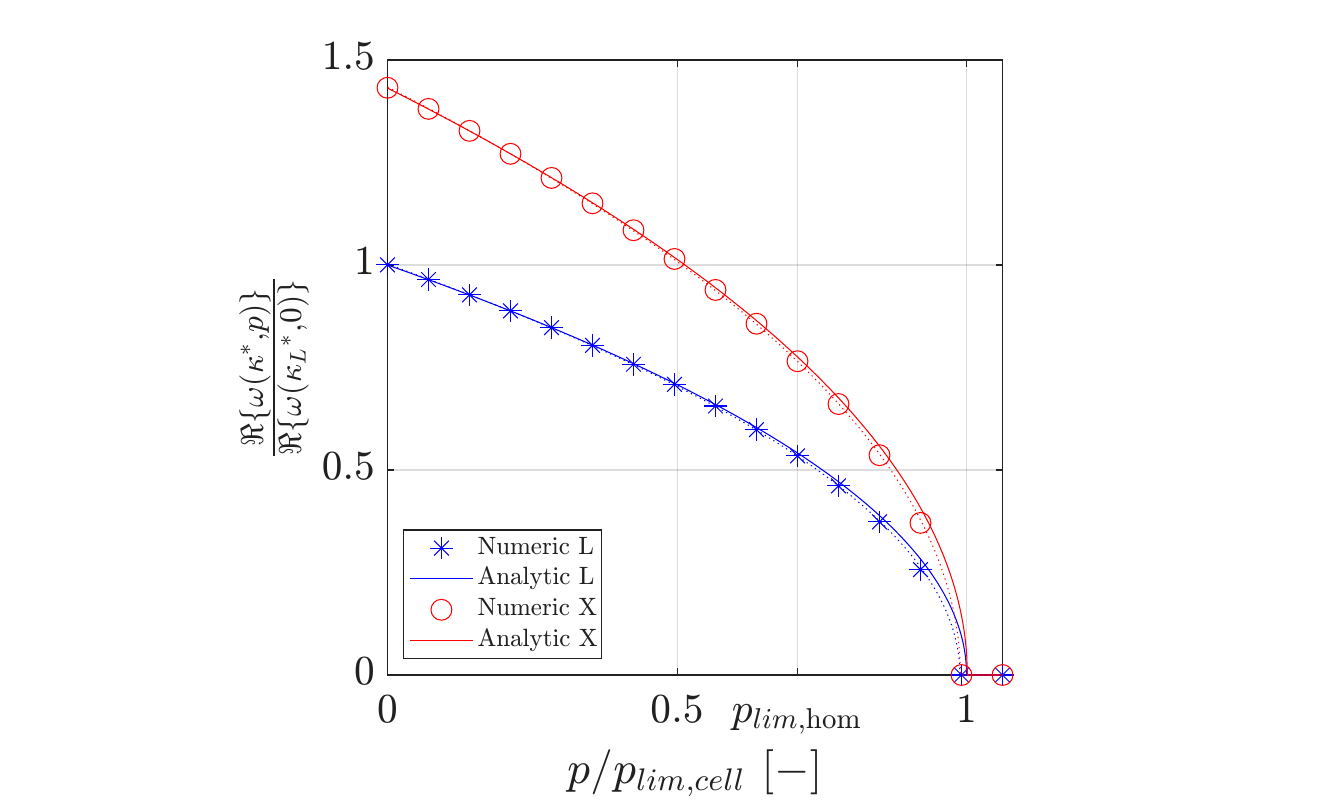}%
    }
    \hfill
    \subfloat[]{%
        \includegraphics[
            width=0.30\linewidth,
            trim=5.45cm 0cm 5.6cm 0.15cm,
            clip
        ]{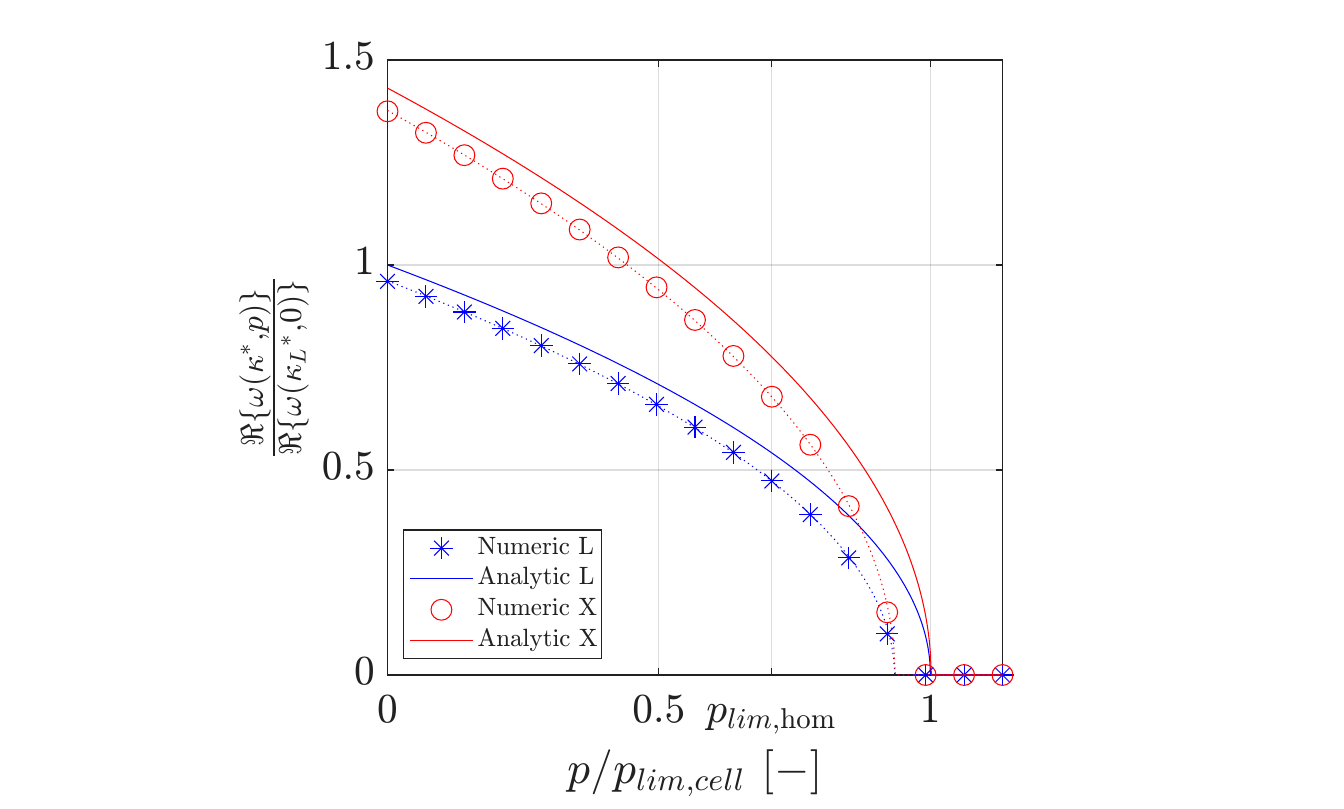}%
    }
    \caption{Lowest eigenfrequency at the two critical wavevectors $\bs {\kappa_L}^\star$ and $\bs {\kappa_X}^\star$, as a function of the applied pressure for the 3D structures. In (a), results for the single-fiber geometry are shown, while (b) and (c) correspond to the sheaf-of-beams geometry with rigid and compliant connections, respectively. The results are normalized with the analytical eigenfrequency at $\bs {\kappa_L}^\star$ at zero applied pressure. The dotted lines represent the fitting curve of the numerical results.}
    \label{fig:3D_OmegaComparison}
\end{figure}

Furthermore, the most restrictive instability mechanism may vary between the considered geometries. In some cases, $p_{\mathrm{lim,loc}}$ is the limiting condition, whereas in others $p_{\mathrm{lim,glob}}$ is more restrictive.
As discussed in Section~\ref{sec:Instability}, this difference is mainly related to the dependence of $p_{\mathrm{lim,loc}}$ on $d_{K}$, whereas $p_{\mathrm{lim,glob}}$ is independent of this parameter. 
Consequently, in the sheaf-of-beams structure, the parameter $\lambda$, which governs $d_{K}$, can be adjusted to increase $p_{\mathrm{lim,loc}}$ and thereby reduce the relevance of this constraint. 
By contrast, this tuning is not possible in the single-fiber structure, where $d_K$ cannot be adjusted independently of the other stiffness parameters. 
This suggests that multi-fiber links may provide additional design freedom to increase the local buckling limit while controlling the axial stiffness of the link, potentially enabling a better trade-off between the two limiting conditions while retaining the pentamode behavior.
% In such cases, one possible design strategy is to transition from a single-fiber geometry to a multi-fiber geometry, thereby increasing $p_{\mathrm{lim,loc}}$ and potentially achieving a more balanced trade-off between the two limiting conditions.
% is shown in Figures \todo{vel e omega@k* compl}. As we can expect, 
% \marco{; modeling of rigid bodies, especially the linearized rotations.}
\\
\\
A similar procedure is used to evaluate the 3D results, collected in Figures~\ref{fig:3D_VelocitiesComparison} and \ref{fig:3D_OmegaComparison}.
Some differences in the evaluation procedure are worth noting:
\begin{itemize}
    \item due to the cubic symmetry, the homogenized wave velocities depend on the propagation direction. Hence, they are evaluated along the $\Gamma-X$ direction, for which the most critical condition is found (see Appendix \ref{sec:AppendixDispersions}). 
    \item Second, the local instability is evaluated at the two critical wavevectors, $\bs{\kappa}_L^\star$ and $\bs{\kappa}_X^\star$, identified in Section~\ref{sec:Instability}. The more restrictive of the two corresponding pressure limits is then taken as the local buckling limit.
\end{itemize}
Table~\ref{tab:3D_results} summarizes the geometric parameters used to define the 3D lattices, the stiffness coefficients employed in the analytical calculations, and the resulting analytical and numerical buckling pressures.
\begin{table}[ht]
    \centering
    \caption{Geometrical, stiffness and buckling parameters for 3D the single fiber geometry (A), and the sheaf geometries (B).}
    \label{tab:3D_results}
    \begin{tabular}{l||l||c|cc}
        \toprule
        \multicolumn{2}{c||}{Geometries} & A& \multicolumn{2}{c}{B}\\
        \midrule
        
        \multirow{5}{*}{
            \rotatebox[origin=c]{90}{
                \shortstack{Geometric\\coefficients}
            }
        }
        % & $d_{min}/L$     & $0$   & \multicolumn{2}{c}{\num{2.5e-2}}\\
        & $R_s/L$       & --    & \multicolumn{2}{c}{$0.4$}\\
        % & $A/L_f/L$       & --    & \multicolumn{2}{c}{$0.05$ }\\
        & $\tau$        & --    & \multicolumn{2}{c}{ 0.25 }\\
        & $N_f$         & $1$   & \multicolumn{2}{c}{$37$}\\
        & $d_f/L$         & \num{4e-2}   & \multicolumn{2}{c}{\num{2.93e-2}}\\
        & $\beta$         & \num{4e-2}   & \multicolumn{2}{c}{\num{5.87e-2}}\\
        & $\lambda$     & --    & \multicolumn{2}{c}{ 3.22 }\\
        \midrule
        
        \multirow{5}{*}{%
            \rotatebox[origin=c]{90}{%
                \shortstack{Stiffness\\coefficients}%
            }%
        }
        & $a_K$ [\si{\newton \per \meter}]& \num{ 8.80e5 }& \multicolumn{2}{c}{\num{3.5e7}}\\
        & $b_{K,e}$ [\si{\newton \per \meter}]& \num{1.06e3}& \multicolumn{2}{c}{\num{9.03e4}}\\
        & $b_{K,g}$ [\si{\per \meter}]& \num{120}& \multicolumn{2}{c}{\num{240}}\\
        & $c_{K,e}$ [\si{\newton\meter}]& \num{6.77e-3}& \multicolumn{2}{c}{\num{0.549}}\\
        & $c_{K,g}$ [\si{\meter}]& \num{0}& \multicolumn{2}{c}{\num{1.07e-3}}\\
        & $d_{K,e}$ [\si{\newton\meter}]& \num{3.52e-2}& \multicolumn{2}{c}{\num{80.7}}\\
        & $d_{K,g}$ [\si{\meter}]  & \num{1.33e-3}& \multicolumn{2}{c}{\num{2.67e-3}}\\
        \midrule
        
         \multicolumn{2}{c||}{Link connection}& & Rigid & Compliant \\
        \midrule
        % \cmidrule(lr){2-5}
        \multirow{6}{*}{
            \rotatebox[origin=c]{90}{
                \shortstack{Pressure\\limits}
            }
        }
        & Global (analytic) [\si{\mega\pascal}]   & \num{2.08e-2}& \multicolumn{2}{c}{\num{1.15}} \\
        & Global (numeric)  [\si{\mega\pascal}]   & \num{2.12e-2}& 1.30& 1.08 \\
        & Error              & $-2.2\%$&  $-11\%$& $  +6.2\%$ \\
        \cmidrule(lr){2-5}
        & Local $L$ (analytic)[\si{\mega\pascal}] & \num{1.79e-2}& \multicolumn{2}{c}{\num{1.62}} \\
        & Local $X$ (analytic)[\si{\mega\pascal}] & \num{1.73e-2}& \multicolumn{2}{c}{\num{1.63}} \\
        & Local $L$ (numeric) [\si{\mega\pascal}] & \num{1.79e-2}& 1.61 & 1.52 \\
        & Local $X$ (numeric) [\si{\mega\pascal}] & \num{1.72e-2}& 1.61 & 1.53 \\
        & Error             & $+0.57\%$& $  +1.2\%$ & $+7.0\%$ \\
        
    \end{tabular}
\end{table}
The agreement between the analytical and numerical predictions is again very good for the single-fiber geometry, with errors below $3\%$. For the sheaf-of-beams geometry, larger discrepancies are observed for the global buckling pressure, particularly when rigid connections are considered.
In this case, the numerical pressure limit is higher than the analytical prediction, whereas for compliant connections the analytical prediction slightly overestimates the numerical result. The discrepancy is therefore partly compensated when compliant connections are considered, resulting in a smaller overall error. Across all the considered 3D cases, the maximum discrepancy between the analytical and numerical homogenized buckling pressures is $11\%$. Regarding the local instability, the numerical results obtained at $\bs{\kappa}_L^\star$ and $\bs{\kappa}_X^\star$ are very similar. In particular, for the single-fiber geometry, the two numerical limits differ by less than $1\%$, while for the sheaf-of-beams geometry they coincide within the reported numerical accuracy. This confirms that both wavevectors provide nearly equivalent critical conditions for the considered structures.
% Higher values are found for homogenized velocity of the fiber-sheaf geometry with rigid connections between the links. This discrepancy is overcompensated in when compliant connection are considered due to the reduction of the overall stiffness. The final results show that using a fully analytical model, the maximum error found remains bounded to a limit of $11\%$.Regarding the local buckling limit, they also show that there is not much difference numerically in the results found using $\bs \kappa_L^\star$ or $\bs \kappa_X^\star$ 

The 3D results lead to the same overall conclusions as those obtained for the 2D case, with two additional considerations.
First, the local buckling limit is governed not only by the bending stiffness $d_K$ but also by the torsional stiffness $c_K$, owing to the three-dimensionality of the problem. However, as in the 2D case, the single-fiber geometry does not provide sufficient geometric freedom to tune the bending and torsional stiffnesses independently of the axial stiffness.
Second, the single-fiber and the sheaf-of-beams geometries exhibit buckling pressure limits that differ by approximately two orders of magnitude. This difference is directly related to the different stiffness levels of their links, which are substantially lower in the single-fiber case. Therefore, when designing a pentamode lattice, moving from a single-fiber to a sheaf-of-beams geometry does not necessarily result in lower buckling resistance. In contrast, the additional geometric freedom provided by the sheaf may allow higher link stiffnesses to be achieved while retaining the pentamode behavior, resulting in a higher buckling limit.

\section{Conclusions}
\label{sec:conclusions}

In this work, a general analytical framework has been developed to assess the buckling stability of prestressed isotropic PM lattices. The formulation is based on a generic link model with appropriate symmetry assumptions, whose condensed stiffness is used to derive the mechanical response of the lattice. This provides a common framework for different link geometries and applies to both two- and three-dimensional PM structures.

The stability analysis distinguishes between two distinct instability mechanisms. Global instability is associated with the loss of stability of the effective medium and occurs in the long-wavelength limit, whereas local instability occurs at finite wavelengths comparable to the lattice cell size. The buckling limit is therefore determined by the lower of the corresponding critical pressures, showing that the homogenized response alone is not sufficient to assess the stability of a pentamode lattice. The analytical expressions further reveal that the two instability mechanisms depend on different mechanical characteristics of the links: while the global limit is primarily controlled by their axial and transverse stiffnesses, the local limit is additionally affected by bending stiffness in 2D and by both bending and torsional stiffness in 3D.

The analytical predictions are validated against numerical simulations of different 2D and 3D architectures, showing good agreement over the configurations considered. The results also highlight a fundamental trade-off between PM behavior and stability, since reducing the transverse stiffness of the links, which is essential for approaching the PM limit, also reduces the buckling pressure. Interestingly, however, the sheaf-of-beams architecture does not necessarily lead to lower buckling resistance, despite being specifically designed to reduce the transverse-to-axial stiffness ratio. The additional geometric freedom provided by multiple fibers allows this reduction to be achieved while maintaining substantially higher absolute stiffness levels than those attainable with a single fiber, ultimately resulting in higher buckling pressures while retaining the desired PM behavior.

Overall, the proposed framework provides closed-form estimates of the buckling limits and relates the stability of PM lattices directly to the mechanical characteristics of their links. It therefore offers a simple analytical basis for designing PM architectures with improved combinations of mechanical performance and buckling resistance.

% ============================================================
% ACKNOWLEDGEMENTS
% ============================================================

% \section*{Acknowledgements}

% Grazia, graziella e

% ============================================================
% APPENDIX
% ============================================================

\appendix

\section{Stiffness and  mass matrices for the sheaf of beams}
\label{sec:AppendixSheafOfBeams}

In both 2D and 3D, we consider link geometries consisting of sheaves of beams connected to rigid bodies, as presented in Figure~\ref{fig:Link_and_cell_Models}~\subref{fig:2DUnitCellFasci} and \subref{fig:3DUnitCellFasci}. It is worth noting that by choosing the right parameters, we encompass the classic hexagonal honeycomb and pentamode lattices (Figure~\ref{fig:Link_and_cell_Models}~\subref{fig:2DUnitCell} and \subref{fig:3DUnitCell}). To compute the coefficients for the matrix in Eq.~\eqref{eq:M_or_K_beam_3D}, we first derive the effect of the sheaf of beams, assuming negligible rigid-body dimensions and Euler-Bernoulli models. We then account for finite rigid bodies through a coordinate transformation and an additional geometric stiffness and inertial contribution. 
We remark that considering the connections between links as rigid has already been shown by Malek \textit{et~al.} \cite{Malek2015EffectiveHoneycombs} to be effective in 2D honeycombs; this is even more valid here since we expect greater differences in stiffness between the connections and the link due to the shape of the sheaf.

These considerations are carried out for both stiffness and mass matrices, although the latter are not used for the definition of the buckling limits, but only for model validation.

\subsection{Stiffness Matrices}
Let us start from the 2D case. We consider a single fiber that composes the sheaf, with length $L_f$ and height $h_f$; we define its aspect ratio as $\beta = h_f/L_f$, and its compressive load as $P_f = P/N_f$, where $N_f$ denotes the number of beams composing the sheaf. Its stiffness matrix shows the structure presented in Equation~\eqref{eq:K_beam_2D}, where the coefficients are 
 $a_{K,f} = E_{ps} \beta$, $b_{K,f} =  E_{ps}\beta^3-6P_f/(5L_f)$, and $d_{K,f} = E_{ps}\beta^3L_f^2/3-2P_fL_f/15$, the subscript $f$ denoting the single fiber. 

Let us first consider the case with no rigid body, highlighted by the subscript $\tau = 0 $. We determine the coefficient $a_{K,\tau = 0}$ by applying a unit displacement $u_1$. Each fiber moves of $u_{1,i} = u_1$ and the axial force that appears in each fiber tip is $a_f u_1$. Therefore, the total axial force is 
\begin{align}
    a_{K,\tau = 0}u_1=N_fa_fu_1    
\end{align}
yielding the result 
\begin{align}
    a_{K,\tau = 0} = N_f a_f.
\end{align}
By analogous reasoning, a unit displacement $v_1$ is applied, obtaining 
\begin{align}
    b_{K,\tau = 0} = N_f b_f.
\end{align}
Finally, we need to find $d_{K,\tau = 0}$. We can activate a unit rotation $\psi_1$, which induces a rotation of $\psi_{1,i} = \psi_1$ and an axial displacement $u_{1,i} = \psi_1y_i$ for each fiber, where $y_i$ is the coordinate of the neutral axis  of the $i^{th}$ fiber.  Therefore, the resulting moment is \cite{Brambilla2025HighWater}
\begin{align}
d_{K,\tau = 0}\psi_1 = (N_fd_{K,f}+a_{K,f}\sum_iy_i^2)\psi_1, 
\end{align}leading to 
\begin{align}
    d_{K,\tau = 0}= (N_fd_{K,f}+a_{K,f}\sum_iy_i^2) .     
\end{align}
We can compute the contribution $\sum_i y_i^2$. We assume that the adjacent fibers are symmetrically spaced by a distance $\lambda h_f$, with $\lambda>1$. Their coordinates from the neutral axis are 
\begin{align}
    y_i & = \left(-\frac{N_f-1}{2}+i-1\right)\lambda h_f & i &\in [1,\dots, N_f]
\end{align}
yielding 
\begin{align}
    \sum_{i=1}^{N_f}y_i^2 = \frac{1}{12}N_f(N_f^2-1)\lambda^2h_f^2
\end{align}
and the bending stiffness coefficient is 
\begin{align}
    \begin{aligned}
    d_{K,\tau = 0}  &= N_f d_f  + \frac{1}{12}a_f N_f(N_f^2-1)\lambda^2h_f^2, \\
                    & = N_fd_f + \frac{1}{12}a_{K,\tau = 0}(N_f^2-1)\lambda^2\beta^2L_f^2.
\end{aligned}
\end{align}

The resulting equation shows that the bending stiffness of a sheaf of beams is not simply the sum of the effect of each fiber. In fact, there is also a factor introduced by the spacing of the fibers: the farther the fibers are disposed from one another, the higher the bending stiffness. This knowledge is also important because it can be used to improve the local buckling limit, as it depends on the stiffness parameters $d_K$.  

Finally, we need to consider the effect of the rigid bodies connected to the fibers. We consider the total length of the link equal to $L$ and the rigid offset length to be $t = \tau L$, $\tau\in (0,1/2)$. Therefore, $L_f = (1-2\tau)L$. 
The main effect of the rigid bodies is the induced change of coordinates
\begin{align}
    \begin{Bmatrix}
        u_1'  \\ v_1' \\ \psi_1'
    \end{Bmatrix}  & = 
    \begin{bmatrix}
        1 & 0&0  \\
        0 & 1  & \tau L \\
        0 & 0 & 1
    \end{bmatrix}
    \begin{Bmatrix}
        u_1  \\ v_1 \\ \psi_1
    \end{Bmatrix}= [\mathsf T]\begin{Bmatrix}
        u_1  \\ v_1 \\ \psi_1
    \end{Bmatrix},
\end{align} 
where 
% the variables are represented in  Figure \todo{fare figura} and
$[\mathsf T]$ is the transformation operator. The opposite link tip is symmetric; hence, its transformation matrix is simply $[\mathsf T]^{-1 }$.  
In addition, under compressive loading, rigid-body rotations produce an additional geometric stiffness contribution due to the eccentricity of the load path equal to $P t = P\tau L$ \cite{Landau1959TheoryElasticity}.

\begin{figure}
    \centering
    \includegraphics[width=0.4\linewidth]{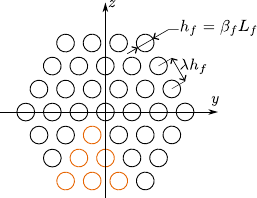}
    \caption{Example of fiber disposition for the 3D sheaf. Here, $n = 4$ and $N_f = 37$.}
    \label{fig:3DDisposizioneFasci}
\end{figure}

The resulting stiffness matrix of the link reads:
\begin{align}
    \mathsf K = 
    \begin{bmatrix}
            [\mathsf T] & [0] \\
            [0] &[\mathsf T]^{-1}
    \end{bmatrix}^\top 
    \mathsf K_{\tau=0}
    \begin{bmatrix}
            [\mathsf T] & [0] \\
            [0] &[\mathsf T]^{-1}
    \end{bmatrix}-P
    \begin{bmatrix}
           0& & & & &\\
           & 0& & & [0]& \\
           & & \tau L& & & \\
           & & & 0& & \\
           & [0]& & & 0& \\
           & & & & & \tau L
    \end{bmatrix}
\end{align}
We note that the final formulation will still follow the shape found in Eq.~\eqref{eq:K_beam_2D}. Particularly, after some algebra, the final coefficients will be: 
% \begin{align}
%     & a_K =E_{ps}N_f\beta\\
%     & b_K =  E_{ps}N_f\beta^3-\frac65\frac P {L (1-2\tau)} \\
%     \begin{split}
%         d_K  = E_{ps}N_f\left[ \left( \frac13+\frac1{12}(N_f^2-1)\lambda^2\right)(1-2\tau)^2+\tau(1-\tau)\right]\beta^3L^2 - 
%         \\
%         P \left( \frac{2+10\tau-10\tau^2}{15(1-2\tau)} \right)L.
%     \end{split}
% \end{align}
\begin{align}
    \begin{aligned}
         a_K & =E_{ps}N_f\beta &
         b_K & = b_{K,e}- Pb_{K,g}  &
         d_K & = d_{K,e}- Pd_{K,g} 
    \end{aligned}
\end{align}
with 
\begin{align}
    \begin{aligned}
        b_{K,e} &=  E_{ps}N_f\beta^3  &
        b_{K,g} & = \frac6 {5L (1-2\tau)} \\
        d_{K,e} &= E_{ps}N_f\left[ \frac13\left(1 +\frac{N_f^2-1}{4}\lambda^2\right)(1-2\tau)^2+\tau(1-\tau)\right]\beta^3L^2  &
        d_{K,g} &= \frac{2+10\tau-10\tau^2}{15(1-2\tau)} L
    \end{aligned}
\end{align}
where we clearly identified the elastic and geometric contributions. \\ 
\\
We can perform a similar process for 3D geometries
% as that shown in Figure \todo{figura fasci e disposizione esagonale}. 
In this case, we still denote by $L_f$ the fiber length, while $h_f$ is its diameter and $\beta = h_f/L_f$ its aspect ratio. Each fiber is subject to a load $P_f = P/N_f$, where $N_f$ is the total number of fibers in the sheaf. 
The total number of fibers $N_f$ is related to the number of fibers on each edge of the hexagon $n$ with the relation $N_f = 3n(n-1)+1$\cite{Brambilla2025HighWater}. We consider Euler-Bernoulli beams, which satisfy Equation~\eqref{eq:K_beam_3D} with coefficient 
% $a_{K,f} = \pi E\beta^2L_f/4$, $b_{K,f} = 3\pi E\beta^4L_f/16-6P_f/(5L_f)$, $c_{K,f} = \pi G\beta^4L_f^3/32 $, and $d_{K,f} =\pi E\beta^4L_f^3/16-2P_fL_f/15$, 
\begin{align}
    \begin{aligned}
        a_{K,f} &= \pi E\beta^2L_f/4 &
        b_{K,f} &= 3\pi E\beta^4L_f/16-6P_f/(5L_f)\\
        c_{K,f} &= \pi G\beta^4L_f^3/32  &
        d_{K,f} &=\pi E\beta^4L_f^3/16-2P_fL_f/15 
    \end{aligned}
\end{align}
where $G = E/(2(1+\nu))$ and $\nu$ is the Poisson's ratio.
The sheaf  is arranged in a hexagonal pattern, as shown in Figure~\ref{fig:3DDisposizioneFasci}. 
This ensures that all the hypotheses made in Section~\ref{sec:LatticeDefinition} are satisfied. 
Therefore, the sheaf stiffness matrix, both with and without rigid bodies, respects Equation~\eqref{eq:K_beam_3D}, and its coefficients must be found. 
We follow the same path traced for the 2D case; starting with a rigid body of negligible dimensions, we can immediately define $ a_{K,\tau = 0 } = N_fa_{K,f} $ and $b_{K, \tau = 0} = N_fb_{K,f}$. 
The term $d_{K,\tau = 0}$ can be found by activating $\varphi_1$ or $\psi_1$, as the stiffness associated with either displacement is the same. 
Using the same reasoning shown in 2D we obtain $d_{K,\tau = 0}= (N_fd_{K,f}+a_{K,f}\sum_iy_i^2)$ or $d_{K,\tau = 0}= (N_fd_{K,f}+a_{K,f}\sum_iz_i^2)$, which shows that $\sum_iy_i^2 = \sum_i z_i^2$; hence, we need to compute $\sum_iz_i^2$ only. 
In addiction, we can activate a unit angular displacement $\vartheta_1$ to find $c_{K,\tau = 0}$, not present in the 2D case. 
In this case, each fiber experiences the same angular displacement $\vartheta_{1,i} = \vartheta_1$, and a transversal displacement $v_{1,i} = z_i\vartheta_1$, $w_{1,i} = y_i\vartheta_1$, which generate, respectively, a torque equal to $c_{K,f}\vartheta_1$ and shear loads $b_{K,f}v_{1,i}$ and  $b_{K,f}w_{1,i}$, centered on the beams neutral axis. 
Therefore, the total torque of the sheaf can be computed as $c_{K,\tau = 0}\vartheta_1 = \sum_i(c_{K,f}+b_{K,f}(y_i^2+z_i^2))\vartheta_1$, yielding $c_{K,\tau = 0} = N_fc_{K,f}+2b_{K,f}\sum_iz_i^2$.

% We need to compute $\sum_i z_i^2$ 
For the pattern in Figure~\ref{fig:3DDisposizioneFasci},
% We consider 
\begin{align}
    \sum_{i=1}^{N_f}z_i^2 = \sum_{m=2}^n\sum_{i_m=1}^{6(m-1)}z_{i_m}^2
\end{align}
where $z_{i_m}$ are the $z$ coordinates of the $6(m-1)$ fibers that form the $m^{\text{th}}$ hexagonal row. Note that $m=1$ represents the central fiber.
Consider the fibers shown in orange and take only the $m^{\text{th}}$ row. The coordinates of the neutral axis of these fibers will be  
\begin{align}
\label{eq:coordinates_fibers}
\begin{aligned}
    y_{i_m} & = \lambda h_f\left(i_m-\frac{m+1}{2}\right) \\
    z_{i_m} & = -\frac{\sqrt{3}}{2}\lambda h_f(m-1)
\end{aligned}
\end{align}
with $i_m \in [1, m-1]$. The neutral axis of the other fibers can be found by applying rotations of \SI{60}{\degree}. 
% \todo{check su come si scrive.} 
We can exploit the symmetry about the origin and only compute the coordinates of the beam at \SI{60}{\degree} and \SI{120}{\degree}, which are
\begin{align}
    \begin{Bmatrix}
        y_{i_m+m-1} \\ z_{i_m+m-1} 
    \end{Bmatrix} & = 
    \frac 12
    \begin{bmatrix}
        1 & -\sqrt 3 \\
        \sqrt 3  & 1
    \end{bmatrix}
    \begin{Bmatrix}
        y_{i_m} \\ z_{i_m}
    \end{Bmatrix},
    \\
    \begin{Bmatrix}
        y_{i_m+2(m-1)} \\ z_{i_m+2(m-1)} 
    \end{Bmatrix} & = 
    \frac 12
    \begin{bmatrix}
        -1 & -\sqrt 3 \\
        \sqrt 3  & -1
    \end{bmatrix}
    \begin{Bmatrix}
        y_{i_m} \\ z_{i_m}
    \end{Bmatrix}.
\end{align}
Therefore,
\begin{align}
    \begin{aligned}
        \sum_{i_m=1}^{6(m-1)}z_{i_m}^2 & = 2\sum_{i_m=1}^{3(m-1)}z_{i_m}^2 \\
        % \sum_{i_m=1}^{6(m-1)}z_{i_m}^2 
        & = 2\sum_{i_m = 1}^{m-1}\left(z_{i_m}^2+\left( \frac 12
                \begin{bmatrix}
                    \sqrt 3 & 1
                \end{bmatrix}
                \begin{Bmatrix}
                     y_{i_m} \\ z_{i_m}
                \end{Bmatrix}
                \right)^2+\left(\frac 12
                \begin{bmatrix}
                    \sqrt 3 & -1
                \end{bmatrix}
                \begin{Bmatrix}
                     y_{i_m} \\ z_{i_m}
                \end{Bmatrix}
                \right)^2\right) \\
        & 
        = 3\sum_{i_m = 1}^{m-1}(y_{i_m}^2+z_{i_m}^2)
    \end{aligned}
\end{align}
and by incorporating  Eq.~\eqref{eq:coordinates_fibers} we obtain
\begin{align}
    \sum_{i_m=1}^{6(m-1)}z_{i_m}^2 = \frac 12 \lambda^2h_f^2(m-1)(5m^2-10m+6).
\end{align}
Thus, 
\begin{align}
    \begin{aligned}
          \sum_{i=1}^{N_f}z_i^2& = \frac 12 \lambda^2h_f^2\sum_{m=2}^n(m-1)(5m^2-10m+6) \\
          & = \frac 18 \lambda^2\beta^2n(n-1)(5n^2-5n+2)L_f^2 
    \end{aligned}
\end{align}
and applying the relationship between $N_f$ and $n$ we finally obtain
\begin{align}
    \sum_{i=1}^{N_f}z_i^2 =
    % \frac {1}{8} \frac{N_f-1}{3}\frac{5N_f+1}{3}\lambda^2\beta^2L_f^2
     \frac {1}{72} (N_f-1)(5N_f+1)\lambda^2\beta^2L_f^2 
\end{align}
which is the required result. Then, the stiffness coefficients are  
% \begin{align}
%     d_{K,\tau = 0} &=\frac{\pi}{16}E\left(N_f+\frac{1}{18}(N_f-1)(5N_f+1)\lambda^2\right)\beta^4L_f^3  - \frac{2}{15}PL_f\\
%      c_{K,\tau = 0} &=\frac{\pi}{32}E\left(\frac{N_f}{2(1+\nu)}+\frac{1}{6}(N_f-1)(5N_f+1)\beta^2\lambda^2  \right)\beta^4L_f^3 - \frac{(N_f-1)(5N_f+1)\beta^2\lambda^2}{30N_f}PL_f
% \end{align}
\begin{align}
    \begin{aligned}
        d_{K,\tau = 0} &=d_{Ke, \tau = 0}  -Pd_{Kg, \tau = 0} &
         c_{K,\tau = 0} &=c_{Ke, \tau = 0}- Pc_{Kg, \tau = 0}
    \end{aligned}
\end{align}
with
% \begin{align}
%     d_{Ke,\tau = 0} &=\frac{\pi}{16}N_fE\left(1+\frac{(N_f-1)(5N_f+1)}{18N_f}\lambda^2\right)\beta^4L_f^3  \\
%     d_{Kg,\tau = 0} & =\frac{2}{15}L_f\\
%      c_{Ke,\tau = 0} &=\frac{\pi}{32}N_fE\left(\frac{1}{2(1+\nu)}+\frac{(N_f-1)(5N_f+1)}{6N_f}\beta^2\lambda^2  \right)\beta^4L_f^3 \\
%      c_{Kg,\tau = 0} &=\frac{(N_f-1)(5N_f+1)\beta^2\lambda^2}{30N_f}L_f
% \end{align}
\begin{align}
    \begin{aligned}
        d_{Ke,\tau = 0} &=\frac{\pi}{16}N_fE\left(1+\frac{\Delta}{18}\lambda^2\right)\beta^4L_f^3,  &
        d_{Kg,\tau = 0} & =\frac{2}{15}L_f,\\
         c_{Ke,\tau = 0} &=\frac{\pi}{32}N_fE\left(\frac{1}{2(1+\nu)}+\frac{\Delta}{6}\beta^2\lambda^2  \right)\beta^4L_f^3, &
         c_{Kg,\tau = 0} &=\frac{\Delta}{30}\beta^2\lambda^2L_f,
    \end{aligned}
\end{align}%\todo{sistemare come 2D}
where 
\begin{align}
\label{eq:Delta_N_f}
    \Delta = \frac{(N_f-1)(5N_f+1)}{N_f}.
\end{align}
Finally, we add the effect of the rigid body. As in the 2D case, its effects are twofold: the introduction of a change of variables and an additional geometric stiffness contribution.  

The change of variables is given by 
\begin{align}
    \begin{Bmatrix}
        u_1' \\
        v_1' \\
        w_1' \\
        \vartheta_1' \\
        \varphi_1' \\
        \psi_1'
    \end{Bmatrix}^\top
    &= \mathsf T
    \begin{Bmatrix}
        u_1 \\
        v_1 \\
        w_1 \\
        \vartheta_1 \\
        \varphi_1 \\
        \psi_1
    \end{Bmatrix}^\top
    &\qquad \text{where}\quad\mathsf T &= 
    \begin{bmatrix}
        1 & 0 & 0 & 0 & 0 & 0 \\
        0 & 1 & 0 & 0 & 0 & t \\
        0 & 0 & 1 & 0 & -t & 0 \\
        0 & 0 & 0 & 1 & 0 & 0 \\
        0 & 0 & 0 & 0 & 1 & 0 \\
        0 & 0 & 0 & 0 & 0 & 1
    \end{bmatrix}
\end{align}
% where
% \begin{align}
%     \mathsf T = 
%     \begin{bmatrix}
%         1 & 0 & 0 & 0 & 0 & 0 \\
%         0 & 1 & 0 & 0 & 0 & t \\
%         0 & 0 & 1 & 0 & -t & 0 \\
%         0 & 0 & 0 & 1 & 0 & 0 \\
%         0 & 0 & 0 & 0 & 1 & 0 \\
%         0 & 0 & 0 & 0 & 0 & 1
%     \end{bmatrix}
% \end{align}
% 
while, following the procedure of the 2D case, the geometric stiffness contribution is given by 
\begin{align}
    \mathsf K_{\tau, g} = 
    \text{diag}(
    \begin{Bmatrix}
        0 & 0 & 0 & 0 & \tau L & \tau L &0 & 0 & 0 & 0 & \tau L & \tau L
    \end{Bmatrix}
    ).
\end{align}
The stiffness matrix of the links is computed as
\begin{align}
    \mathsf K = 
    \begin{bmatrix}
            [\mathsf T] & [0] \\
            [0] &[\mathsf T]^{-1}
    \end{bmatrix}^\top 
    \mathsf K_{\tau=0}
    \begin{bmatrix}
            [\mathsf T] & [0] \\
            [0] &[\mathsf T]^{-1}
    \end{bmatrix}-P
   \mathsf K_{\tau,g}
\end{align}
and the resulting stiffness coefficients are
\begin{align}
    \left\{
    \begin{aligned}
        a_K & = a_{K,\tau = 0} =  E \frac{\pi}{4}N_f\beta^2L(1-2\tau) \\
        b_K & = b_{K,e}-Pb_{K,g}\\
         c_K & = c_{K,e}- Pc_{K,g} \\
         d_K & = d_{K,e}- Pd_{K,g}
    \end{aligned}
    \right.
\end{align}
with 
\begin{align}
    &\left\{
    \begin{aligned}
        b_{K,e} & =b_{Ke, \tau = 0}=  \frac {3\pi}{16} N_f E\beta^4L(1-2\tau) \\
        b_{K,g} & =b_{Kg, \tau = 0} = \frac{6}{5L(1-2\tau)}
    \end{aligned}\right.\\
    &\left\{
    \begin{aligned}
        c_{K,e} & =c_{Ke, \tau = 0}=\frac{\pi}{32}N_fE\left(\frac{1}{2(1+\nu)}+\frac{\Delta}{6}\beta^2\lambda^2  \right)\beta^4L_f^3 \\
        c_{K,g} & =c_{Kg, \tau = 0}=\frac{\Delta}{30}\beta^2\lambda^2L_f
    \end{aligned}\right.\\
    &\left\{
    \begin{aligned}
        d_{K,e} & =  \frac{\pi }{16}N_fE\left[ 1+\frac{\Delta}{18}\lambda^2+\frac{3\tau(1-\tau)}{(1-2\tau)^2}\right](1-2\tau)^3\beta^4L^3 \\
        d_{K,g} & = \frac{2-5\tau(1-4\tau)}{15(1-2\tau)}L
    \end{aligned}\right.
\end{align}

\subsection{Mass matrices}
Although not strictly necessary for the definition of the limit pressure, we also define the matrices and coefficients related to the inertial terms. This was used to obtain the analytic dispersions of the studied lattices, which were then compared with the numerical ones in correspondence with the critical wave number. 

To obtain the mass matrices more coefficients need to be defined. An analogous procedure to that used for the stiffness has been followed, by activating one degree of freedom at a time with unitary acceleration.\\
\\
Let us start from the 2D geometries. The coefficients of Equation~\eqref{eq:M_beam_2D} for a single fiber are 
\begin{align}
    \begin{aligned}
        a_{M,f}& =\frac{1}{3}\rho \beta {L_f}^2  & 
        b_{M,f}& =\frac{13}{35}\rho \beta {L_f}^2  &
        d_{M,f}& =\frac{1}{105}\rho \beta {L_f}^4  &
        e_{M,f}& =\frac{1}{6}\rho \beta {L_f}^2  \\
        f_{M,f}& =\frac{11}{210}\rho \beta {L_f}^3  &
        g_{M,f}& =\frac{9}{70}\rho \beta {L_f}^2  &
        i_{M,f}& =-\frac{1}{140}\rho \beta {L_f}^4  &
        j_{M,f}& =-\frac{13}{420}\rho \beta {L_f}^3  \\
    \end{aligned}
\end{align}
Following the same idea shown for the stiffness, we compute these coefficients for a sheaf of fiber with negligible rigid body dimensions: 
\begin{align}
    \begin{aligned}
        a_{M,\tau = 0} & =  N_f a_{M,f}  & 
        b_{M,\tau = 0} & =  N_f b_{M,f}  &
        d_{M,\tau = 0} & =  N_f d_{M,f} + a_{M,f}\sum_iy_i^2  &
        e_{M,\tau = 0} & =  N_f e_{M,f}  \\
        f_{M,\tau = 0} & =  N_f f_{M,f}  &
        g_{M,\tau = 0} & =  N_f g_{M,f}  &
        i_{M,\tau = 0} & =  N_f i_{M,f} + e_{M,f}\sum_iy_i^2  &
        j_{M,\tau = 0} & =  N_f j_{M,f}  \\
    \end{aligned}
\end{align}
On top of that, we account for the rigid body by applying the same change of coordinates seen for the stiffness, yielding 
\begin{align}
    \begin{aligned}
        d_{M} & = d_{M,\tau = 0}+\tau L (2 f_{M,\tau = 0}+\tau L b_{M,\tau = 0} ), &
        f_{M} & = f_{M,\tau = 0}+\tau L b_{M,\tau = 0}  ,\\
        i_{M} & = i_{M,\tau = 0}+\tau L (2 j_{M,\tau = 0}-\tau L g_{M,\tau = 0} ), &
        j_{M} & = j_{M,\tau = 0}-\tau L  g_{M,\tau = 0},
    \end{aligned}
\end{align}
while the other coefficients remain constant. We do not add the mass of the rigid body at this stage, but rather to the full unit cell matrix, for simplicity. We can write the final result in extended form: 
\begin{align}
    \begin{aligned}
        \frac{a_M}{\rho} &= \frac{1}{3}     N_f \beta   \delta^2 L^2,&
        \frac{b_M}{\rho} &= \frac{13}{35}   N_f \beta   \delta^2 L^2,\\
        \frac{d_M}{\rho} &= \frac{1}{105}   N_f \beta   \delta^4 L^4 \left[ 1     +\frac{35(N_f^2-1)}{12}\lambda^2\beta^2 +\frac{11\tau}{\delta}+\frac{39\tau^2}{\delta^2}\right],&
        \frac{e_M}{\rho} &= \frac{1}{6}     N_f \beta   \delta^2 L^2,\\
        \frac{f_M}{\rho} &= \frac{1}{210}   N_f \beta   \delta^2 L^3(11+56\tau),    &
        \frac{g_M}{\rho} &= \frac{9}{70}    N_f \beta   \delta^2 L^2, \\
        \frac{i_M}{\rho} &= -\frac{1}{140}  N_f \beta   \delta^4 L^4
                                                    \left[ 1-\frac{35(N_f^2-1)}{18}\lambda^2\beta^2 +\frac{26\tau}{3\delta}+\frac{18\tau^2}{\delta^2}\right],&
        \frac{j_M}{\rho} &= -\frac{1}{420}  N_f \beta   \delta^2 L^3 (13+28\tau).
    \end{aligned}
\end{align}
where $\delta = (1-2\tau)$.

To ease computations, the inertia of the rigid bodies connecting the sheaf has been added directly in the unit cell mass matrix in Equation~\eqref{eq:UnitCellMatrix} in correspondence with node \# 1 and \# 2, rather than into the link matrices. 
% inwe add the mass and rotational inertial terms related to the connection between the links, modeled as a rigid body
Particularly, the mass and rotational inertial terms related to the connection have been evaluated. Their value has been assessed by removing the influence of the segments of a circle cut to lay the sheaves from the regular circle. 
The mass of the regular circle is $M_{c,reg} = \rho \pi {R_c}^2$ and its polar inertia $I_{c,reg} =\rho \pi /2 {R_c}^4$. The surface area of a segment of a circle is 
\begin{align}
    S_{seg} = {R_c}^2 \arccos\left(\frac{\tau L}{R_c}\right) - \tau L\frac{h_{tot}}{2}
\end{align}
with $ h_{tot}= 2\sqrt{{R_c}^2-(\tau L)^2} $. Therefore, the mass of the rigid block is
\begin{align}
    M_c = M_{c,reg}-3\rho S_{seg}.
\end{align}
The rotational inertia of the segment of a circle can be computed using its definition
\begin{align}
    \begin{aligned}
        % I_{seg}& = \int_{\tau L}^{R_c} \int_{-\sqrt{R_c^2-x^2}}^{\sqrt{R_c^2-x^2}} x^2+y^2 \mathrm d\ \\
        I_{seg}& = \rho \iint_{S_{seg}} x^2+y^2\mathrm dS\\
        & = \rho \frac{1}{4}{R_c}^4
            \left(
                \pi-
                \frac{2}{3}\frac{\tau L }{R_c} \left(1+2\left(\frac{\tau L}{R_c}\right)^2\right) \sqrt{1-\left(\frac{\tau L }{R_c}\right)^2} -
                2\arcsin{\frac{\tau L}{R_c}}
            \right)
    \end{aligned}
\end{align}
yielding 
\begin{align}
    I_c = I_{c,reg}-3I_{seg},
\end{align}
which is the parameter used for the analytical modeling.
\\
\\
The same procedure is followed for a 3D sheaf. For a single fiber, the mass matrix in Equation~\eqref{eq:M_or_K_beam_3D} has coefficients
\begin{align}
    \begin{aligned}
        a_{M,f} & =\frac1 3\rho A L_f,             & b_{M,f} & =\frac{13}{35}\rho A L_f  ,&   
        c_{M,f} & =\frac1 3 \rho J L_f,            & d_{M,f} & =\frac1 {105} \rho A L_f^3,  \\    
        e_{M,f} & =\frac{1}{6} \rho A L_f,         & f_{M,f} & =\frac{11}{210}\rho A L_f^2,  &    
        g_{M,f} & =\frac{9}{70}\rho A L_f,         & h_{M,f} & =\frac{1}{6}\rho J L_f  ,\\    
        i_{M,f} & =-\frac{1}{140}\rho A L^3_f,     & j_{M,f} & =-\frac{13}{420}\rho A L_f^2.  \\    
    \end{aligned}
\end{align}
with $ A = \pi/4\beta^2L_f^2$ and $ J = \pi/32\beta^4L_f^4$. The coefficient for the sheaf of fiber with negligible rigid body dimension becomes
\begin{align}
    \begin{aligned}
        a_{M,\tau = 0} &=  N_f a_{M,f},                                  & b_{M,\tau = 0} &= N_f b_{M,f} , &        c_{M,\tau = 0} &=  N_f c_{M,f} + 2b_{M,f} \sum_{i=1}^{N_f}z_i^2,  \\
        d_{M,\tau = 0} &= N_f d_{M,f} + a_{M,f} \sum_{i=1}^{N_f}z_i^2, &        e_{M,\tau = 0} &=  N_f e_{M,f},     \\
        f_{M,\tau = 0} &= N_f f_{M,f} , &        g_{M,\tau = 0} &=  N_f g_{M,f},      & h_{M,\tau = 0} &= N_f h_{M,f} + 2g_{M,f}\sum_{i=1}^{N_f}z_i^2,\\  
        i_{M,\tau = 0} &=  N_f i_{M,f} + e_{M,f}\sum_{i=1}^{N_f}z_i^2,           & j_{M,\tau = 0} &= N_f j_{M,f}.\\    
    \end{aligned}
\end{align}
To consider the rigid body, we apply the same change of coordinates already seen for the stiffness matrix, obtaining the following: 
\begin{align}
    \begin{aligned}
        d_{M} &= d_{M,\tau = 0}+\tau L (2f_{M,\tau = 0}+\tau L b_{M,\tau = 0}), & f_{M} &= f_{M,\tau = 0}+\tau L b_{M,\tau = 0} \\ i_{M} &=i_{M,\tau = 0}+\tau L (2j_{M,\tau = 0} - \tau L g_{M,\tau = 0}),         & j_{M} &= j_{M,\tau = 0}- \tau L g_{M,\tau = 0}.   
    \end{aligned}
\end{align}
while keeping the remaining coefficients equal. In extended form: 
\begin{align}
    \begin{aligned}
    \frac{a_M}{\rho} & =    \frac{\pi}{12}      N_f \beta^2 L^3\delta^3, &
    \frac{b_M}{\rho} & =    \frac{13\pi}{140}   N_f \beta^2 L^3\delta^3,\\
    \frac{c_M}{\rho} & =    \frac{\pi}{96}      N_f \beta^4 L^5\delta^5\left[1+\frac{26}{105}\Delta\lambda^2 \right],&
    \frac{d_M}{\rho} & =    \frac{\pi}{420}     N_f \beta^2 L^5\delta^5 \left[ 1+\frac{\tau (11+17\tau)}{\delta^2} +\frac{35}{72}\Delta \beta^2\lambda^2 \right],\\
    \frac{e_M}{\rho} & =    \frac{\pi}{24}      N_f \beta^2 L^3\delta^3, &
    \frac{f_M}{\rho} & =    \frac{\pi}{840}     N_f \beta^2 L^4\delta^3(11+56\tau),\\
    \frac{g_M}{\rho} & =    \frac{9\pi}{280}    N_f \beta^2 L^3\delta^3,\\
    \frac{h_M}{\rho} & =    \frac{\pi}{192}     N_f \beta^4 L^5\delta^5 \left[ 1+\frac{\Delta}{9}\lambda^2 \right],&
    \frac{i_M}{\rho} & = -  \frac{\pi}{840}     N_f \beta^2 L^5\delta^5 \left[ 1 + \frac{2}{3}\frac{\tau (13+\tau)}{\delta^2} -\frac{13}{9}\Delta\beta^2 \lambda^2 \right],\\
    \frac{j_M}{\rho} & = -  \frac{\pi}{1680}    N_f \beta^2 L^4\delta^3(13+28\tau)
    \end{aligned}
\end{align}
As in the 2D case, we add the influence of rigid blocks in the analytical models to nodes \#1 and \#2 of the unit cell matrix, rather than in the link local matrix, considering them as lumped parameters. In particular, the mass $M_s$ and rotational inertia $I_s$ have been found and added to Eq.~\eqref{eq:3DUnitCellMatrix}.

These quantities are obtained by subtracting the contributions of the spherical caps from those of a regular sphere. For a regular sphere, $M_{s,reg}/\rho = 4/3\pi R_s^3$ and $I_{s,reg}/\rho = 8/15\pi R_s^5$. 

The volume of a spherical cap is
\begin{align}
    V_{cap} = \frac{1}{3} \pi(R_s-\tau L)^2(2R_s+\tau L)
\end{align}
yielding 
\begin{align}
    M_s = M_{s,reg}-4\rho V_{cap}.
\end{align}
The influence of the caps on the rotational inertia is computed using the definition 
\begin{align}
    I = \rho \iiint_{V} \varpi ^2 \mathrm dV,
\end{align}
where $\varpi$ is the radius from the rotation axis. Without loss of generality, we take this axis to be the $z$-axis in the following computations. We consider a cap oriented at an angle $\alpha$ with respect to it and apply a change of coordinates to describe the cap in a polar frame. Assuming that the axis of the cap is placed in the $xz$ plane, the change of coordinates reads:
\begin{align}
    x & = r \cos{\alpha}\cos \theta + z'\sin\alpha\\
    y & = r \sin \theta\\
    z & = -r \sin{\alpha}\cos \theta + z'\cos\alpha
\end{align}
where $r$, $\theta$, and $z'$ are the polar coordinates. Therefore, the rotational inertia of the cap is 
\begin{align}
    \frac{I_{cap}(\alpha)}{\rho} & =\iiint_{V_{cap}} x^2+y^2 \mathrm{d}V \\
     & =\int_{\tau L}^{R_s}\int_0^{2\pi}\int_0^{\sqrt{{R_s}^2-t^2}}
    \left(( r \cos{\alpha}\cos \theta + z'\sin\alpha)^2+(r\sin \theta)^2\right) r \mathrm d r \mathrm d \theta \mathrm d z'\\
    & = \frac{1}{120}\pi\left[32 {R_s}^5 +t\left( 3 t^4+10t^2{R_s}^2 -45{R_s}^4-15({R_s}^2-t^2)^2\cos{2\alpha}\right)\right].
\end{align}
Having the direction of each cap, we can define 
\begin{align}
    I_s = I_{s,reg}-\sum_{i=1}^4I_{cap}(\alpha_i)
\end{align}
where $i$ are the four caps removed to accommodate the sheaves. We highlight that for the geometries considered 
\begin{align}
    \sum_{i=1}^4\cos{(2\alpha_i)} = -\frac{4}{3}
\end{align}
independently on the chosen axis and on their orientation. This justifies the usage of a unique rotational inertia parameter $I_s$.

\section{Computed dispersion diagrams}
\label{sec:AppendixDispersions}
This section presents the numerical dispersion diagrams obtained at different load levels for the structures considered in the numerical validation and discusses the main features observed in the results.

Due to the similarity of the results, the dispersion diagrams for the cases with rigid connections between the fibers of the sheaf are not reported.

\begin{figure}
    \centering
    \subfloat[]{%
        \raisebox{0mm}{
            \includegraphics[
                width=0.5\linewidth,
                % trim=4cm 0.15cm 5cm 0.15cm,
                % clip
            ]{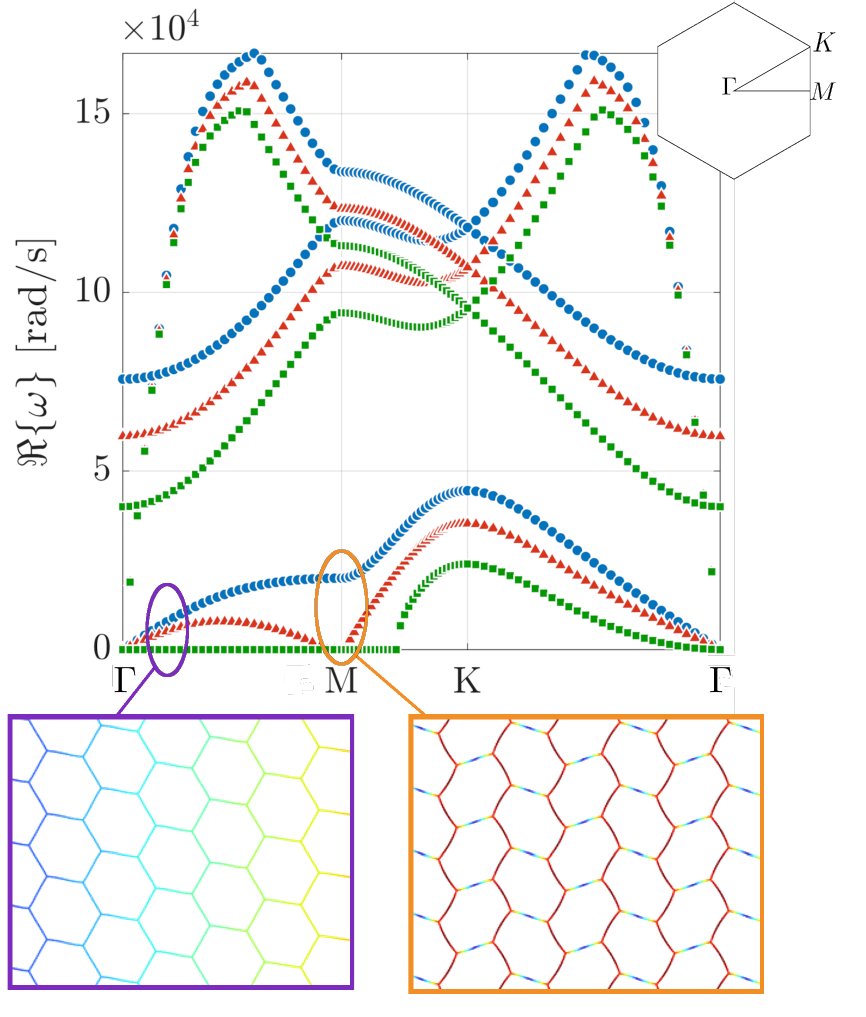}%
        }
    }
    % \hfill
    \subfloat[]{%
        \raisebox{2mm}{
            \includegraphics[
                width=0.5\linewidth,
                % trim=5.4cm 0cm 5.5cm 0.15cm,
                % clip
            ]{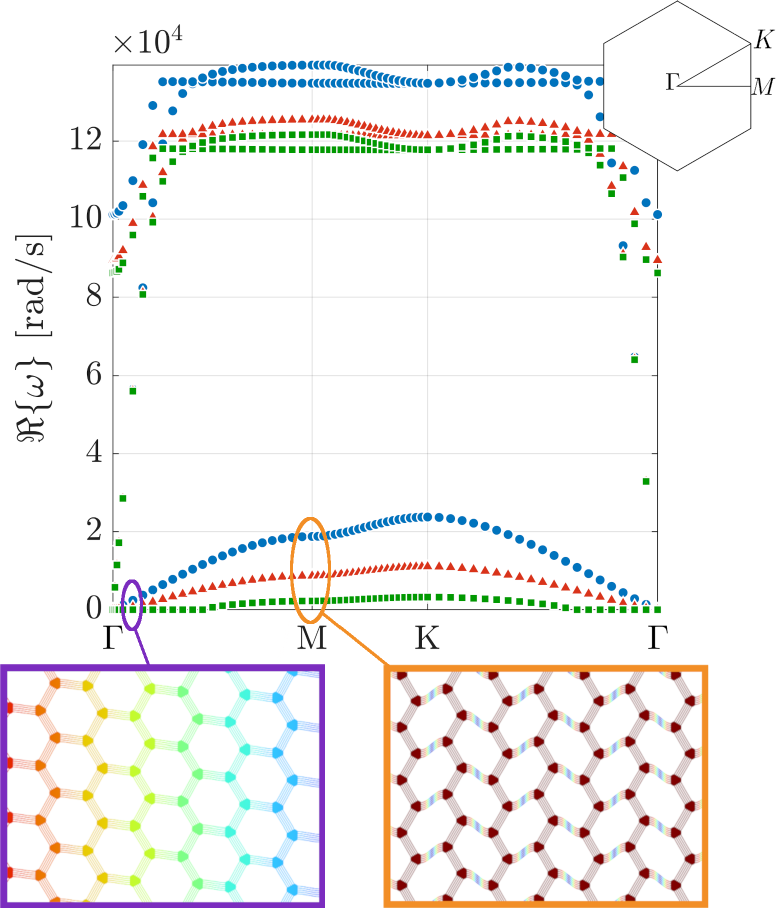}%
        }
    }
    
    \caption{Dispersion diagrams for the 2D geometries with (a) single-fiber links and (b) sheaf-of-beams links under different load conditions. In (a) the unloaded case (blue), the onset of local instability (red), and onset of global instability (green) are depicted. In (b) the unloaded case (blue), an intermediate load level (red) and the onset of global instability (green) are shown. The critical mode shapes, which are almost invariant with the applied load, are depicted for both geometries, illustrating the global or local nature of the corresponding deformations.}
    \label{fig:2DDispersions}
\end{figure}

Let us consider the 2D geometries. Figure~\ref{fig:2DDispersions} shows 
the dispersion diagrams for the single-fiber and sheaf-of-beams geometries. For the latter, due to the similarity between the results obtained with rigid and compliant connections, only the case with compliant connections is shown. As discussed in Section~\ref{sec:Instability}, buckling occurs when an additional zero-frequency mode appears at a finite wavevector, in addition to the rigid-body modes at $\bs\kappa=\bs0$. The dispersion diagrams confirm the theoretical predictions. For the single-fiber lattice, local buckling is the governing instability mechanism, as the first additional zero-frequency mode occurs at the point $M$ of the IBZ. Conversely, for the sheaf-of-beams lattice, global buckling occurs first, with the additional zero-frequency mode appearing in the vicinity of $\Gamma$.
% From the mode shapes, we can also infer that while in the global buckling the overall structure collapses, when local instability occurs, the failure mechanism is related to the slip of consecutive planes  

% The mode shapes shown in the figure also illustrate the different nature of the two instability mechanisms. The global instability is characterized by a long-wavelength deformation involving the overall lattice, whereas the local instability is associated with a shorter-wavelength deformation occurring at the scale of the unit cell.

\begin{figure}
    \centering
    \includegraphics[width=0.9\linewidth]{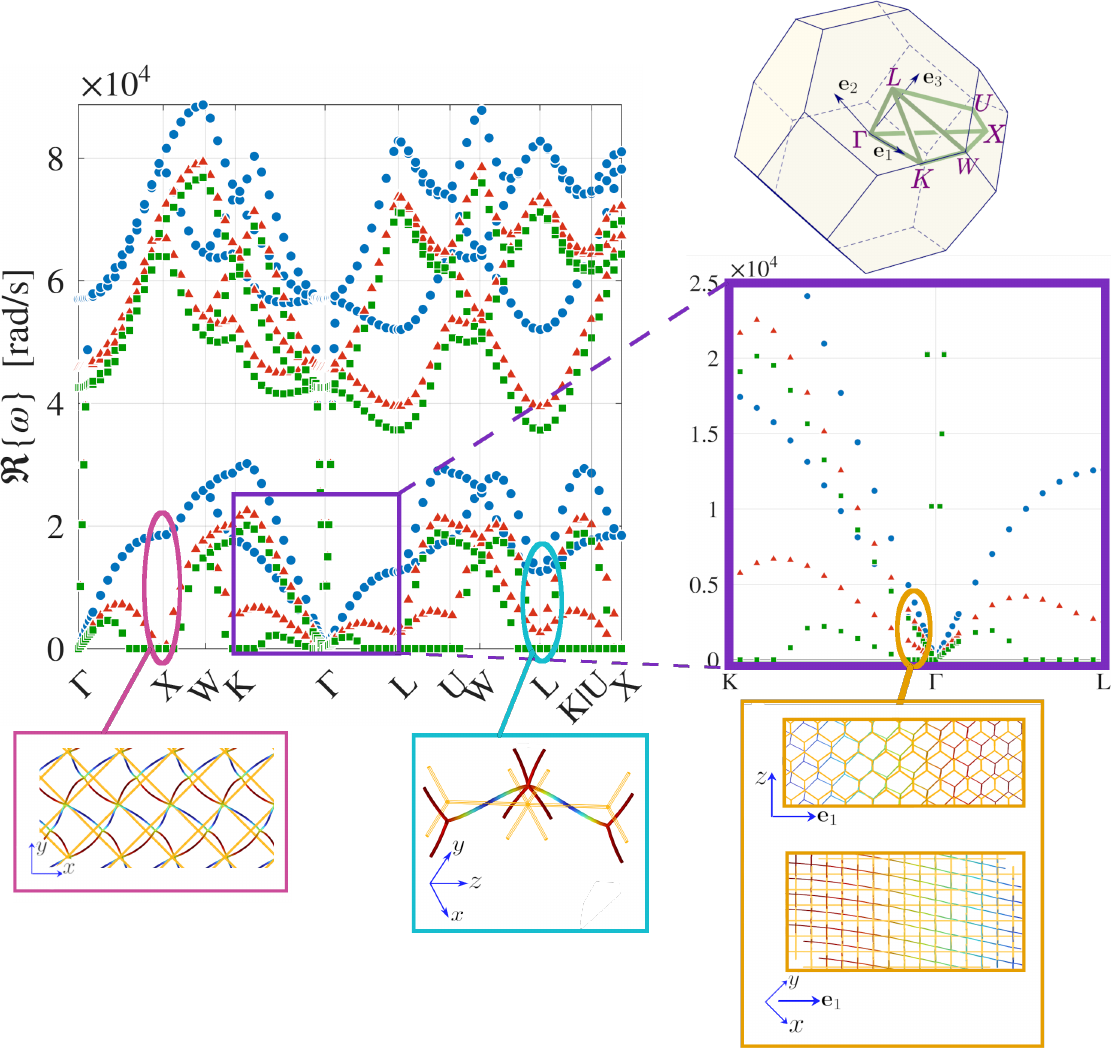}
    \caption{ 
        Dispersion diagrams for the 3D single-fiber geometry under different load conditions. The unloaded case (blue), the onset of local instability (red), and the onset of global instability (green) are shown. The critical mode shapes, which are almost invariant with the applied load, are also shown, illustrating the local or global nature of the corresponding deformations. The mode shapes are represented in an appropriate plane containing the wavevector. At the $X$ and $L$ points, where local instability occurs, only one of the two critical modes is shown because of their similarity, whereas both modes associated with the global shearing behavior are shown.}
    \label{fig:3DDispersions_fibra}
\end{figure}

\begin{figure}
    \centering
    \includegraphics[width=0.9\linewidth]{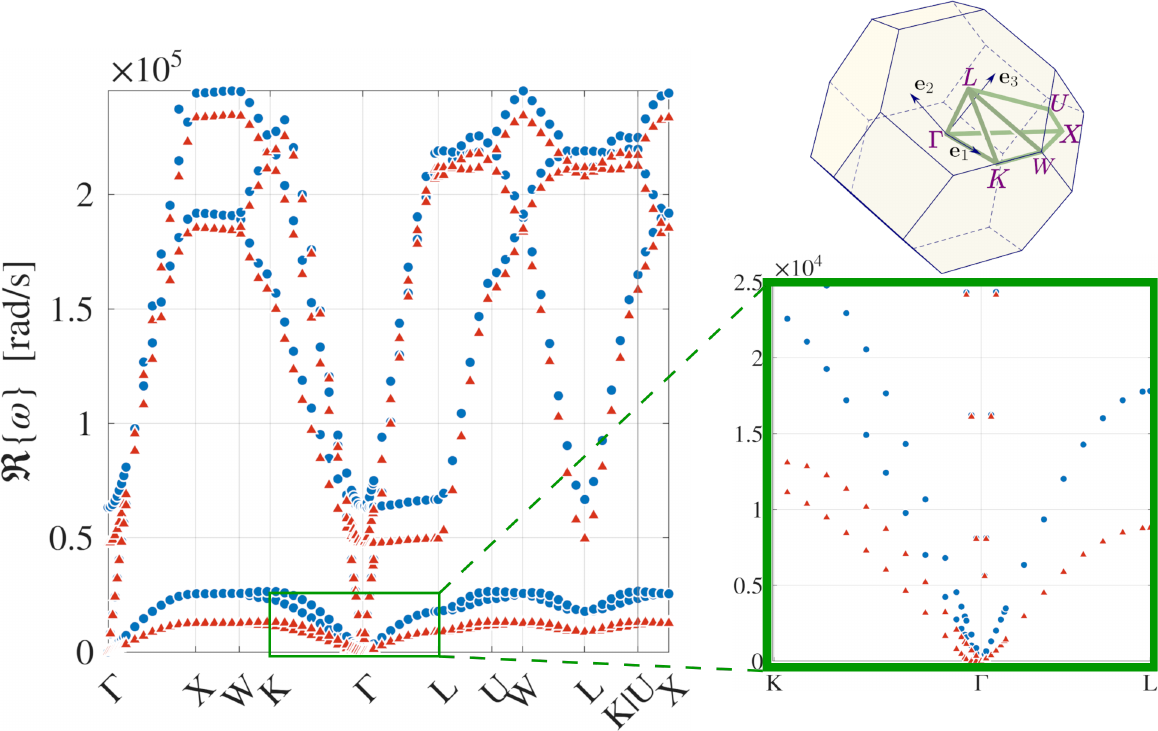}
    \caption{
        Dispersion diagrams for the 3D sheaf-of-beams geometry under different load conditions. The unloaded case (blue) and the onset of global instability (red) are shown. Due to the complexity of the critical mode shapes and their similarity to those of the single-fiber geometry, they are not shown.
    } 
    \label{fig:3DDispersions_fascio}
\end{figure}

Similar considerations apply for the 3D lattices. Their dispersion diagrams are shown in Figure~\ref{fig:3DDispersions_fibra} and \ref{fig:3DDispersions_fascio}.

%
%
%
%
%
%
%
%
%

% Similar considerations apply to the 3D lattices. Their dispersion diagrams are shown in Figures~\ref{fig:3DDispersions_fibra} and \ref{fig:3DDispersions_fascio}.

% \begin{figure}
% \centering
% \includegraphics[width=0.9\linewidth]{Figures/Appendix/Disp3D_fibra.pdf}
% \caption{Dispersion diagrams for the 3D single-fiber geometry under different load conditions. The unloaded case (blue), the onset of local instability (red), and the onset of global instability (green) are shown. The corresponding critical mode shapes are also shown, illustrating the global and local deformation patterns. The mode shapes are represented in an appropriate plane containing the wavevector. At the $X$ and $L$ points, which correspond to local instability, only one of the two critical modes is shown because of their similarity. In contrast, both modes associated with the global shearing behavior are shown. As in the 2D case, the applied load has little influence on the overall mode shapes.}
% \label{fig:3DDispersions_fibra}
% \end{figure}

% \begin{figure}
% \centering
% \includegraphics[width=0.9\linewidth]{Figures/Appendix/Disp3D_fascio.pdf}
% \caption{Dispersion diagrams for the 3D sheaf-of-beams geometry under different load conditions. The unloaded case (blue) and the onset of global instability (red) are shown. Due to the complexity of the mode shapes and their similarity to those of the single-fiber geometry, the corresponding mode shapes are not shown.}
% \label{fig:3DDispersions_fascio}
% \end{figure}

\bibliographystyle{plainnat}
\bibliography{main}

\end{document}